\documentclass[eqsecnum,aps,showpacs,floatfix,superscriptaddress,nofootinbib,twocolumn]{revtex4}
\usepackage{amsmath,amssymb,latexsym}
\usepackage[pdftex]{graphicx}

\begin{document}
\title{Strong-field tidal distortions of rotating black
  holes:\\ Formalism and results for circular, equatorial orbits}
\author{Stephen O'Sullivan}
\affiliation{Department of Physics and MIT Kavli Institute,
  Massachusetts Institute of Technology, Cambridge, MA 02139}
\author{Scott A.\ Hughes}
\affiliation{Department of Physics and MIT Kavli Institute,
  Massachusetts Institute of Technology, Cambridge, MA 02139}
\affiliation{Canadian Institute for Theoretical Astrophysics,
  University of Toronto, 60 St.\ George St., Toronto, ON M5S 3H8,
  Canada}
\affiliation{Perimeter Institute for Theoretical Physics, Waterloo, ON
  N2L 2Y5, Canada}
\begin{abstract}
Tidal coupling between members of a compact binary system can have an
interesting and important influence on that binary's dynamical
inspiral.  Tidal coupling also distorts the binary's members, changing
them (at lowest order) from spheres to ellipsoids.  At least in the
limit of fluid bodies and Newtonian gravity, there are simple
connections between the geometry of the distorted ellipsoid and the
impact of tides on the orbit's evolution.  In this paper, we develop
tools for investigating tidal distortions of rapidly rotating black
holes using techniques that are good for strong-field, fast-motion
binary orbits.  We use black hole perturbation theory, so our results
assume extreme mass ratios.  We develop tools to compute the
distortion to a black hole's curvature for any spin parameter, and for
tidal fields arising from any bound orbit, in the frequency domain.
We also develop tools to visualize the horizon's distortion for black
hole spin $a/M \le \sqrt{3}/2$ (leaving the more complicated $a/M >
\sqrt{3}/2$ case to a future analysis).  We then study how a Kerr
black hole's event horizon is distorted by a small body in a circular,
equatorial orbit.  We find that the connection between the geometry of
tidal distortion and the orbit's evolution is not as simple as in the
Newtonian limit.
\end{abstract}


\maketitle

\section{Introduction}

\subsection{Tidal coupling and binary inspiral}

Tidal coupling in binary inspiral has been a topic of much recent
interest.  A great deal of attention has focused in particular on
systems which contain neutron stars, where tides and their
backreaction on the binary's evolution may allow a new probe of the
equation of state of neutron star matter {\cite{rmsucf2009, hllr2010,
    dnv2012}}.  A great deal of work has been done to rigorously
define the distortion of fluid stars {\cite{dn2009, bp2009}}, the
coupling of the tidal distortion to the binary's orbital energy and
angular momentum {\cite{bdgnr2010}}, and most recently the importance
of nonlinear fluid modes which can be sourced by tidal fields
{\cite{wab2013, vzh2014}}.

Tidal coupling also plays a role in the evolution of binary black
holes.  Indeed, the influence of tidal coupling on binary black holes
has been studied in some detail over the past two decades, but using
rather different language: instead of ``tidal coupling,'' past
literature typically discusses gravitational radiation ``down the
horizon.'' This down-horizon radiation has a dual description in the
tidal deformation of the black hole's event horizon.  A major purpose
of this paper is to explore this dual description, examining
quantitatively how a black hole is deformed by an orbiting companion.

Consider the down-horizon radiation picture first.  The wave equation
governing radiation produced in a black hole spacetime admits two
solutions {\cite{teuk73,tp74}}, one describing outgoing radiation very
far from the hole, and another describing radiation ingoing on the
event horizon.  Both solutions carry energy and angular momentum away
from the binary, and drive (on average) a secular inspiral of the
orbit.  After suitable averaging, we require (for example) the orbital
energy $E_{\rm orb}$ to evolve according to
\begin{equation}
\frac{dE_{\rm orb}}{dt} = -\left(\frac{dE}{dt}\right)^\infty -
\left(\frac{dE}{dt}\right)^{\rm H}\;,
\end{equation}
where $(dE/dt)^\infty$ describes energy carried far away by the waves,
and $(dE/dt)^{\rm H}$ describes energy carried into the event horizon.

The down-horizon flux has an interesting property.  When it is
computed for a small body that is in a circular, equatorial orbit of a
Kerr black hole with mass $M$ and spin parameter $a$, we find that
\begin{equation}
\left(\frac{dE}{dt}\right)^{\rm H} \propto \left(\Omega_{\rm orb} -
\Omega_{\rm H}\right)\;,
\label{eq:horizonflux_prop}
\end{equation}
where $\Omega_{\rm orb} = M^{1/2}/(r^{3/2} + aM^{1/2})$ is the orbital
frequency\footnote{Throughout this paper, we use units with $G = 1 =
  c$.}, and $\Omega_{\rm H} = a/2Mr_+$ is the hole's spin frequency
(Ref.\ {\cite{membrane}}, Sec VIID; see also synopsis in
Sec.\ {\ref{sec:downhoriz}}).  The radius $r_+ = M + \sqrt{M^2 - a^2}$
gives the location of the event horizon in Boyer-Lindquist
coordinates.  We assume that the orbit is prograde, so that the
orbital angular momentum is parallel to the hole's spin angular
momentum.

When $\Omega_{\rm orb} > \Omega_{\rm H}$ (i.e., when the orbit rotates
faster than the black hole spins), we have $(dE/dt)^{\rm H} > 0$ ---
radiation carries energy into the horizon, taking it from the orbital
energy.  This is intuitively sensible, given that an event horizon
generally acts as a sink for energy and matter.  However, when
$\Omega_{\rm orb} < \Omega_{\rm H}$ (the hole spins faster than the
orbit's rotation), we have $(dE/dt)^{\rm H} < 0$.  This means that the
down-horizon component of the radiation {\it augments} the orbital
energy --- energy is transferred from the hole to the orbit.  This is
far more difficult to reconcile with the behavior of an event horizon.

One clue to understanding this behavior is that, when $\Omega_{\rm H}
> \Omega_{\rm orb}$, the modes which contribute to the radiation are
{\it superradiant} {\cite{pt73,chandra}}.  Consider a plane wave which
propagates toward the black hole.  A portion of the wave is absorbed
by the black hole (changing its mass and spin), and a portion is
scattered back out to large radius.  A superradiant mode (see, for
example, Sec.\ 98 of Ref.\ {\cite{chandra}}) is one in which the
scattered wave has higher amplitude than the original ingoing wave.
Some of the black hole's spin angular momentum and rotational energy
has been transferred to the radiation.

\subsection{Tidally distorted strong gravity objects}

Although the condition for superradiance is the same as the condition
under which an orbit gains energy from the black hole, superradiance
does not explain how energy is transferred from the hole to the orbit.
A more satisfying picture of this can be built by invoking the dual
picture of a tidal distortion.  As originally shown by Hartle
{\cite{hartle73,hartle74}}, an event horizon's intrinsic curvature is
distorted by a tidal perturbation.  In analogy with tidal coupling in
fluid systems, the tidally distorted horizon can gravitationally
couple to the orbiting body, transfering energy and angular momentum
from the black hole to the orbit.

Let us examine the fluid analogy in more detail for a moment.
Consider in particular a moon that raises a tide on a fluid body,
distorting its shape from spherical to a prolate ellipsoid.  The tidal
response will produce a bulge that tends to point at the moon.  Due to
the fluid's viscosity, the bulging response will lag the driving tidal
force.  As a consequence, if the moon's orbit is faster than the
body's spin, then the bulge will lag behind.  The bulge will exert a
torque on the orbit that tends to slow down the orbit; the orbit
exerts a torque that tends to speed up the body's spin.  Conversely,
if the spin is faster than the orbit, the bulge will lead the moon's
position, and the torque upon the orbit will tend to speed it up (and
torque from the orbit tends to slow down the spin).  In both cases,
the bulge and moon exert torques on one another in such a way that the
spin and orbit frequencies tend to be equalized\footnote{This is why
  our Moon keeps the same face to the Earth: Tidal coupling has spun
  down the Moon's ``day'' to match its ``year.''  Tidal forces from
  the Moon likewise slow down the Earth's spin, lengthening the day at
  a rate of a few milliseconds per century {\cite{dickeyetal1994}}.
  Given enough time, this effect would drive the Earth to keep the
  same face to the Moon.}.  The action of this torque is such that
energy is taken out of the moon's orbit if the orbit frequency is
larger than the spin frequency, and vice versa.

Since a black hole's shape is changed by tidal forces in a manner
similar to the change in shape of a fluid body, one can imagine that
the horizon's tidal bulge likewise exerts a torque on an orbit.
Examining Eq.\ (\ref{eq:horizonflux_prop}), we see that the sign of
the ``horizon flux'' energy loss is exactly in accord with the tidal
fluid analogy --- energy is lost from the orbit if the orbital
frequency exceeds the black hole's spin frequency, and vice versa.
Using the membrane paradigm {\cite{membrane}}, one can assign a
viscosity to the horizon, making the fluid analogy even more
compelling.

However, as was first noted by Hartle {\cite{hartle73}}, the geometry
of a black hole's tidal bulge behaves in a rather counterintuitive
manner.  At least using a weak-field, slow spin analysis, the bulge
{\it leads} the orbit when $\Omega_{\rm orb} > \Omega_{\rm H}$, and
{\it lags} when $\Omega_{\rm orb} < \Omega_{\rm H}$.  This is opposite
to the geometry which the fluid analogy would lead us to expect.  This
is because an event horizon is a teleological object: Whether an event
in spacetime is inside or outside a horizon depends on that event's
null future.  At some moment in a given time slicing, an event horizon
arranges itself in anticipation of the gravitational stresses it will
be feeling in the future.  This is closely related to the manner in
which the event horizon of a spherical black hole expands outward when
a spherical shell falls into it.  See Ref.\ {\cite{membrane}},
Sec.\ VI\,C\,6 for further discussion.

Much of this background has been extensively discussed in past
literature
{\cite{hartle73,hartle74,membrane,tp08,dl2009,bp2009,pv2010,vpm11}}.
Recent work on this problem has examined in detail how one can
quantify the tidal distortion of a black hole, demonstrating that the
``gravitational Love numbers'' which characterize the distortion of
fluid bodies vanish for non-rotating black holes {\cite{bp2009}}, but
that the geometry's distortion can nonetheless be quantified assuming
particularly useful coordinate systems {\cite{dl2009,pv2010}} and in a
fully covariant manner {\cite{vpm11}}.  Indeed, one can define
``surficial Love numbers,'' which quantify the distortion of a body's
surface, for Schwarzschild black holes {\cite{lp14}}.  These
techniques have been used to study horizon distortion in the
Schwarzschild and slow spin limits, and for slow orbital velocities
{\cite{fl05,dl2009,vpm11}}.

\subsection{Our analysis: Strong-field, rapid spin tidal distortions}

The primary goal of this paper is to develop tools to explore the
distorted geometry of a black hole in a binary which are good for fast
motion, strong field orbits.  We use techniques originally developed
by Hartle {\cite{hartle74}} to compute the Ricci scalar curvature
$R_{\rm H}$ associated with the 2-surface of the distorted horizon;
this is closely related to the intrinsic horizon metric developed in
Ref.\ {\cite{vpm11}}.  We will restrict our binaries to large mass
ratios in order to use the tools of black hole perturbation theory.
We also develop tools to embed the horizon in a 3-dimensional space in
order to visualize the tidal distortions.  In this paper, we restrict
our embeddings to black hole spins $a/M \le \sqrt{3}/2$.  This is the
largest spin at which the horizon can be embedded in a global
Euclidean space; black holes with spins in the range $\sqrt{3}/2 < a/M
\le 1$ must either be embedded in a space that is partially Euclidean,
partially Lorentzian {\cite{smarr}}, or be embedded in another space
altogether {\cite{frolov,gibbons}}.  Although no issue of principle
prevents us for examining larger spins, it does not add very much to
the physics we wish to study here, so we defer embeddings for $a/M >
\sqrt{3}/2$ to a later paper.

A secondary goal of this paper is to investigate whether there is a
simple connection between the geometry of the tidal bulge and the
orbit's evolution.  In particular, we wish to see if the sign of
$dE^{\rm H}/dt$, which is determined by $\Omega_{\rm orb} -
\Omega_{\rm H}$, is connected to the bulge's geometry relative to the
orbit.  This turns out to be somewhat tricky to investigate.  The
orbit and the horizon are at different locations, so we must map the
orbit's position onto the horizon.  There is no unique way to do
this\footnote{Indeed, the behavior of the map depends on the gauge
  used for the calculation, and the time slicing that is used, neither
  of which we investigate in this paper.}, so the results depend at
least in part on how we make the map.  We present two maps from orbit
to horizon.  One, based on ingoing zero-angular momentum light rays,
is useful for comparing with past literature.  The other, based on the
geometry of the horizon's embedding and the orbit at an instant of
constant ingoing time, is useful for describing our numerical data (at
least for small spin).  Another way to characterize the bulge geometry
is to examine the relative phase of the bulge's curvature to the tidal
field which distorts the black hole.  Both of these quantities are
defined at $r = r_+$, so no mapping is necessary.

We find that, at the extremes, the response of a black hole to a
perturbing tide follows Newtonian logic (modulo a swap of ``lag'' and
``lead,'' thanks to the horizon's teleological nature).  In
particular, when $\Omega_{\rm orb} \gg \Omega_{\rm H}$ (so that
$dE^{\rm H}/dt > 0$), the bulge leads the orbit, no matter how we
compare the bulge to the orbit.  When $\Omega_{\rm orb} \ll
\Omega_{\rm H}$ ($dE^{\rm H}/dt < 0$), the bulge lags the orbit.
However, relations between lag, lead, and $dE^{\rm H}/dt$ are not so
clear cut when $\Omega_{\rm orb} \sim \Omega_{\rm H}$.  Consider, in
particular the case $\Omega_{\rm orb} = \Omega_{\rm H}$, for which
$dE^{\rm H}/dt = 0$.  For Newtonian, fluid bodies, the tidal bulge
points directly at the orbiting body in this case, with no exchange of
torque between the body and the orbit.  For black holes, we find no
particular relation between the horizon's bulge and the orbit's
position.  The relation between tidal coupling and tidal distortion is
far more complicated in black hole systems than it is for fluid bodies
in Newtonian gravity --- which is not especially surprising.

Soon after we submitted this paper and posted a preprint to the arXiv,
Cabero and Krishnan posted an analysis of tidally deformed spinning
black holes {\cite{ck14}}.  Although their techniques and analysis
differ quite a bit from ours (focusing on the Bowen-York {\cite{by80}}
initial data set, and using the framework of isolated horizons), their
results seem broadly consistent with ours.  It may be useful in future
work to explore this apparent consistency more closely, and to borrow
some of the tools that they have developed for the systems that we
analyze here.

\subsection{Outline of this paper, units, and conventions}

The remainder of this paper is organized as follows.  Our formalism
for computing the geometry of distorted Kerr black holes is given in
Sec.\ {\ref{sec:formalism}}.  We show how to compute the curvature of
a tidally distorted black hole, and how to quantify the relation of
the geometry of this distortion to the geometry of the orbit which
produces the tidal field.  We also discuss how to compute $dE^{\rm
  H}/dt$, demonstrating that the information which determines this
down-horizon flux is identical to the information which determines the
geometry of the distorted event horizon.

Sections {\ref{sec:schw_results}} and {\ref{sec:kerr_results}} present
results for Schwarzschild and Kerr, respectively.  In both sections,
we first look at the black hole's curvature in a slow motion, slow
spin expansion (slow motion only for Schwarzschild).  This allows us
to develop analytic expressions for the curvature, which are useful
for comparing to the fast motion, rapid spin numerical results that we
then compute.  We visualize tidally distorted black holes by embedding
their horizons in a 3-dimensional space.  This provides a useful way
to see how tides change the shape of a black hole.  In
Sec.\ {\ref{sec:leadlag}}, we examine in some detail whether there is
a simple connection between a black hole's tidally distorted geometry
and the coupling between the hole and the orbit.  In short, the answer
we find is ``no'' --- Newtonian, fluid intuition breaks down for black
holes and strong-field orbits.

Concluding discussion is given in Sec.\ {\ref{sec:conclude}}, followed
by certain lengthy technical details which we relegate to appendices.
Appendix {\ref{app:ethdetails}} describes in detail how to compute
$\bar\eth$, a Newman-Penrose operator which lowers the spin-weight of
quantities needed for our analysis.  Appendix {\ref{app:embed}}
describes how to embed a distorted black hole's event horizon in a
3-dimensional Euclidean space.  As mentioned above, one cannot embed
black holes with $a/M > \sqrt{3}/2$ in Euclidean space, but must use a
either a mixed Euclidean/Lorentzian space {\cite{smarr}}, or something
altogether different {\cite{frolov,gibbons}}.  We will examine the
range $a/M > \sqrt{3}/2$ in a later paper.  Appendix
{\ref{app:spheroidal_lin}} computes, to leading order in spin, the
spheroidal harmonics which are used as basis functions in black hole
perturbation theory.  This is needed for the slow-spin expansions we
present in Sec.\ {\ref{sec:kerr_results}}.  Finally, Appendix
{\ref{app:glossary}} summarizes certain changes in notation that we
have introduced versus previous papers that use black hole
perturbation theory.  These changes synchronize our notation with that
used in the literature from which we have recently adopted our core
numerical method {\cite{ft04,ft05}}.

All of our calculations are done in the background of a Kerr black
hole.  Two coordinate systems, described in detail in
Ref.\ {\cite{poisson}}, are particularly useful for us.  The
Boyer-Lindquist coordinates ($t, r, \theta, \phi$) yield the line
element
\begin{eqnarray}
ds^2 &=& -\left(1 - \frac{2Mr}{\Sigma}\right)dt^2 -
\frac{4Mar\sin^2\theta}{\Sigma}dt\,d\phi + \frac{\Sigma}{\Delta}dr^2
\nonumber\\
&+& \!\! \Sigma\,d\theta^2 + \frac{(r^2 + a^2)^2 -
  a^2\Delta\sin^2\theta}{\Sigma}\sin^2\theta\,d\phi^2\;,
\label{eq:Kerr_BL}
\end{eqnarray}
where
\begin{equation}
\Delta = r^2 - 2Mr + a^2\;,\qquad
\Sigma = r^2 + a^2\cos^2\theta\;.
\label{eq:DeltaSigma}
\end{equation}
The function $\Delta$ has two roots, $r_\pm = M \pm \sqrt{M^2 - a^2}$;
$r_+$ is the location of the event horizon.  We will also often find
it useful to use ingoing coordinates $(v, r', \theta, \psi)$, related
to the Boyer-Lindquist coordinates by {\cite{poisson}}
\begin{eqnarray}
dv &=& dt + \frac{(r^2 + a^2)}{\Delta}\,dr\;,
\label{eq:vdef}
\\
d\psi &=& d\phi + \frac{a}{\Delta}\,dr\;.
\label{eq:psidef}
\\
dr' &=& dr\;,
\label{eq:rpdef}
\end{eqnarray}
These coordinates are well-behaved on the event horizon, and so are
useful tools for describing fields that fall into the hole.  Although
the relation between $r$ and $r'$ is trivial, it can be useful to
distinguish the two as a bookkeeping device when transforming between
the two coordinate systems.  When there is no ambiguity, we will drop
the prime on the ingoing radial coordinate.  The Kerr metric in
ingoing coordinates is given by
\begin{eqnarray}
ds^2 &=& -\left(1 - \frac{2Mr'}{\Sigma}\right)dv^2 + 2dv\,dr' -
2a\sin^2\theta\,dr'\,d\psi
\nonumber\\
&-& \!\! \frac{4Mar'\sin^2\theta}{\Sigma}dv\,d\psi
\nonumber\\
&+&\!\! \Sigma\,d\theta^2 + \frac{[(r')^2 + a^2]^2 -
  a^2\Delta\sin^2\theta}{\Sigma}\sin^2\theta\,d\psi^2\;.
\nonumber\\
\label{eq:Kerr_IN}
\end{eqnarray}
The quantities $\Sigma$ and $\Delta$ here are exactly as in
Eq.\ (\ref{eq:DeltaSigma}), but with $r \to r'$.

It is not difficult to integrate up Eqs.\ (\ref{eq:vdef}) and
(\ref{eq:psidef}) to find
\begin{equation}
v = t + r^*\;,\qquad
\psi = \phi + \bar r\;,
\label{eq:v_and_psi}
\end{equation}
where {\cite{poisson}}
\begin{eqnarray}
r^* &=& r + \frac{Mr_+}{\sqrt{M^2 - a^2}}\ln\left(\frac{r}{r_+} -
1\right)
\nonumber\\
& & \qquad- \frac{Mr_-}{\sqrt{M^2 - a^2}}\ln\left(\frac{r}{r_-} -
1\right)\;,
\label{eq:rstar}\\
\bar r &=& \frac{a}{2\sqrt{M^2 - a^2}}\ln\left(\frac{r - r_+}{r -
  r_-}\right)\;.
\label{eq:rbar}
\end{eqnarray}
Notice that $\psi = \phi$ when $a = 0$.

For $r = r_+ + \delta r$, $\delta r \ll M$,
\begin{equation}
\bar r - \Omega_{\rm H}r^* = K(a) + O(\delta r)\;,
\label{eq:nearhorizradialbehavior}
\end{equation}
where
\begin{eqnarray}
K(a) &=& \frac{a}{2M(Mr_+ - a^2)}\biggl\{a^2 - Mr_+
\nonumber\\
&+& 2M^2{\rm arctanh}\left(\sqrt{1 - a^2/M^2}\right)
\nonumber\\
&+& M\sqrt{M^2 - a^2} \ln\left[\frac{a^2}{4(M^2 - a^2)}\right]\biggr\}
\nonumber\\
&=& -\frac{a}{2M} + \left[1 - 2\ln\left(\frac{a}{2M}\right)\right]
  \left(\frac{a}{2M}\right)^3 + O(a^5)\;.
\nonumber\\
\label{eq:K_of_a}
\end{eqnarray}
This means that, near the horizon, the combination $\bar r -
\Omega_{\rm H}r^*$ cancels out the logarithms in both $r^*$ and $\bar
r$, trending to a constant $K(a)$ that depends only on spin.  The
quantity $K(a)$ plays an important role in setting the phase of tidal
fields on the event horizon.

\section{Formalism}
\label{sec:formalism}

In this section, we develop the formalism we use to study the geometry
of deformed event horizons.  The details of this calculation are
presented in Sec.\ {\ref{sec:geometry}}.  Two pieces of this
calculation are sufficiently involved that we present them separately.
First, in Sec.\ {\ref{sec:ZH}}, we give an overview of how one solves
the radial perturbation equation to find the amplitude that sets the
magnitude of the tidal distortion.  This material has been discussed
at great length in many other papers, so we present just enough detail
to illustrate what is needed for our analysis.  We include in our
discussion the static limit, mode frequency $\omega = 0$.  Since
static modes do not carry energy or angular momentum, they have been
neglected in almost all previous analyses.  However, these modes
affect the shape of a black hole, so they must be included here.
Second, in Sec.\ {\ref{sec:barethbareth}} we provide detailed
discussion of the angular operator $\bar\eth\bar\eth$ and its action
upon the spin-weighted spheroidal harmonic.

Section {\ref{sec:bulge}} describes how we characterize the bulge in
the event horizon which is raised by the orbiting body's tide.  The
bulge is a simple consequence of the geometry, but this discussion
deserves separate treatment in order to properly discuss certain
choices and conventions we must make.  We conclude this section by
briefly reviewing down-horizon fluxes in Sec.\ {\ref{sec:downhoriz}}.
Although this discussion is tangential to our main focus in this
paper, we do this to explicitly show that the deformed geometry and
the down-horizon flux are just different ways of presenting the same
information about the orbiting body's perturbation to the black hole.

\subsection{The geometry of an event horizon}
\label{sec:geometry}

We will characterize the geometry of distorted black holes using the
Ricci scalar curvature $R_{\rm H}$ associated with their event
horizon's 2-surface.  The scalar curvature of an undistorted Kerr
black hole is given by\footnote{Reference {\cite{smarr}} actually
  computes the horizon's Gaussian curvature ${\cal R}_{\rm H}$.  The
  Gaussian curvature ${\cal R}$ of any 2-surface is exactly half that
  surface's scalar curvature $R$, so $R_{\rm H} = 2{\cal R}_{\rm H}$.}
{\cite{smarr}}
\begin{equation}
R_{\rm H} = R^{(0)}_{\rm H} = \frac{2}{r_+^2}\frac{(1 + a^2/r_+^2)(1 -
  3a^2\cos^2\theta/r_+^2)}{(1 + a^2\cos^2\theta/r_+^2)^3}\;.
\label{eq:Kerr_curvature}
\end{equation}
For $a = 0$, $R^{(0)}_{\rm H} = 2/r_+^2$, the standard result for a
sphere of radius $r_+$.  For $a/M \ge \sqrt{3}/2$, $R^{(0)}_{\rm H}$
changes sign near the poles.  This introduces important and
interesting complications to how we represent the tidal distortions of
a rapidly rotating black hole's horizon.

To first order in the mass ratio, tidal distortions leave the horizon
at the coordinate $r = r_+$, but change the scalar curvature on that
surface (at least in all ``horizon-locking gauges'' {\cite{vpm11}},
which we implicitly use in our analysis).  Using the Newman-Penrose
formalism {\cite{np62}}, Hartle {\cite{hartle74}} shows that the
perturbation $R^{(1)}_{\rm H}$ to the curvature is simply related to
the perturbing tidal field $\psi_0$:
\begin{eqnarray}
R^{(1)}_{\rm H} &=& -4\,{\rm Im}\sum_{lmkn}\frac{ \bar\eth\bar\eth
  \psi^{\rm HH}_{0,lmkn}}{p_{mkn}(ip_{mkn} + 2\epsilon)}
\nonumber\\
&\equiv& \sum_{lmkn} R^{(1)}_{{\rm H},lmkn}\;,
\label{eq:horizcurv1}
\end{eqnarray}
with all quantities evaluated at $r = r_+$.  The quantity $\psi^{\rm
  HH}_{0,lmkn}$ is a term in a multipolar and harmonic expansion of
the Newman-Penrose curvature scalar $\psi_0$, computed using the
Hawking-Hartle tetrad {\cite{hh72}}:
\begin{eqnarray}
\psi^{\rm HH}_0 &\equiv& -C_{\alpha\beta\gamma\delta}(l^\alpha)^{\rm HH}
(m^\beta)^{\rm HH} (l^\gamma)^{\rm HH} (m^\delta)^{\rm HH}
\nonumber\\
&=& \sum_{lmkn}\psi_{0,lmkn}^{\rm HH}\;.
\label{eq:psi0_def}
\end{eqnarray}
The tensor $C_{\alpha\beta\gamma\delta}$ is the Weyl curvature, and
the vectors $(l^\alpha)^{\rm HH}$ and $(m^\alpha)^{\rm HH}$ are
Newman-Penrose tetrad legs in the Hawking-Hartle representation.  See
Appendix {\ref{app:ethdetails}} for detailed discussion of this tetrad
and related quantities.

We assume that $\psi_0$ arises from an object in a bound orbit of the
Kerr black hole.  This object's motion can be described using the
three fundamental frequencies associated with such orbits: an axial
frequency $\Omega_\phi$, a polar frequency $\Omega_\theta$, and a
radial frequency $\Omega_r$.  The indices $m$, $k$, and $n$ label
harmonics of these frequencies:
\begin{equation}
\omega_{mkn} = m\Omega_\phi + k\Omega_\theta + n\Omega_r\;.
\end{equation}
The index $l$ labels a spheroidal harmonic mode, and is discussed in
more detail below.  The remaining quantities appearing in
Eq.\ (\ref{eq:horizcurv1}) are the wavenumber for ingoing
radiation\footnote{This wavenumber is often written $k$ in the
  literature; we use $p$ to avoid confusion with harmonics of the
  $\theta$ frequency.}
\begin{equation}
p_{mkn} = \omega_{mkn} - m\Omega_{\rm H}\;,
\label{eq:pmkndef}
\end{equation}
and
\begin{equation}
\epsilon = \frac{\sqrt{M^2 - a^2}}{4Mr_+} \equiv \frac{\kappa}{2}\;.
\label{eq:epsilondef}
\end{equation}
The quantity $\kappa$ is the Kerr surface gravity.  We will find this
interpretation of $\epsilon$ to be useful when discussing the geometry
of the horizon's tidal distortion.  We discuss the operator
$\bar\eth\bar\eth$ in detail in Sec.\ {\ref{sec:barethbareth}}.  For
now, note that it involves derivatives with respect to $\theta$.

The calculation of $R^{(1)}_{\rm H}$ involves several computations
that use the Newman-Penrose derivative operator $D \equiv
l^\alpha\partial_\alpha$.  Using the Hawking-Hartle form of $l^\alpha$
and ingoing Kerr coordinates (see Appendix {\ref{app:ethdetails}}), we
find that
\begin{equation}
D \to \frac{\partial}{\partial v} + \Omega_{\rm
  H}\frac{\partial}{\partial\psi}
\end{equation}
as $r \to r_+$.  The fields to which we apply this operator have the
form $e^{i(m\psi - \omega_{mkn} v)}$ near the horizon, so
\begin{equation}
D{\cal F} = i(m\Omega_{\rm H} - \omega_{mkn}){\cal F} = -ip_{mkn}{\cal
  F}
\end{equation}
for all relevant fields ${\cal F}$.  Hartle choses a time coordinate
$t$ such that $D \equiv \partial/\partial t$ near the horizon,
effectively working in a frame that corotates with the black hole.  As
a consequence, his Eq.\ (2.21) [equivalent to our
  Eq.\ (\ref{eq:horizcurv1})] has $\omega$ in place of $p$.  Hartle's
(2.21) also corresponds to a single Fourier mode, and so is not summed
over indices.

The Hawking-Hartle tetrad is used in Eq.\ (\ref{eq:psi0_def}) because
it is well behaved on the black hole's event horizon {\cite{hh72}}.
In many discussions of black hole perturbation theory based on the
Teukolsky equation, we instead use the Kinnersley tetrad, which is
well designed to describe distant radiation
{\cite{teuk73,kinnersley}}.  The Kinnersley tetrad is described
explicitly in Appendix {\ref{app:ethdetails}}.  The relation between
$\psi_0$ in these two tetrads is [cf.\ Ref.\ {\cite{tp74}},
  Eq.\ (4.43)]
\begin{equation}
\psi_0^{\rm HH} = \frac{\Delta^2}{4(r^2 + a^2)^2} \psi_0^{\rm K}\;.
\label{eq:psi0_convert}
\end{equation}
Further, we know that $\psi_0^{\rm K}$ on the horizon can be written
{\cite{tp74}}
\begin{equation}
\psi_{0,lmkn}^{\rm K} = \frac{W^{\rm
    H}_{lmkn}\,{_{+2}S}_{lm}(\theta;a\omega_{mkn})}
    {\Delta^2}e^{i(m\phi - \omega_{mkn} t - p_{mkn} r^*)}\;.
\label{eq:psi0_K}
\end{equation}
We have introduced $W^{\rm H}_{lmkn}$, a complex
amplitude\footnote{This amplitude is written $Y$ rather than $W$ in
  Ref.\ {\cite{tp74}}; we have changed notation to avoid confusion
  with the spherical harmonic.} which we will discuss in more detail
below, as well as the spheroidal harmonic of spin-weight $+2$,
${_{+2}S}_{lm}(\theta;a\omega_{mkn})$.  Spheroidal harmonics are often
used in black hole perturbation theory, since the equations governing
a field of spin-weight $s$ in a black hole spacetime separate when
these harmonics are used as a basis for the $\theta$ dependence.  In
the limit $a\omega_{mkn} \to 0$, they reduce to the spin-weighted
spherical harmonics:
\begin{equation}
{_sS}_{lm}(\theta;a\omega_{mkn}) \to {_sY}_{lm}(\theta) \quad\mbox{as}\quad
  a\omega_{mkn} \to 0\;.
\end{equation}
${_sY}_{lm}(\theta)$ denotes the spherical harmonic without the axial
dependence: ${_s}Y_{lm}(\theta,\phi) = {_s}Y_{lm}(\theta)e^{im\phi}$.
In what follows, we will abbreviate:
\begin{equation}
{_{+2}S}_{lm}(\theta;a\omega_{mkn}) \equiv S^+_{lmkn}(\theta)\;.
\end{equation}
We will likewise write the spin-weight $-2$ spheroidal harmonic as
$S^{-}_{lmkn}(\theta)$.

Combining Eqs.\ (\ref{eq:psi0_convert}) and (\ref{eq:psi0_K}), we find
\begin{equation}
\psi^{\rm HH}_{0,lmkn} = \frac{W^{\rm H}_{lmkn}
  S^+_{lmkn}(\theta)}{4(r^2 + a^2)^2} \, e^{i(m\phi - \omega_{mkn} t -
  p_{mkn}r^*)}\;.
\label{eq:psi0HHexpand}
\end{equation}
Using Eqs.\ (\ref{eq:v_and_psi}) and (\ref{eq:pmkndef}), we can
rewrite the phase factor using coordinates that are well-behaved on
the horizon:
\begin{eqnarray}
m\phi - \omega_{mkn}t - p_{mkn}r^*\! &=& m(\psi - \bar r)
- \omega_{mkn}(v - r^*)
\nonumber\\
& & - (\omega_{mkn} - m\Omega_{\rm H})r^*
\nonumber\\
&=& m\psi - \omega_{mkn} v - m(\bar r - \Omega_{\rm H}r^*)\;.
\nonumber\\
\end{eqnarray}
Taking the limit $r \to r_+$ and using
Eq.\ (\ref{eq:nearhorizradialbehavior}), we find
\begin{equation}
\psi^{\rm HH}_{0,lmkn} = \frac{W^{\rm
    H}_{lmkn}S^+_{lmkn}(\theta)}{16M^2r_+^2}e^{i\Phi_{mkn}(v,\psi)}\;,
\label{eq:psi0HHexpand2}
\end{equation}
where
\begin{equation}
\Phi_{mkn}(v,\psi) = m\psi - \omega_{mkn}v - m K(a)\;,
\label{eq:Phi_mkn}
\end{equation}
with $K(a)$ defined in Eq.\ (\ref{eq:K_of_a}).  We finally find
\begin{eqnarray}
R^{(1)}_{{\rm H},lmkn} = -{\rm Im}\left[\frac{W^{\rm H}_{lmkn}
    e^{i\Phi_{mkn}(v,\psi)}\bar\eth\bar\eth
    S^+_{lmkn}(\theta)}{4M^2r_+^2p_{mkn}(ip_{mkn} +
    2\epsilon)}\right]\;.  \nonumber\\
\label{eq:gauss2}
\end{eqnarray}

We will use a Teukolsky equation solver
{\cite{h2000,dh2006,thd_inprep}} which computes the curvature scalar
$\psi_4$ rather than $\psi_0$.  Although $\psi_4$ is usually used to
study radiation far from the black hole, one can construct $\psi_0$
from it using the Starobinsky-Churilov identities {\cite{tp74,sc73}}.
In the limit $r \to r_+$,
\begin{eqnarray}
\psi_4 &=& \frac{\Delta^2}{(r - ia\cos\theta)^4} \sum_{lmkn}
Z^{\rm H}_{lmkn} S^-_{lmkn}(\theta)
\nonumber\\
& & \qquad\qquad\times\,e^{i(m\phi - \omega_{mkn} t - p_{mkn}r^*)}\;.
\end{eqnarray}
We briefly summarize how we compute $Z^{\rm H}_{lmkn}$ in
Sec.\ {\ref{sec:ZH}}.  Using the Starobinsky-Churilov identities, we
find that $Z^{\rm H}_{lmkn}$ and $W^{\rm H}_{lmkn}$ are related by
\begin{equation}
W^{\rm H}_{lmkn} = \beta_{lmkn} Z^{\rm H}_{lmkn}\;,
\label{eq:WlmknZHlmkn}
\end{equation}
where
\begin{equation}
\beta_{lmkn} = \frac{64(2Mr_+)^4p_{mkn}(p_{mkn}^2 + 4\epsilon^2)(p_{mkn} +
  4i\epsilon)}{c_{lmkn}}\;,
\label{eq:betalmkn}
\end{equation}
and where the complex number $c_{lmkn}$ is given by
\begin{eqnarray}
|c_{lmkn}|^2 &=& \left\{\left[(\lambda + 2)^2 + 4ma\omega_{mkn} -
    4a^2\omega_{mkn}^2\right]\right.
\nonumber\\
& & \left. \qquad\times \left(\lambda^2 + 36ma\omega_{mkn} -
  36a^2\omega_{mkn}^2\right)\right.
\nonumber\\
& &\left. \qquad +\,(2\lambda + 3)(96a^2\omega_{mkn}^2 -
  48ma\omega_{mkn})\right\}
\nonumber\\
& & + 144\omega_{mkn}^2(M^2 - a^2)\;,
\label{eq:clmknsqr}
\\
{\rm Im}\,c_{lmkn} &=& 12 M \omega_{mkn}\;,
\\
{\rm Re}\,c_{lmkn} &=& +\sqrt{|c_{lmkn}|^2 - 144M^2\omega_{mkn}^2}\;.
\end{eqnarray}
The real number $\lambda$ appearing here is
\begin{equation}
\lambda = {\cal E}_{lmkn} - 2 a m \omega_{mkn} + a^2\omega_{mkn}^2 - 2\;,
\end{equation}
with ${\cal E}_{lmkn}$ the eigenvalue of $S^-_{lmkn}(\theta)$.  In the
limit $a\omega_{mkn} \to 0$, ${\cal E}_{lmkn} \to l(l+1)$.  For our
later weak-field expansion, it will be useful to have $\lambda$ as an
expansion in $a\omega_{mkn}$.  See Appendix {\ref{app:spheroidal_lin}}
for discussion of this.

Using these results, we can write the tidal distortion of the
horizon's curvature as
\begin{eqnarray}
R^{(1)}_{{\rm H},lmkn} &=& -{\rm Im}\left[
  \frac{\beta_{lmkn}Z^{\rm H}_{lmkn}
    e^{i\Phi_{mkn}(v,\psi)}\bar\eth\bar\eth S^+_{lmkn}(\theta)}
       {4M^2r_+^2p_{mkn}(ip_{mkn} + 2\epsilon)}\right]
\nonumber\\
&\equiv& {\rm Im}\left[{\cal C}_{lmkn}Z^{\rm H}_{lmkn}e^{i \Phi_{mkn}(v,\psi)}
\bar\eth\bar\eth S^+_{lmkn}(\theta)\right]\;, \nonumber\\
\label{eq:gauss_kerr}
\end{eqnarray}
where
\begin{equation}
{\cal C}_{lmkn} = 256M^2r_+^2 c_{lmkn}^{-1}(p_{mkn} + 4i\epsilon)(ip_{mkn} -
  2\epsilon)\;.
\label{eq:Clmkn}
\end{equation}
Equation (\ref{eq:gauss_kerr}) is the workhorse of our analysis.  We
use a slightly modified version of the code described in Refs.
{\cite{h2000,dh2006,thd_inprep}} to compute the complex numbers
$Z^{\rm H}_{lmkn}$ and the angular function $\bar\eth\bar\eth
S^{+}_{lmkn}$.  We briefly describe these calculations in the next two
subsections.

\subsection{Computing $Z^{\rm H}_{lmkn}$}
\label{sec:ZH}

Techniques for computing the amplitude $Z^{\rm H}_{lmkn}$ have been
discussed in great detail in other papers, so our discussion here will
be very brief; our analysis follows that given in
Ref.\ {\cite{dh2006}}.  The major change versus previous works is that
we need the solution for static modes ($\omega = 0$).  Our goal here
is to present enough detail to see how earlier studies can be modified
fairly simply to include these modes.  It is worth noting that we have
changed notation from that used in previous papers by our group in
order to more closely follow the notation of Fujita and Tagoshi
{\cite{ft04,ft05}}.  Appendix {\ref{app:glossary}} summarizes these
changes.

The complex number $Z^{\rm H}_{lmkn}$ is the amplitude of solutions to
the Teukolsky equation for spin-weight $s = -2$, so we begin there:
\begin{equation}
\Delta^2\frac{d}{dr}\left(\frac{dR_{lm\omega}}{dr}\right) -
V_{lm}(r)R_{lm\omega} = {\cal T}_{lm\omega}(r)\;.
\label{eq:teuk}
\end{equation}
This is the frequency-domain version of this equation, following the
introduction of a modal and harmonic decomposition which separates the
original time-domain equation; see {\cite{teuk73}} for further
details.  The potential $V_{lm}$ is discussed in Sec.\ IIIA of
Ref.\ {\cite{dh2006}}; the source term ${\cal T}_{lm\omega}$ is
discussed in Sec.\ IIIB of that paper.

Equation (\ref{eq:teuk}) has two homogeneous solutions relevant to our
analysis: The ``in'' solution is purely ingoing on the horizon, but is
a mixture of ingoing and outgoing at future null infinity; the ``up''
solution is purely outgoing at future null infinity, but is a mixture
of ingoing and outgoing on the horizon.  We discuss these solutions in
more detail below.  For now, it is enough that these solutions allow
us to build a Green's function {\cite{poisson93}},
\begin{eqnarray}
G(r|r') &=& \frac{1}{{\cal W}}R^{\rm up}_{lm\omega}(r)R^{\rm
  in}_{lm\omega}(r')\;, \quad r' < r\;,
\nonumber\\
&=& \frac{1}{{\cal W}}R^{\rm in}_{lm\omega}(r)R^{\rm
  up}_{lm\omega}(r')\;, \quad r' > r\;,
\label{eq:Green}
\end{eqnarray}
where
\begin{equation}
{\cal W} =
\frac{1}{\Delta}\left[R^{\rm in}_{lm\omega}\frac{dR^{\rm up}_{lm\omega}}{dr}
- R^{\rm up}_{lm\omega}\frac{dR^{\rm in}_{lm\omega}}{dr}\right]
\end{equation}
is the equation's Wronskian.  This is then integrated against the
source to build the general inhomogeneous solution:
\begin{eqnarray}
R_{lm\omega}(r) &=& \int_{r_+}^\infty G(r|r'){\cal T}_{lm\omega}(r')dr'
\nonumber\\
&\equiv& Z^{\rm in}_{lm\omega}(r)R^{\rm up}_{lm\omega}(r) + 
Z^{\rm up}_{lm\omega}(r)R^{\rm in}_{lm\omega}(r)\;.
\nonumber\\
\label{eq:inhomogeneous}
\end{eqnarray}
We have defined
\begin{eqnarray}
Z^{\rm in}_{lm\omega}(r) &=& \frac{1}{\cal
  W}\int_{r_+}^r\frac{R^{\rm in}_{lm\omega}(r'){\cal
    T}_{lm\omega}(r')}{\Delta(r')^2}dr'\;,
\label{eq:Zin1}
\\
Z^{\rm up}_{lm\omega}(r) &=& \frac{1}{\cal
  W}\int_r^\infty\frac{R^{\rm up}_{lm\omega}(r'){\cal
    T}_{lm\omega}(r')}{\Delta(r')^2}dr'\;.
\label{eq:Zup1}
\end{eqnarray}

A key property of ${\cal T}_{lm\omega}$ is that it is the sum of three
terms, one proportional to $\delta[r - r_{\rm orb}(t)]$, one
proportional to $\delta'[r - r_{\rm orb}(t)]$, and one proportional to
$\delta''[r - r_{\rm orb}(t)]$ (where $'$ denotes $d/dr$).  Putting
this into Eqs.\ (\ref{eq:Zin1}) and (\ref{eq:Zup1}), we find that
\begin{eqnarray}
Z^\star_{lm\omega}(r) &=& \frac{1}{\cal W}\Biggl\{ {\cal
  I}^0_{lm\omega}\left[R^\star_{lm\omega}(r)\right] + {\cal
  I}^1_{lm\omega}\left[\frac{dR^\star_{lm\omega}}{dr}\biggr|_r\right]
\nonumber\\
& & + {\cal I}^2_{lm\omega}\left[\frac{d^2R^\star_{lm\omega}}{dr^2}\biggr|_r
\right]\Biggr\}\;,
\label{eq:Zstargeneric}
\end{eqnarray}
(where $\star$ can stand for ``up'' or ``in'').  The factors ${\cal
  I}^{0,1,2}_{lm\omega}$ are operators which act on $R^\star_{lm\omega}$
and its derivatives.  These operators integrate over the $r$ and
$\theta$ motion of the orbiting body.


In this analysis, we are concerned with the solution of the
perturbation equation on the event horizon, so we want $R_{lm\omega}$
as $r \to r_+$.  In this limit, $Z^{\rm in}_{lm\omega} = 0$.  We
define
\begin{equation}
Z^{\rm H}_{lm\omega} \equiv Z^{\rm up}_{lm\omega}(r_+)\;.
\end{equation}
For a source term corresponding to a small body in a bound Kerr orbit,
we find that Eq.\ (\ref{eq:Zstargeneric}) has the form
\begin{equation}
Z^{\rm H}_{lm\omega} = \sum_{kn}Z^{\rm H}_{lmkn}\delta(\omega - \omega_{mkn})\;.
\end{equation}
It is then not difficult to read off $Z^{\rm H}_{lmkn}$.  See
Ref.\ {\cite{dh2006}} for detailed discussion of how to evaluate
Eq.\ (\ref{eq:Zstargeneric}) and read off these amplitudes.

Key to computing $Z^{\rm H}_{lmkn}$ is computing the homogeneous
solutions $R^{\rm up}_{lm\omega}(r)$, $R^{\rm in}_{lm\omega}(r)$, and
their derivatives.  Our methods for doing this depend on whether
$\omega_{mkn}$ is zero or not.

\subsubsection{$\omega_{mkn} \ne 0$}

The homogeneous solutions for $\omega_{mkn} \ne 0$ have been amply
discussed in the literature; our analysis is based on that of
Ref.\ {\cite{dh2006}}.  In brief, the two homogeneous solutions of
Eq.\ (\ref{eq:teuk}) have the following asymptotic behavior:
\begin{eqnarray}
R^{\rm in}_{lm\omega}(r \to r_+) &=& B^{\rm
  trans}_{lm\omega}\Delta^2e^{-ipr^*}\;,
\label{eq:Rinr+}
\\
R^{\rm in}_{lm\omega}(r \to \infty) &=& B^{\rm
  ref}_{lm\omega}r^3e^{i\omega r^*} + \frac{B^{\rm
    inc}_{lm\omega}}{r}e^{-i\omega r^*}\;;
\nonumber\\
\label{eq:Rininf}
\\
R^{\rm up}_{lm\omega}(r \to r_+) &=& C^{\rm up}_{lm\omega}e^{ipr^*}
+ C^{\rm ref}_{lm\omega}\Delta^2e^{-ip r^*}\;,
\nonumber\\
\label{eq:Rupr+}\\
R^{\rm up}_{lm\omega}(r \to \infty) &=& C^{\rm
  trans}_{lm\omega}r^3e^{i\omega r^*}\;.
\label{eq:Rupinf}
\end{eqnarray}
These asymptotic solutions yield the Wronskian:
\begin{equation}
{\cal W} = 2i\omega B^{\rm inc}_{lm\omega}C^{\rm trans}_{lm\omega}\;.
\end{equation}
An effective algorithm for computing all of the quantities which we
need is described by Fujita and Tagoshi {\cite{st03,ft04,ft05}}.  It
is based on expanding the solution in a basis of hypergeometric and
Coulomb wave functions, with the coefficients of the expansion
determined by solving a recurrence relation; see Secs.\ 4.2 -- 4.4 of
Ref.\ {\cite{st03}} for detailed discussion.  We use a code based on
these methods {\cite{thd_inprep}} for all of our $\omega_{mkn} \ne 0$
calculations; the analytic limits we present in
Secs.\ {\ref{sec:schw_analytic}} and {\ref{sec:kerr_analytic}} are
also based on these methods.

\subsubsection{$\omega_{mkn} = 0$}

Static modes have been neglected in much past work.  They do not carry
any energy or angular momentum, and so are not important for many
applications.  These modes do play a role in setting the shape of the
distorted event horizon, however, and must be included here.

It turns out that homogeneous solutions for $\omega_{mkn} = 0$ are
available as surprisingly simple closed form expressions.  Teukolsky's
Ph.D.\ thesis {\cite{teukphd}} presents two solutions that satisfy
appropriate boundary conditions.  Defining
\begin{equation}
x = \frac{r - r_+}{r_+ - r_-}\;,\quad\gamma = \frac{iam}{r_+ - r_-}\;,
\end{equation}
the two solutions of the radial Teukolsky equation for $s = -2$ are
\begin{eqnarray}
R^{\rm in}_{lm0}(r) &=& (r_+ - r_-)^4 x^2(1+x)^2\left(\frac{x}{1 +
  x}\right)^\gamma\,\times
\nonumber\\
& & _2F_1(2-l, l+3; 3 + 2\gamma, -x)\;,
\label{eq:Rin_omega0}\\
R^{\rm up}_{lm0}(r) &=& (r_+ - r_-)^{(1 - l)}x^{(1 - l)}(1 +
1/x)^{(2-\gamma)}\,\times
\nonumber\\
& & _2F_1(l+3,l+1-2\gamma;2l+2,-1/x)\;.
\nonumber\\
\label{eq:Rup_omega0}
\end{eqnarray}
In these equations, $_2F_1(a,b;c,x)$ is the hypergeometric function.
These solutions satisfy regularity conditions at infinity and on the
horizon: $R^{\rm in}_{lm0}(r\to r_+) \propto \Delta^2$, and $R^{\rm
  up}_{lm0}(r\to\infty)\propto 1/r^{l-1}$ \cite{teukphd}.  We have
introduced powers of $r_+ - r_-$ to insure that we have the correct
asymptotic behavior in $r$, rather than in the dimensionless variable
$x$.  The Wronskian corresponding to these solutions is
\begin{equation}
{\cal W} = -\frac{(2l+1)!}{(l+2)!}\frac{\Gamma(3 + 2\gamma)}{\Gamma(l
  + 1 + 2\gamma)}(r_+ - r_-)^{(2 - l)}\;.
\label{eq:Wronskian_omega0}
\end{equation}
Using Eqs.\ (\ref{eq:Rin_omega0}), (\ref{eq:Rup_omega0}), and
(\ref{eq:Wronskian_omega0}), it is simple to adapt existing codes to
compute $Z^{\rm H}_{lmkn}$ for $\omega_{mkn} = 0$.

The results we present in Secs.\ {\ref{sec:schw_results}} and
{\ref{sec:kerr_results}} will focus on circular, equatorial orbits,
for which $k = n = 0$.  The zero-frequency modes in this limit have $m
= 0$, for which $\gamma = 0$.  The Wronskian simplifies further:
\begin{equation}
{\cal W}_{(m = 0)} = -\frac{2(2l+1)!}{l!(l+2)!}(r_+ - r_-)^{(2 -
  l)}\;.
\end{equation}
For generic orbit geometries, there will exist cases that have
$\omega_{mkn} = 0$ with $m \ne 0$, akin to the ``resonant'' orbits
studied at length in Refs.\ {\cite{fh12,fhr14}}.  We defer discussion
of this possibility to a later analysis which will go beyond circular
and equatorial orbits.

\subsection{The operator $\bar\eth\bar\eth$}
\label{sec:barethbareth}

The operator $\bar\eth$, when acting on a quantity $\eta$ of
spin-weight $s$, takes the following form:
\begin{equation}
\bar\eth\eta = \left[\bar\delta - (\alpha - \bar\beta)\right]\eta\;;
\end{equation}
$\bar\eth\eta$ is then a quantity of spin-weight $s-1$.  The
quantities $\alpha$ and $\beta$ are both Newman-Penrose spin
coefficients, and $\bar\delta$ is a Newman-Penrose derivative
operator.  These quantities are all related to the tetrad legs ${\bf
  m}$, $\bar{\bf m}$:
\begin{eqnarray}
\bar\delta &=& \bar m^\mu\partial_\mu\;,
\\
\alpha - \bar\beta &=& \frac{1}{2}\bar m^\nu\left(\bar m^\mu\nabla_\nu
m_\mu - m^\mu\nabla_\nu \bar m_\mu\right)\;.
\end{eqnarray}
We do this calculation using the Hawking-Hartle tetrad; details are
given in Appendix {\ref{app:ethdetails}}.  The result for general
black hole spin $a$ is
\begin{eqnarray}
\bar\eth\eta &=& \frac{1}{\sqrt{2}(r_+ - ia\cos\theta)} \Biggl(L^s_- -
  am\Omega_{\rm H}\sin\theta
\nonumber\\
& &\qquad\qquad\qquad\quad - \frac{isa\sin\theta}{r_+ -
    ia\cos\theta}\Biggr)\eta\;.
\label{eq:kerr_eth1}
\end{eqnarray}
The operator\footnote{This operator is denoted $\bar\eth_0$ in
  Ref.\ {\cite{hartle74}}.  We will use the symbol $\bar\eth_0$ to
  instead denote the Schwarzschild limit of $\bar\eth$.} $L^s_-$
lowers the spin-weight of the spherical harmonics by 1:
\begin{eqnarray}
L^s_- {_sY}_{lm} &=& \left(\partial_\theta + s\cot\theta +
m\csc\theta\right) {_sY}_{lm}
\nonumber\\
&=& \sqrt{(l+s)(l-s+1)}
{_{s-1}Y}_{lm}\;.
\label{eq:sphericalharmoniclower}
\end{eqnarray}
In a few places, we will need to evaluate
$L^s_-\left[\cos\theta\eta\right]$ and
$L^s_-\left[\sin\theta\eta\right]$.  This requires that we rewrite
$\cos\theta$ and $\sin\theta$ in a form that properly indicates their
spin weight.  We treat $\cos\theta$ as spin-weight zero, writing
\begin{equation}
\cos\theta = \sqrt{\frac{4\pi}{3}}\, {_0}Y_{10}\;.
\end{equation}
Likewise, we treat $\sin\theta$ as spin-weight $-1$, writing
\begin{equation}
\sin\theta = -\sqrt{\frac{8\pi}{3}}\, {_{-1}}Y_{10}\;.
\end{equation}
This accounts for the fact that $\sin\theta$ always appears in our
calculation inside operators that lower spin-weight.

With this, we find the following identities:
\begin{eqnarray}
L^s_-\left[\cos\theta\eta\right] &=&
\sqrt{\frac{4\pi}{3}} L^s_-\left[{_0}Y_{10}\,\eta\right]
\nonumber\\
&=& \sqrt{\frac{4\pi}{3}}\left({_0}Y_{10}\,L^s_-\eta + 
\eta\, L^s_-\,{_0}Y_{10}\right)
\nonumber\\
&=& \sqrt{\frac{4\pi}{3}}\left({_0}Y_{10}\,L^s_-\eta + 
\eta\,\sqrt{2}\,{_{-1}}Y_{10}\right)
\nonumber\\
&=&  \cos\theta\,L^s_-\eta - \sin\theta\eta\;;
\label{eq:Lminus_s_costheta}
\end{eqnarray}
\begin{eqnarray}
L^s_-\left[\sin\theta\eta\right] &=&
-\sqrt{\frac{8\pi}{3}} L^s_-\left[_{-1}Y_{10}\eta\right]
\nonumber\\
&=& -\sqrt{\frac{8\pi}{3}}\left(_{-1}Y_{10}\,L^s_-\eta + 
\eta\,L^s_-\,{_{-1}}Y_{10}\right)
\nonumber\\
&=& -\sqrt{\frac{8\pi}{3}}\,{_{-1}Y}_{10}L^s_-\,\eta
\nonumber\\
&=&  \sin\theta\,L^s_-\eta\;.
\label{eq:Lminus_s_sintheta}
\end{eqnarray}
We used the fact that $L^s_-$ applied to ${_{-1}Y}_{10}$ yields zero.

Using these results, it follows that
\begin{eqnarray}
L^s_-\left(1 - \frac{ia\cos\theta}{r_+}\right)^{-s}\eta &=& \left(1 -
\frac{ia\cos\theta}{r_+}\right)^{-s}
\nonumber\\
& &
\!\!\!\!\!\!\!\!
\!\!\!\!\!\!\!\!
\!\!\!\!\times
\left(L^s_- - \frac{ias\sin\theta}{r_+ -
    ia\cos\theta}\right)\eta\;.
\nonumber\\
\label{eq:usefulidentity}
\end{eqnarray}
We can next rewrite Eq.\ (\ref{eq:kerr_eth1}) as
\begin{eqnarray}
\bar\eth\eta &=& \frac{1}{\sqrt{2}r_+}\left(1 -
\frac{ia\cos\theta}{r_+}\right)^{s-1}
\nonumber\\
& &
\times\left(L^s_- -
am\Omega_{\rm H}\sin\theta\right)\left(1 -
\frac{ia\cos\theta}{r_+}\right)^{-s}\eta\;.
\nonumber\\
\label{eq:kerr_eth}
\end{eqnarray}
When $a = 0$, this reduces to
\begin{equation}
\bar\eth\eta = \frac{1}{2\sqrt{2}M}L^s_-\eta \equiv \bar\eth_0\;.
\end{equation}
When $\eta$ is of spin-weight 2, Eq.\ (\ref{eq:kerr_eth}) tells us that
\begin{equation}
\bar\eth\bar\eth\eta = \frac{1}{2r_+^2} \left(L^s_- -
am\Omega_{\rm H}\sin\theta\right)^2\left(1 -
\frac{ia\cos\theta}{r_+}\right)^{-2}\eta\;.
\label{eq:barethsqr_s=2}
\end{equation}
For $a \ll M$, Eq.\ (\ref{eq:barethsqr_s=2}) reduces to
\begin{equation}
\bar\eth\bar\eth\eta = \frac{1}{8M^2}\,L^s_-L^s_-\left(1 +
\frac{ia\cos\theta}{M}\right)\eta\;,
\label{eq:barethbareth_small_a}
\end{equation}
which reproduces Eq.\ (4.19) of Ref.\ {\cite{hartle74}}.

We will apply $\bar\eth\bar\eth$ to the spheroidal harmonic
$S^{+}_{lm}(\theta)$.  Following Ref.\ {\cite{h2000}}, we compute this
function by expanding it using a basis of spherical harmonics, writing
\begin{equation}
S^+_{lm}(\theta) = \sum_{q = q_{\rm min}}^\infty b^l_q(a\omega_{mkn})
_{+2}Y_{qm}(\theta)\;,
\label{eq:spheroidexpand}
\end{equation}
where $q_{\rm min} = {\rm min}(2,|m|)$.  Efficient algorithms exist to
compute the expansion coefficients $b^l_q(a\omega_{mkn})$
(cf.\ Appendix A of Ref.\ {\cite{h2000}}).  Expanding
Eq.\ (\ref{eq:barethsqr_s=2}) puts it into a form very useful for our
purposes:
\begin{equation}
\bar\eth\bar\eth\eta = \frac{1}{2(r_+ -
  ia\cos\theta)^2}\left[L^s_-L^s_- + {\cal A}_1L^s_- + {\cal
    A}_2\right]\eta\;,
\label{eq:barethsqr_s=2_expand}
\end{equation}
where
\begin{eqnarray}
{\cal A}_1 &=& -2a\sin\theta\left[m\Omega_{\rm H} + \frac{2i}{r_+ -
  ia\cos\theta}\right]\;,
\label{eq:barethbareth_A1}\\
{\cal A}_2 &=& a^2\sin^2\theta\biggl[m^2\Omega^2_{\rm H}
+ \frac{4im\Omega_{\rm H}}{r_+ - ia\cos\theta}
\nonumber\\
& &\qquad\qquad\qquad
-\frac{6}{(r_+ - ia\cos\theta)^2}\biggr]\;.
\label{eq:barethbareth_A2}
\end{eqnarray}
Combining Eqs.\ (\ref{eq:spheroidexpand}) and
(\ref{eq:barethsqr_s=2_expand}), and making use of
Eq.\ (\ref{eq:sphericalharmoniclower}), we finally obtain
\begin{eqnarray}
\bar\eth\bar\eth S^+_{lm} &=& \frac{1}{2(r_+ - ia\cos\theta)^2}
\sum_{q=q_{\rm min}}^\infty
b^l_q(a\omega_{mkn})
\nonumber\\
& &\!\!\!\!\!\!\!\!\!\!\!\!\!\!\!
\times
\biggl[\sqrt{(q+2)(q+1)q(q-1)}\,{_0}Y_{qm}
\nonumber\\
& &\!\!\!\!\!\!\!\!\!\!\!
+ {\cal A}_1\sqrt{(q+2)(q-1)}\,{_1}Y_{qm} + {\cal
  A}_2\,{_2}Y_{qm}\biggr]\;.
\nonumber\\
\label{eq:barethbarethS}
\end{eqnarray}
This equation is simple to evaluate using the techniques presented in
Appendix A of Ref.\ {\cite{h2000}}.

\subsection{The phase of the tidal bulge}
\label{sec:bulge}

As we will see when we examine the geometry of distorted event
horizons in detail in Secs.\ {\ref{sec:schw_results}} and
{\ref{sec:kerr_results}}, a major effect of tides on a black hole is
to cause the horizon to bulge.  As has been described in detail in
past literature (e.g., {\cite{membrane}}), the result is not so
different from the response of a fluid body to a tidal driving force,
albeit with some counterintuitive aspects thanks to the teleological
nature of the event horizon.

In this section, we describe three ways to characterize the tidal
bulge of the distorted event horizon.  Two of these methods are based
on comparing the position at which the horizon is most distorted to
the position of the orbit.  Because the orbit and the horizon are at
different locations, comparing their positions requires us to map from
one to the other.  The notion of bulge phase that follows then depends
on the choice of map we use.  As such, any notion of bulge phase built
from comparing orbit position to horizon geometry must be somewhat
arbitrary, and can only be understood in the context of the mapping
that has been used.

We use two maps from orbit to horizon.  The first is a ``null map.''
Following Hartle {\cite{hartle74}}, we connect the orbit to the
horizon using an inward-going, zero-angular-momentum null geodesic.
This choice is commonly used in the literature, and so is useful for
comparing our results with past work.  The second is an
``instantaneous map.''  We compare the horizon geometry to the orbit
position on a slice of constant ingoing time coordinate $v$.  This is
particularly convenient for showing figures of the distorted horizon.

The third method of computing bulge phase directly compares the
horizon's response to the applied tidal field.  Since both quantities
are defined on the horizon, no mapping is necessary, and no arbitrary
choices are needed.  We do not use this notion of bulge phase very
much in this analysis, but anticipate using it in future work which
will examine more complicated cases than the circular, equatorial
orbits that are our focus here.

\subsubsection{Relative position of orbit and bulge I: Null map}
\label{sec:nullmap}

In his original examination of black hole tidal distortion, Hartle
{\cite{hartle74}} connects the orbit to the horizon with a zero
angular momentum ingoing light ray.  Choosing our origins
appropriately, the orbiting body is at angle
\begin{equation}
\phi_{\rm o} = \Omega_{\rm orb}t
\end{equation}
in Boyer-Lindquist coordinates.  We convert to ingoing coordinates
using Eq.\ (\ref{eq:v_and_psi}):
\begin{eqnarray}
\psi_{\rm o} &=& \Omega_{\rm orb}(v - r^*_{\rm o}) + \bar r_{\rm o}
\nonumber\\
&\equiv& \Omega_{\rm orb}v + \Delta\psi(r_{\rm o})\;,
\end{eqnarray}
where $\bar r_{\rm o} \equiv \bar r(r_{\rm o})$ and $r^*_{\rm o}
\equiv r^*(r_{\rm o})$ are given by Eqs.\ (\ref{eq:rbar}) and
(\ref{eq:rstar}), and where
\begin{equation}
\Delta\psi(r_{\rm o}) \equiv \bar r_{\rm o} - \Omega_{\rm orb} r^*_{\rm o}
\label{eq:Deltapsi_orb}
\end{equation}
is, for each orbital radius $r_{\rm o}$, a fixed angular offset
associated with the transformation from Boyer-Lindquist to ingoing
coordinates.

The orbit's location mapped onto the horizon is then
\begin{equation}
\psi^{\rm NM}_{\rm o} = \Omega_{\rm orb}v + \Delta\psi(r_{\rm o}) +
\delta\psi^{\rm null}\;,
\label{eq:psi_o_nullmap}
\end{equation}
where $\delta\psi^{\rm null}$ is the axial shift accumulated by the
ingoing null ray as it propagates from the orbit to the horizon.  This
shift must in general be computed numerically, but to leading order in
$a$ (which will be sufficient for our purposes) it is given by
\begin{equation}
\delta\psi^{\rm null} = -\frac{a}{2M} + \frac{a}{r_{\rm o}}
= 2M\Omega_{\rm H}\left(\frac{2M}{r_{\rm o}} - 1\right)\;.
\label{eq:deltapsinull_expand}
\end{equation}
The second form uses $\Omega_{\rm H} = a/4M^2$ for small $a$ to
rewrite this formula, which will be useful when we compare our results
to previous literature for small spin.  (One should also correct the
ingoing time, $v \to v + \delta v$, to account for the time it takes
for the ingoing null ray to propagate from the orbit to the horizon.
However, at leading order $\delta v \propto a^2$, so we can neglect it
for the applications we will use in this paper.  The time shift is
also neglected in all previous papers we are aware of which examine
the angular offset of the tidal bulge {\cite{hartle74,fl05}}, since
they only consider $a = 0$ or $a/M \ll 1$.)

Let $\psi^{\rm bulge}$ be the angle at which $R^{(1)}_{\rm H}$ is
maximized.  This value varies from mode to mode, but is easy to read
off once $R^{(1)}_{\rm H}$ is computed.  The offset of the orbit and
bulge using the null map is then
\begin{eqnarray}
\delta\psi^{\rm OB-NM} &\equiv& \psi^{\rm bulge} - \psi^{\rm NM}_{\rm
  o}
\nonumber\\
&=& \psi^{\rm bulge} - \Omega_{\rm orb}v - \Delta\psi(r_{\rm o}) -
\delta\psi^{\rm null}\;.
\nonumber\\
\label{eq:bulge_vs_orbit_null}
\end{eqnarray}
A positive value for $\delta\psi^{\rm OB-NM}$ means that the bulge
leads the orbit.

\subsubsection{Relative position of orbit and bulge II: Instantaneous map}
\label{sec:instmap}

Consider next a mapping that is instantaneous in ingoing time
coordinate $v$.  This choice is useful for making figures that show
both bulge and orbit, since we simply show their locations at a given
moment $v$.  This mapping neglects the term $\delta\psi^{\rm null}$,
but is otherwise identical to the null map:
\begin{equation}
\psi^{\rm IM}_{\rm o} = \psi_{\rm o} = \Omega v + \Delta\psi(r_{\rm o})\;.
\label{eq:psi_o_inst}
\end{equation}
The offset of the orbit and bulge in this mapping is
\begin{eqnarray}
\delta\psi^{\rm OB-IM} &\equiv& \psi^{\rm bulge} - \psi^{\rm IM}_{\rm
  o}
\nonumber\\
&=& \psi^{\rm bulge} - \Omega_{\rm orb}v - \Delta\psi(r_{\rm o})\;.
\label{eq:bulge_vs_orbit_inst}
\end{eqnarray}
Since $\delta\psi^{\rm null} = 0$ for $a = 0$, the null and
instantaneous maps are identical for Schwarzschild black holes.

Before concluding our discussion of the tidal bulge phase, we
emphasize again that the phase in both the null map and the
instantaneous map follow from arbitrary choices, and must be
interpreted in the context of those choices.  Other choices could be
made.  For example, one could make a map that is instantaneous in a
different time coordinate, or that is based on a different family of
ingoing light rays (e.g., the principle ingoing null congruence, along
which $v$, $\psi$, and $\theta$ are constant; such a map would be
identical to the instantaneous map).  These two maps are good enough
for our purposes --- the null map allows us to compare with other
papers in the literature, and the instantaneous map is excellent for
characterizing the plots we will show in
Secs.\ {\ref{sec:schw_results}} and {\ref{sec:kerr_results}}.

\subsubsection{Relative phase of tidal field and response}
\label{sec:phase_tide_horiz}

Our third method of characterizing the tidal bulge is to use the
relative phase of the horizon distortion $R^{(1)}_{\rm H}$ and
distorting tidal field $\psi_0$.  For our frequency-domain study, this
phase is best understood on a mode-by-mode basis.  Begin by
re-examining Eq.\ (\ref{eq:horizcurv1}):
\begin{eqnarray}
R^{(1)}_{{\rm H},lmkn} &=& -4\,{\rm Im}\left[\frac{\bar\eth\bar\eth
  \psi^{\rm HH}_{0,lmkn}}{p_{mkn}(ip_{mkn} + 2\epsilon)}\right]
\nonumber\\
&\equiv& {\rm Im}\left[R^{\rm c}_{lmkn}\right]\;.
\end{eqnarray}
Let us define the phase $\delta\psi^{\rm TB}_{lmkn}$ by
\begin{equation}
\frac{R^{\rm c}_{lmkn}}{\psi^{\rm HH}_{0,lmkn}} =
\frac{|R^{\rm c}_{lmkn}|}{|\psi^{\rm
    HH}_{0,lmkn}|}e^{-i\delta\psi^{\rm TB}_{lmkn}}\;.
\label{eq:deltapsiTB_def}
\end{equation}
As with $\delta\psi^{\rm OB-NM}$ and $\delta\psi^{\rm OB-IM}$,
$\delta\psi^{\rm TB}_{lmkn} > 0$ means that the horizon's response
leads the tidal field.

Using Eq.\ (\ref{eq:psi0HHexpand2}), we see that
\begin{equation}
\frac{R^{\rm c}_{lmkn}}{\psi^{\rm HH}_{0,lmkn}} =
-\frac{4}{p_{mkn}(ip_{mkn} + 2\epsilon)}\frac{\bar\eth\bar\eth
  S^+_{lmkn}}{S^+_{lmkn}}\;.
\label{eq:deltapsiTB1}
\end{equation}
With a few definitions, this form expedites our identification of
$\delta\psi^{\rm TB}_{lmkn}$.  First, note that $p_{mkn}$ and
$S^+_{lmkn}$ are both real, so the phase arises solely from the factor
$1/(ip_{mkn} + 2\epsilon)$ and the operator $\bar\eth\bar\eth$.  The
first factor is easily rewritten in a more useful form:
\begin{equation}
\frac{1}{ip_{mkn} + 2\epsilon} =
\frac{e^{-i\arctan(p_{mkn}/2\epsilon)}}{\sqrt{p_{mkn}^2 +
    4\epsilon^2}}\;.
\label{eq:deltapsiTB2}
\end{equation}
To clean up the phase associated with $\bar\eth\bar\eth$, we make a
definition:
\begin{equation}
\frac{\bar\eth\bar\eth S^+_{lmkn}}{S^+_{lmkn}} \equiv
\Sigma_{lmkn}(\theta) e^{-i{\cal S}_{lmkn}(\theta)}\;.
\label{eq:spheroidratio}
\end{equation}
The amplitude ratio $\Sigma_{lmkn}(\theta)$ and phase ${\cal
  S}_{lmkn}(\theta)$ must in general be determined numerically.  We
will show expansions for small $a$ and slow motion in
Sec.\ {\ref{sec:kerr_results}}.  We include $S^+_{lmkn}$ in this
definition because it may pass through zero at a different angle than
$\bar\eth\bar\eth S^+_{lmkn}$ passes through zero.  This will appear
as a change by $\pi$ radians in the phase ${\cal S}_{lmkn}$.

Combining Eqs.\ (\ref{eq:deltapsiTB_def}) -- (\ref{eq:spheroidratio})
and using the fact that $\epsilon = \kappa/2$ (where $\kappa$ is the
black hole surface gravity), we at last read out
\begin{equation}
\delta\psi^{\rm TB}_{lmkn} = \arctan\left(p_{mkn}/\kappa\right) + {\cal
  S}_{lmkn}(\theta)\;.
\label{eq:deltapsiTB}
\end{equation}
Recall that the wavenumber $p_{mkn} = \omega_{mkn} - m\Omega_{\rm H}$.
In geometrized units, $\kappa^{-1}$ is a timescale which characterizes
how quickly the horizon adjusts to an external disturbance
(cf.\ Sec.\ VI C 5 of Ref.\ {\cite{membrane}} for discussion).  The
first term in Eq.\ (\ref{eq:deltapsiTB}) is thus determined by the
wavenumber times this characteristic horizon time.  For a circular,
equatorial orbit which has $\Omega_{\rm orb} = \Omega_{\rm H}$, this
term is zero, in accord with the Newtonian intuition that the tide and
the response are exactly aligned when the spin and orbit frequencies
are identical.  This intuition does not quite hold up thanks to the
correcting phase ${\cal S}_{lmkn}(\theta)$.  We will examine the
impact of this correction in Sec.\ {\ref{sec:kerr_results}}.

The phase $\delta\psi^{\rm TB}_{lmkn}$ is particularly useful for
describing the horizon's response to complicated orbits where the
relative geometry of the horizon and the orbit is dynamical.  For
example, Vega, Poisson, and Massey {\cite{vpm11}} use a measure
similar to $\delta\psi^{\rm TB}_{lmkn}$ to describe how a
Schwarzschild black hole responds to a body that comes near the
horizon on a parabolic encounter, demonstrating that the horizon's
response leads the applied tidal field (cf.\ Sec.\ 5.2 of
Ref.\ {\cite{vpm11}}).  We will examine $\delta\psi^{\rm TB}_{lmkn}$
briefly for the circular, equatorial orbits we focus on in this paper,
but will use it in greater depth in a follow-up analysis that looks at
tides from generic orbits.

When $a = 0$, the operator $\bar\eth\bar\eth$ is real, and ${\cal
  S}_{lmkn}(\theta) = 0$.  We have $p_{mkn} = \omega_{mkn}$ and
$\kappa = 1/4M$ in this limit, so
\begin{equation}
\delta\psi^{\rm TB}_{lmkn}\Bigr|_{a = 0} \to\quad \delta\phi^{\rm TB}_{mkn}
= \arctan\left(4M\omega_{mkn}\right)\;.
\label{eq:schw_bulge_vs_tide}
\end{equation}
We will show in Sec.\ {\ref{sec:schw_results}} that this agrees with
the phase shift obtained by Fang and Lovelace {\cite{fl05}}.  It also
agrees with the results of Vega, Poisson, and Massey {\cite{vpm11}},
though in somewhat different language.  They work in the time domain,
showing that a Schwarzschild black hole's horizon response leads the
field by a time interval $\kappa_{\rm Schw}^{-1} = 4M$.  For a field
that is periodic with frequency $\omega$, this means that we expect
the response to lead the field by a phase angle $4M\omega$, exactly as
Eq.\ (\ref{eq:schw_bulge_vs_tide}) says.

\subsection{The down-horizon flux}
\label{sec:downhoriz}

Although not needed for this paper, we now summarize how one computes
the down-horizon flux.  Our purpose is to show that the coefficients
$Z^{\rm H}_{lmkn}$ which characterize the geometry of the deformed
event horizon also characterize the down-horizon gravitational-wave
flux, showing that the ``deformed horizon'' and ``down-horizon flux''
pictures are just different ways of interpreting how the horizon
interacts with the orbit.

Our discussion follows Teukolsky and Press {\cite{tp74}}, which in
turn follows Hawking and Hartle {\cite{hh72}}, modifying the
presentation slightly to follow our notation.  The starting point is
to note that a tidal perturbation shears the generators of the event
horizon.  This shear, $\sigma$, causes the area of the event horizon
to grow:
\begin{equation}
\frac{d^2A}{d\Omega dt} = \frac{2Mr_+}{\epsilon}|\sigma|^2\;.
\label{eq:areagrowth}
\end{equation}
We also know the area of a black hole's event horizon,
\begin{equation}
A = 8\pi\left(M^2 + \sqrt{M^4 - S^2}\right)\;,
\end{equation}
where $S = aM$ is the black hole's spin angular momentum.  Using this,
we can write the area growth law as
\begin{equation}
\frac{d^2A}{d\Omega dt} = \frac{8\pi}{\sqrt{M^4 - S^2}}
\left(2M^2r_+\frac{d^2M}{d\Omega dt} -
S\frac{d^2S}{d\Omega dt}\right)\;.
\end{equation}

Consider now radiation going down the horizon.  Radiation carrying
energy $dE^{\rm H}$ and angular momentum $dL_z^{\rm H}$ into the hole
changes its mass and spin by
\begin{equation}
dM = dE^{\rm H}\;,\qquad dS = dL_z^{\rm H}\;.
\end{equation}
Angular momentum and energy carried by the radiation are related
according to
\begin{equation}
dL_z = \frac{m}{\omega_{mkn}} dE\;.
\end{equation}
Putting all of this together and using Eq.\ (\ref{eq:pmkndef}), we
find
\begin{eqnarray}
\frac{d^2E^{\rm H}}{dt d\Omega} &=& \frac{\omega_{mkn} Mr_+}{2\pi
  p_{mkn}}|\sigma|^2\;,
\label{eq:dEHdtdOmega}
\\
\frac{d^2L_z^{\rm H}}{dt d\Omega} &=& \frac{mMr_+}{2\pi
  p_{mkn}}|\sigma|^2\;.
\label{eq:dLzHdtdOmega}
\end{eqnarray}

So to compute the down-horizon flux, we just need to know the shear
$\sigma$.  It is simply computed from the tidal field $\psi_0^{\rm
  HH}$.  First, expand $\sigma$ as
\begin{equation}
\sigma = \sum_{lmkn}\sigma_{lmkn}S^+_{lmkn}(\theta)e^{i[m\psi -
    \omega_{mkn}v - mK(a)]}\;.
\end{equation}
The shear mode amplitudes $\sigma_{lmkn}$ are related to the tidal
field mode $\psi^{\rm HH}_{0,lmkn}$ by {\cite{tp74}}:
\begin{equation}
\sigma_{lmkn} = \frac{i\psi_{0,lmkn}^{\rm HH}}{p_{mkn} - 2i\epsilon}\;.
\label{eq:sigmapsi0relation}
\end{equation}
Combine Eq.\ (\ref{eq:sigmapsi0relation}) with
Eqs.\ (\ref{eq:psi0HHexpand}), (\ref{eq:WlmknZHlmkn}), and
(\ref{eq:betalmkn}).  Integrate over solid angle, using the
orthogonality of the spheroidal harmonics.  Equations
(\ref{eq:dEHdtdOmega}) and (\ref{eq:dLzHdtdOmega}) become
\begin{eqnarray}
\left(\frac{dE}{dt}\right)^{\rm H} &=& \sum_{lmkn} \alpha_{lmkn}
\frac{|Z^{\rm H}_{lmkn}|^2}{4\pi\omega_{mkn}^2}\;,
\label{eq:dEHdt}\\
\left(\frac{dL_z}{dt}\right)^{\rm H} &=& \sum_{lmkn} \alpha_{lmkn}
\frac{m|Z^{\rm H}_{lmkn}|^2}{4\pi\omega_{mkn}^3}\;.
\label{eq:dLzHdt}
\end{eqnarray}
The coefficient
\begin{eqnarray}
\alpha_{lmkn} &=& \frac{256(2Mr_+)^5p_{mkn}\omega_{mkn}^3}{|c_{lmkn}|^2}
\nonumber\\
& &\times\,
(p_{mkn}^2 + 4\epsilon^2)(p_{mkn}^2 + 16\epsilon^2)\;,
\end{eqnarray}
with $|c_{lmkn}|^2$ given by Eq.\ (\ref{eq:clmknsqr}), comes from
combining the various prefactors in the relations that lead to
Eqs.\ (\ref{eq:dEHdt}) and (\ref{eq:dLzHdt}).  Notice that
$\alpha_{lmkn} \propto p_{mkn}$.  This means that $\alpha_{lmkn} = 0$
when $\omega_{mkn} = m\Omega_{\rm H}$.  The down-horizon fluxes
(\ref{eq:dEHdt}) and (\ref{eq:dLzHdt}) are likewise zero for modes
which satisfy this condition.

It is interesting to note that the shear $\sigma_{lmkn}$ and the tidal
field $\psi^{\rm HH}_{0,lmkn}$ are both proportional to $p_{mkn}$, and
hence both vanish when $\omega_{mkn} = m\Omega_{\rm H}$.  The
horizon's Ricci curvature $R^{(1)}_{{\rm H},lmkn}$ does not, however,
vanish in this limit.  Mathematically, this is because $R^{(1)}_{{\rm
    H},lmkn}$ includes a factor of $1/p_{mkn}$ which removes this
proportionality [cf.\ Eq.\ (\ref{eq:horizcurv1})].  Physically, this
is telling us that when $\Omega_{\rm H} = \Omega_{\rm orb}$, the
horizon is deformed, but the deformation is static in the horizon's
reference frame.  This static deformation does not shear the
generators, and does not carry energy or angular momentum into the
hole.

Equations (\ref{eq:dEHdt}) and (\ref{eq:dLzHdt}) illustrate the point
of this section: The fluxes of $E$ and $L_z$ into the horizon are
determined by the same numbers $Z^{\rm H}_{lmkn}$ used to compute the
horizon's deformed geometry, Eq.\ (\ref{eq:gauss_kerr}).

\section{Results I: Schwarzschild}
\label{sec:schw_results}

Using the formalism we have assembled, we now examine the tidally
deformed geometry of black hole event horizons.  In this paper, we
will only study the circular, equatorial limit: The orbiting body is
at $r = r_{\rm o}$, $\theta = \pi/2$, and $\phi = \Omega_{\rm orb}t$.
Harmonics of $\Omega_\theta$ and $\Omega_r$ can play no role in any
physics arising from these orbits, so the index set $\{lmkn\}$ reduces
to $\{lm\}$, and the mode frequency $\omega_{mkn}$ to $\omega_m$.  We
will consider general orbits in a later analysis.

Before tackling general black hole spin, it is useful to examine
Eq.\ (\ref{eq:gauss_kerr}) for Schwarzschild black holes.  Several
simplifications occur when $a = 0$:

\begin{itemize}

\item The radius $r_+ = 2M$; the frequency $\Omega_{\rm H} = 0$, so
  the wavenumber $p_m = \omega_m$; the factor $\epsilon = 1/8M$; the
  phase factor $K(a) = 0$ [cf.\ Eq.\ (\ref{eq:K_of_a})]; and the ingoing
  axial coordinate $\psi = \phi$.

\item The spin-weighted spheroidal harmonic becomes a spin-weighted
  spherical harmonic: ${_{+2}S}_{lm}(\theta)
  \to\ {_{+2}Y}_{lm}(\theta)$.  The eigenvalue of the angular function
  therefore simplifies, as does the complex number $c_{lm}$: ${\cal E}
  = l(l+1)$, and $c_{lm} = (l+2)(l+1)l(l-1) + 12iM\omega_m$.

\item The angular operator $\bar\eth \equiv \bar\eth_0 =
  1/(2\sqrt{2}M)L^s_-$.  Using Eq.\ (\ref{eq:sphericalharmoniclower}),
  we have
\begin{equation}
L^s_-L^s_-\, {_{+2}}Y_{lm}(\theta) =
\sqrt{(l+2)(l+1)l(l-1)}\,{_0}Y_{lm}\;,
\end{equation}
which tells us that
\begin{equation}
\bar\eth\bar\eth S^+_{lm}(\theta) =
\frac{1}{8M^2} \sqrt{(l+2)(l+1)l(l-1)}\,{_0}Y_{lm}
\label{eq:barethbarethS_schw}
\end{equation}
for $a = 0$.
\end{itemize}

Putting all of this together, for $a = 0$ we have
\begin{eqnarray}
R^{(1)}_{{\rm H},lm} &=& {\rm Im}\left[ {\cal C}_{lm} Z^{\rm H}_{lm}
e^{i\Phi_m}\right] \times
\nonumber\\
& &\!\!\!\!\!\!\!\!\!\!\!
\frac{1}{8M^2}\sqrt{(l+2)(l+1)l(l-1)}\,{_0}Y_{lm}(\theta)\;,
\label{eq:gauss_schw}
\end{eqnarray}
where
\begin{eqnarray}
{\cal C}_{lm} &=& \frac{1024M^2(iM\omega_m - 1/4)(M\omega_m +
  i/2)}{(l+2)(l+1)l(l-1) + 12iM\omega_m}\;,
\label{eq:Clm_schw}\\
\Phi_m &=& m\phi - \omega_m v\;.
\end{eqnarray}

\subsection{Slow motion: Analytic results}
\label{sec:schw_analytic}

We begin our analysis of the Schwarzschild tidal deformations by
expanding all quantities in orbital speed $u \equiv (M/r_{\rm
  o})^{1/2}$.  We take all relevant quantities to $O(u^5)$ beyond the
leading term; this is far enough to see how the curvature behaves for
multipole index $l \le 4$.  These results should be accurate for
weak-field orbits, when $u \ll 1$.  In the following subsection, we
will compare with numerical results that are good into the strong
field.

Begin with ${\cal C}_{lm}$.  Expanding Eq.\ (\ref{eq:Clm_schw}), we
find
\begin{eqnarray}
{\cal C}_{2m} &=& -\frac{16i}{3}M^2\exp\left(-\frac{13}{2}imu^3\right)\;,
\label{eq:C2m_schw}\\
{\cal C}_{3m} &=& -\frac{16i}{15}M^2\exp\left(-\frac{61}{10}
imu^3\right)\;,
\label{eq:C3m_schw}\\
{\cal C}_{4m} &=& -\frac{16i}{45}M^2
\exp\left(-\frac{181}{30}imu^3\right)\;.
\label{eq:C4m_schw}
\end{eqnarray}
To perform this expansion, we used the fact that, for $a = 0$,
$M\Omega_{\rm orb} = u^3$, so $M\omega_m = mu^3$.

Next, we construct analytic expansions for the amplitudes $Z^{\rm
  H}_{lm}$, following the algorithm described in Sec.\ {\ref{sec:ZH}}.
All the results which follow are understood to neglect contributions
of $O(u^6)$ and higher.  We also introduce $\mu$, the mass of the
small body whose tides deform the black hole.

For $l = 2$, the amplitudes are
\begin{eqnarray}
Z^{\rm H}_{20} &=& \sqrt{\frac{3\pi}{10}}\frac{\mu}{r_{\rm o}^3} \left(1 +
  \frac{7}{2}u^2 + \frac{561}{56}u^4\right)\;,
\label{eq:ZHschw20}\\
Z^{\rm H}_{21} &=& -3 i \sqrt{\frac{\pi }{5}}\frac{\mu}{r_{\rm o}^3}
\left(u + \frac{8}{3} u^3 + \frac{10i}{3} u^4
+ \frac{152}{21} u^5\right)
\nonumber\\
&=& -3i\sqrt{\frac{\pi}{5}}\frac{\mu}{r_{\rm o}^3}
\left(u + \frac{8}{3} u^3 + \frac{152}{21} u^5\right)
\nonumber\\
& &\qquad\qquad\quad\times\,
\exp\left(\frac{10}{3}iu^3\right)\;,
\label{eq:ZHschw21}
\\
Z^{\rm H}_{22} &=& -\frac{3}{2}\sqrt{\frac{\pi }{5}}\frac{\mu}{r_{\rm o}^3}
\biggl(1 + \frac{3}{2}u^2 + \frac{23i}{3}u^3 + \frac{1403}{168}u^4
\nonumber\\
& &\qquad\qquad\qquad\qquad
+ \frac{473i}{30}u^5\biggr)
\nonumber\\
&=& -\frac{3}{2}\sqrt{\frac{\pi }{5}}\frac{\mu}{r_{\rm o}^3}
\left(1 + \frac{3}{2}u^2 + \frac{1403}{168}u^4\right)
\nonumber\\
& &\qquad\qquad\quad \times\,
\exp\left[i\left(\frac{23}{3}u^3 + \frac{64}{15}u^5\right)\right]\;.
\nonumber\\
\label{eq:ZHschw22}
\end{eqnarray}
For $l = 3$,
\begin{eqnarray}
Z^{\rm H}_{30} &=& -i \sqrt{\frac{30 \pi }{7}}\frac{\mu}{r_{\rm o}^3}
\left(u^3 + 4 u^5\right)\;,
\label{eq:ZHschw30}\\
Z^{\rm H}_{31} &=& -\frac{3}{2}\sqrt{\frac{5\pi}{14}}\frac{\mu}{r_{\rm o}^3}
\left(u^2 + \frac{13}{3} u^4 + \frac{43i}{30}u^5\right)
\nonumber\\
&=& -\frac{3}{2}\sqrt{\frac{5\pi}{14}}\frac{\mu}{r_{\rm o}^3}
\left(u^2 + \frac{13}{3} u^4\right)
\exp\left(\frac{43}{30}iu^3\right)\;,
\nonumber\\
\label{eq:ZHschw31}\\
Z^{\rm H}_{32} &=& 5 i \sqrt{\frac{\pi }{7}}\frac{\mu}{r_{\rm o}^3}
\left(u^3 + 4u^5\right)\;,
\label{eq:ZHschw32}\\
Z^{\rm H}_{33} &=& \frac{5}{2}\sqrt{\frac{3 \pi }{14}}\frac{\mu}{r_{\rm o}^3}
\left(u^2 + 3u^4 + \frac{43i}{10}u^5\right)
\nonumber\\
&=& \frac{5}{2}\sqrt{\frac{3 \pi }{14}}\frac{\mu}{r_{\rm o}^3}
\left(u^2 + 3u^4\right)\exp\left(\frac{43}{10}iu^3\right)\;.
\nonumber\\
\label{eq:ZHschw33}
\end{eqnarray}
Finally, for $l = 4$,
\begin{eqnarray}
Z^{\rm H}_{40} &=& -\frac{9}{14}\sqrt{\frac{5\pi}{2}}\frac{\mu}{r_{\rm o}^3}
u^4\;,
\label{eq:ZHschw40}\\
Z^{\rm H}_{41} &=& \frac{45i}{14} \sqrt{\frac{\pi }{2}}\frac{\mu}{r_{\rm o}^3}
u^5\;,
\label{eq:ZHschw41}\\
Z^{\rm H}_{42} &=& \frac{15}{14}\sqrt{\pi}\frac{\mu}{r_{\rm o}^3}
u^4\;,
\label{eq:ZHschw42}\\
Z^{\rm H}_{43} &=& -\frac{15i}{2} \sqrt{\frac{\pi}{14}}\frac{\mu}{r_{\rm o}^3}
u^5\;,
\label{eq:ZHschw43}\\
Z^{\rm H}_{44} &=& -\frac{15}{4}\sqrt{\frac{\pi}{7}}\frac{\mu}{r_{\rm o}^3}
u^4\;.
\label{eq:ZHschw44}
\end{eqnarray}
Note that $Z^{\rm H}_{l-m} = (-1)^l \bar Z^{\rm H}_{lm}$, where
overbar denotes complex conjugation.

\begin{widetext}

It is particularly convenient to combine the modes in pairs, examining
$R^{(1)}_{{\rm H},l-m} + R^{(1)}_{{\rm H},lm}$.  Doing so, we find for
$l = 2$,

\begin{eqnarray}
R^{(1)}_{{\rm H},20} &=& -\frac{\mu}{r_{\rm o}^3}
\left(3\cos^2\theta - 1\right)
\left(1 + \frac{7}{2}u^2 + \frac{561}{56}u^4\right)\;,
\label{eq:R20schw}\\
R^{(1)}_{{\rm H},1-1} + R^{(1)}_{{\rm H},11} &=& 0\;,
\label{eq:R21schw}\\
R^{(1)}_{{\rm H},2-2} + R^{(1)}_{{\rm H},22} &=&
\frac{3\mu}{r_{\rm o}^3}\sin^2\theta \left(1 + \frac{3}{2}u^2 +
\frac{1403}{168}u^4\right) \cos\left[2\left(\phi - \Omega_{\rm orb} v
  - \frac{8}{3}u^3 + \frac{32}{5}u^5\right)\right]\;.
\label{eq:R22schw}
\end{eqnarray}
For $l = 3$, we have
\begin{eqnarray}
R^{(1)}_{{\rm H},30} &=& 0\;,
\label{eq:R30schw}\\
R^{(1)}_{{\rm H},3-1} + R^{(1)}_{{\rm H},31} &=&
\frac{3}{2}\frac{\mu}{r_{\rm o}^3} \sin\theta\left(1 -
5\cos^2\theta\right) u^2\left(1 + \frac{13}{3}u^2\right)
\cos\left[\phi - \Omega_{\rm orb} v - \frac{14}{3}u^3\right]\;,
\label{eq:R31schw}\\
R^{(1)}_{{\rm H},3-2} + R^{(1)}_{{\rm H},32} &=& 0\;,
\label{eq:R32schw}\\
R^{(1)}_{{\rm H},3-3} + R^{(1)}_{{\rm H},33} &=&
\frac{5}{2}\frac{\mu}{r_{\rm o}^3}\sin^3\theta u^2\left(1 +
3u^2\right) \cos\left[3\left(\phi - \Omega_{\rm orb} v -
  \frac{14}{3}u^3\right)\right]\;.  \nonumber\\
\label{eq:R33schw}
\end{eqnarray}
And, for $l = 4$,
\begin{eqnarray}
R^{(1)}_{{\rm H},40} &=& \frac{9}{56}\frac{\mu}{r_{\rm o}^3}
\left(3 - 30\cos^2\theta + 35\cos^4\theta\right)u^4\;,
\label{eq:R40schw}\\
R^{(1)}_{{\rm H},4-1} + R^{(1)}_{{\rm H},41} &=& 0\;,
\label{eq:R41schw}\\
R^{(1)}_{{\rm H},4-2} + R^{(1)}_{{\rm H},42} &=&
\frac{15}{14}\frac{\mu}{r_{\rm o}^3}\sin^2\theta \left(1 -
7\cos^2\theta\right)u^4 \cos\left[2\left(\phi - \Omega_{\rm orb} v -
  \frac{181}{30}u^3\right)\right]\;,
\label{eq:R42schw}\\
R^{(1)}_{{\rm H},4-3} + R^{(1)}_{{\rm H},43} &=& 0\;,
\label{eq:R43schw}\\
R^{(1)}_{{\rm H},4-4} + R^{(1)}_{{\rm H},44} &=&
\frac{15}{8}\frac{\mu}{r_{\rm o}^3}\sin^4\theta u^4
\cos\left[4\left(\phi - \Omega_{\rm orb} v - \frac{181}{30}u^3\right)
  \right]\;.
\label{eq:R44schw}
\end{eqnarray}
\end{widetext}

In the next section, we will compare Eqs.\ (\ref{eq:R20schw}) --
(\ref{eq:R44schw}) with strong-field numerical calculations.  Before
doing so, we examine some consequences of these results and compare
with earlier literature.

\subsubsection{Nearly static limit}

In Ref.\ {\cite{hartle74}}, Hartle examines the deformation of a black
hole due to a nearly static orbiting moon.  To reproduce his results,
consider the $u \to 0$ limit of Eqs.\ (\ref{eq:R20schw}) --
(\ref{eq:R44schw}).  Only the $l = 2$, $m = 0$, $m = \pm2$
contributions remain when $u \to 0$.  Adding these contributions, we
find
\begin{equation}
R^{(1)}_{\rm H} = -\frac{\mu}{r_{\rm o}^3} \left[3\cos^2\theta -
  3\sin^2\theta\cos\left(2\phi'\right) - 1\right]\;,
\label{eq:R1static}
\end{equation}
where $\phi' = \phi - \Omega_{\rm orb} v$ is the azimuthal coordinate
of the orbiting moon.  Hartle writes\footnote{Note that the result
  Hartle presents in Ref.\ {\cite{hartle74}} contains a sign error.
  This can be seen by computing the curvature associated with the
  metric he uses on the horizon [Eqs.\ (5.10) and (5.13) of
    Ref.\ {\cite{hartle73}}].  The embedding surface Hartle uses,
  Eq.\ (4.33) of Ref.\ {\cite{hartle74}} [or (5.14) of
    Ref.\ {\cite{hartle73}}] is correct given this metric.} his result
\begin{equation}
R^{(1)}_{\rm Hartle} = \frac{4\mu}{r_{\rm o}^3}P_2(\cos\chi) =
\frac{2\mu}{r_{\rm o}^3}\left(3\cos^2\chi - 1\right)\;,
\label{eq:R1Hartle}
\end{equation}
where ``$\chi$ is the angle between the point of interest and the
direction to the moon'' [Ref.\ {\cite{hartle74}}, text following
  Eq.\ (4.32)].  The angle $\chi$ can be interpreted as $\theta$ if we
place Hartle's moon at $\theta_{\rm moon} = 0$.  To compare the two
solutions, we must rotate.  One way to do this rotation is to note
that the equatorial plane in our calculation ($\theta = \pi/2$) should
vary with $\phi'$ as Hartle's result varies with $\chi$.  Put $\theta
= \pi/2$ and $\phi' = \chi$ in Eq.\ (\ref{eq:R1static}):
\begin{eqnarray}
R^{(1)}_{\rm H}\bigr|_{\theta = \pi/2, \phi' = \chi}
&=& \frac{\mu}{r_{\rm o}^3}\left(3\cos2\chi + 1\right)
\nonumber\\
&=& \frac{2\mu}{r_{\rm o}^3}\left(3\cos^2\chi - 1\right)\;.
\end{eqnarray}
Another way to compare is to note that the $\phi' = 0$ circle should
vary with angle in a way that duplicates Hartle's result, modulo a
shift in angle, $\theta = \chi + \pi/2$:
\begin{eqnarray}
R^{(1)}_{\rm H}\bigr|_{\theta = \chi + \pi/2, \phi' = 0} &=&
-\frac{\mu}{r_{\rm o}^3}\bigl[3\cos^2(\chi + \pi/2)
\nonumber\\
& &\quad - 3\sin^2(\chi + \pi/2) + 1\bigr]
\nonumber\\
&=& -\frac{\mu}{r_{\rm o}^3}\left(3\sin^2\chi - 3\cos^2\chi + 1\right)
\nonumber\\
&=& \frac{2\mu}{r_{\rm o}^3}\left(3\cos^2\chi - 1\right)\;.
\end{eqnarray}
Both forms reproduce Hartle's static limit.

\subsubsection{Embedding the quadrupolar distortion}
\label{sec:l=2embed}

At various places in this paper, we will examine the geometry of a
distorted horizon by embedding it in a 3-dimensional Euclidean space.
The details of this calculation are given in Appendix
{\ref{app:embed}}; equivalent discussion for Schwarzschild, where the
results are particularly clean, is also given in Ref.\ {\cite{vpm11}}.
Briefly, a Schwarzschild horizon that has been distorted by a tidal
field has the scalar curvature of a spheroid of radius
\begin{equation}
r_{\rm E} = 2M\left[1 + \sum_{lm}\varepsilon_{lm}(\theta,\phi)\right]\;,
\end{equation}
where, as shown in Appendix {\ref{app:embed_schw}} and
Ref.\ {\cite{vpm11}},
\begin{equation}
\varepsilon_{lm} = \frac{2M^2}{(l + 2)(l - 1)}R^{(1)}_{{\rm H},lm}\;.
\label{eq:schw_spheroid}
\end{equation}

By considering a Schwarzschild black hole embedded in a universe
endowed with post-Newtonian tidal fields, Taylor and Poisson
{\cite{tp08}} compute $\varepsilon_{lm}$ in a post-Newtonian
framework.  Specializing to the tides appropriate to a binary system,
they find
\begin{eqnarray}
& &\sum_m \varepsilon_{2m}(\theta,\phi) = \frac{\mu}{b^3}\frac{M^2}{2}
\left(1 + \frac{1}{2}u^2\right)(1 - \cos^2\theta)
\nonumber\\
& & \quad\quad + \frac{3\mu}{b^3}\frac{M^2}{2}\left(1 -
  \frac{3}{2}u^2\right)\sin^2\theta \cos\left[2(\phi - \Omega_{\rm
    orb}v)\right]\;.
\nonumber\\
\label{eq:taylorpoisson1}
\end{eqnarray}
This is Eq.\ (8.8) of Ref.\ {\cite{tp08}}, with $M_2 \to \mu$, $M_1
\to M$, $v_{\rm rel}/c \to u$, and expanded to leading order in $\mu$.
Their parameter $b$ is the separation of the binary in harmonic
coordinates.  Using the fact that $r_{\rm H} = r_{\rm S} - M$ (with
``H'' and ``S'' subscripts denoting harmonic and Schwarzschild,
respectively), it is easy to convert to $r_{\rm o}$, our separation in
Schwarzschild coordinates:
\begin{eqnarray}
\frac{1}{b^3} = \frac{1}{r_{\rm o}^3(1 - M/r_{\rm o})^3}
\simeq \frac{1}{r_{\rm o}^3}\left(1 + 3u^2\right)\;.
\end{eqnarray}
Replacing $b$ for $r_{\rm o}$ and truncating at $O(u^2)$,
Eq.\ (\ref{eq:taylorpoisson1}) becomes
\begin{eqnarray}
& &\sum_m \varepsilon_{2m}(\theta,\phi) = \frac{\mu}{r_{\rm o}^3}\frac{M^2}{2}
\left(1 + \frac{7}{2}u^2\right)(1 - \cos^2\theta)
\nonumber\\
& & \quad\quad + \frac{3\mu}{r_{\rm o}^3}\frac{M^2}{2}\left(1 +
  \frac{3}{2}u^2\right)\sin^2\theta \cos\left[2(\phi - \Omega_{\rm
    orb}v)\right]\;.
\nonumber\\
\label{eq:taylorpoisson2}
\end{eqnarray}
Comparing with Eqs.\ (\ref{eq:R20schw}) and (\ref{eq:R22schw}) and
correcting for the factor $M^2/2$ which converts curvature
$R^{(1)}_{{\rm H},2m}$ to $\varepsilon_{2m}$, we see agreement to
$O(u^2)$.

\subsubsection{Phase of the tidal bulge}
\label{sec:schw_bulge}

Using these analytic results, let us examine the notions of bulge
phase introduced in Sec.\ {\ref{sec:bulge}}.  First consider the
position of the bulge versus the position of the orbit according to
the null and instantaneous maps (which are identical for
Schwarzschild), Eq.\ (\ref{eq:bulge_vs_orbit_null}).  The various
modes which determine the shape of the horizon all peak at angle $\phi
= \Omega_{\rm orb}v + \delta\phi(u)$, where $\delta\phi(u)$ can be
read out of Eqs.\ (\ref{eq:R20schw})--(\ref{eq:R44schw}).  For
Schwarzschild $\bar r = 0$, and the ingoing angle $\psi = \phi$.  The
orbit's position mapped onto the horizon is $\phi_{\rm o}^{\rm NM} =
\phi_{\rm o} = \Omega_{\rm orb} v + \Delta\phi(r_{\rm o})$, where
\begin{equation}
\Delta\phi(r_{\rm o}) = -\Omega_{\rm orb} r^*_{\rm o}
\end{equation}
is Eq.\ (\ref{eq:Deltapsi_orb}) for $a = 0$.  The result for the
bulge's offset from the orbit is
\begin{eqnarray}
\delta\phi^{\rm OB}_{22} &=& \frac{8}{3}u^3 - \frac{32}{5}u^5 -
\Delta\phi(r_{\rm o})\;,
\label{eq:phi_ob_2}\\
\delta\phi^{\rm OB}_{31} &=& \delta\phi^{\rm OB}_{33} =
\frac{14}{3}u^3 - \Delta\phi(r_{\rm o})\;,
\label{eq:phi_ob_3}\\
\delta\phi^{\rm OB}_{42} &=& \delta\phi^{\rm OB}_{44} =
\frac{181}{30}u^3 - \Delta\phi(r_{\rm o})\;.
\label{eq:phi_ob_4}
\end{eqnarray}
For the multipoles which we do not include here, no useful notion of
bulge position exists: for $m = 0$ the bulge is axisymmetric, and for
the others, the bulge's amplitude is zero to this order.  Our results
for $l = |m| = 2$ agree with Fang and Lovelace; cf.\ Eq.\ (4) of
Ref.\ {\cite{fl05}}.

Consider next the relative phase of the tidal bulge and the perturbing
field, Eq.\ (\ref{eq:schw_bulge_vs_tide}).  For small $u$, we have
\begin{equation}
\delta\phi^{\rm TD}_{m} = 4mu^3\;.
\end{equation}
This again agrees with Fang and Lovelace --- compare Eq.\ (6) of
Ref.\ {\cite{fl05}}, bearing in mind that $m$ is built into their
definition of the offset angle [their Eq.\ (50)], and that they fix $m
= 2$.

In both cases, note that the bulge's offset is a positive phase.  This
indicates that the bulge leads both the orbiting body's instantaneous
position, as well as the tidal field that sources the tidal
deformation.  As discussed in the Introduction, this is consistent
with past work, and is a consequence of the horizon's teleological
nature.

\subsection{Fast motion: Numerical results}
\label{sec:schw_numerical}

\begin{figure}[ht]
\includegraphics[width = 0.48\textwidth]{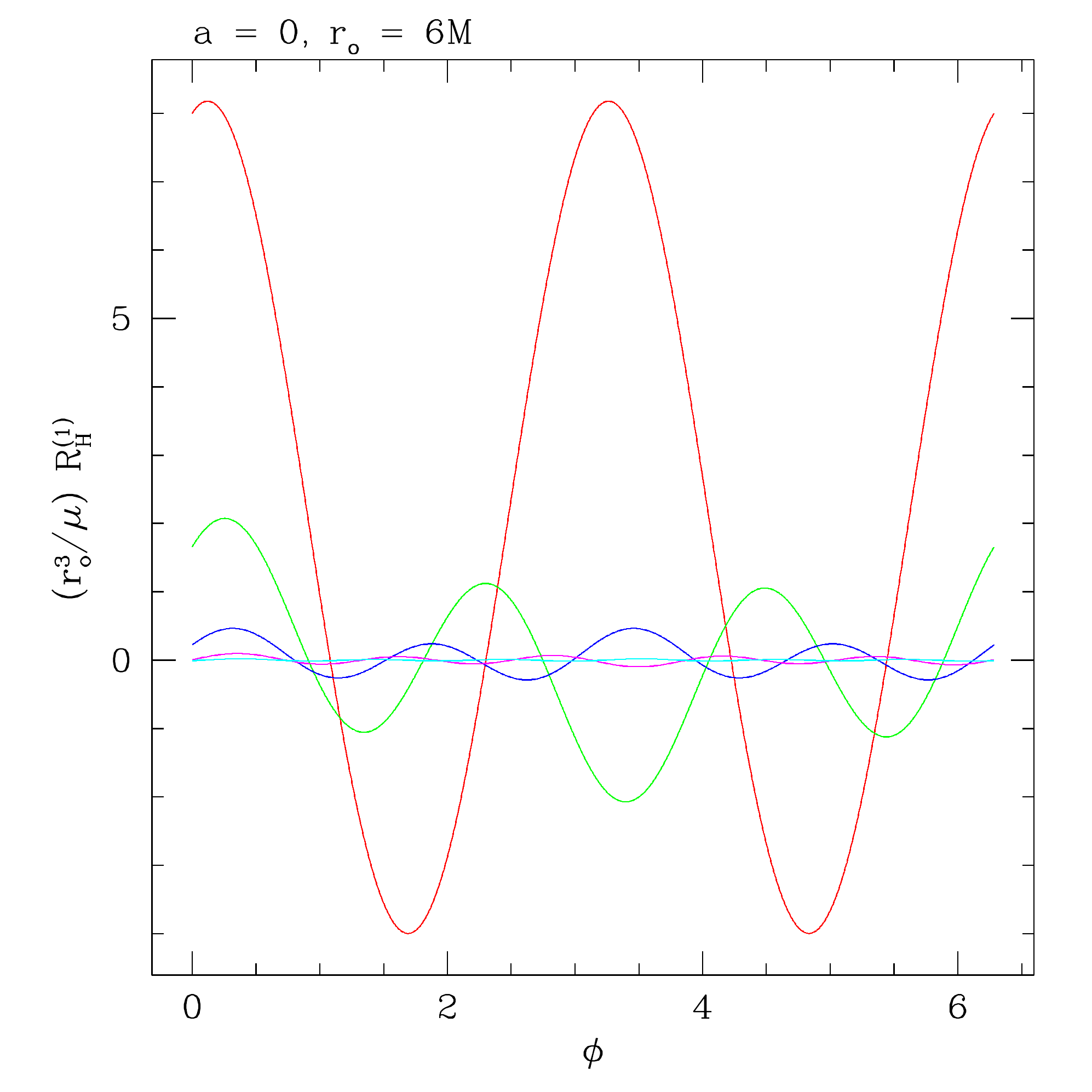}
\caption{Convergence of contributions to the horizon's tidal
  distortion.  We show $R^{(1)}_{{\rm H},lm}$ summed over $m$ for a
  given $l$, scaled by a factor $(r_{\rm o}^3/\mu)$ to account for the
  leading dependence on small body mass and orbital radius.  The
  largest amplitude oscillation is for $l = 2$ (red in color).  The
  next largest is $l = 3$ (green), followed by $l = 4$ (blue), $l = 5$
  (magenta), with the smallest oscillations shown for $l = 6$ (cyan).
  (Higher order contributions are omitted since their variations
  cannot be seen on the scale of this plot.)  These curves are for a
  circular orbit at $r_{\rm o} = 6M$, which has $u = 0.41$, the
  largest value for the Schwarzschild cases we consider.  As such,
  this case has the slowest convergence among Schwarzschild orbits.
  The falloff with $l$ is more rapid for all other cases.}
\label{fig:schw_converge}
\end{figure}

\begin{figure*}[ht]
\includegraphics[width = 0.48\textwidth]{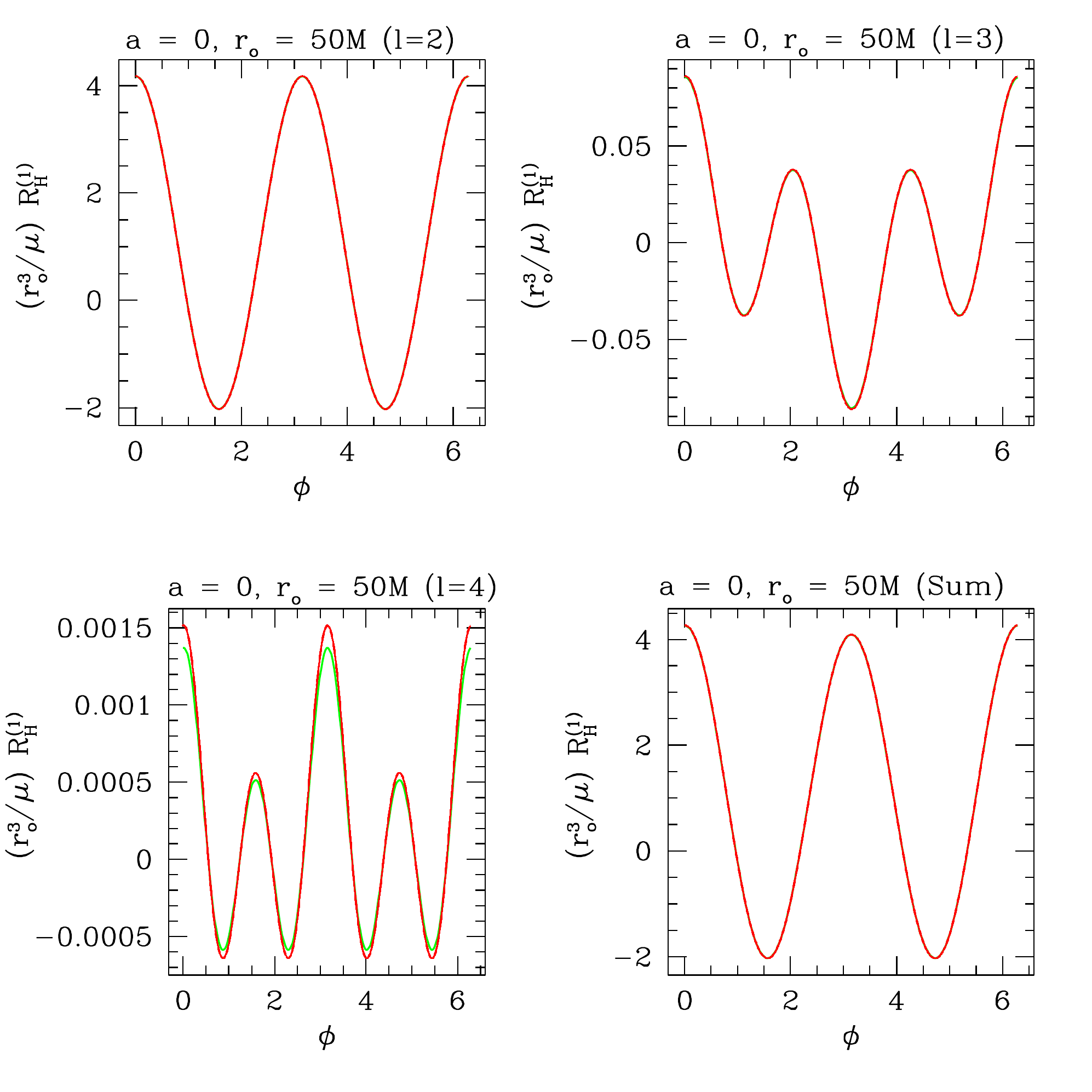}
\includegraphics[width = 0.48\textwidth]{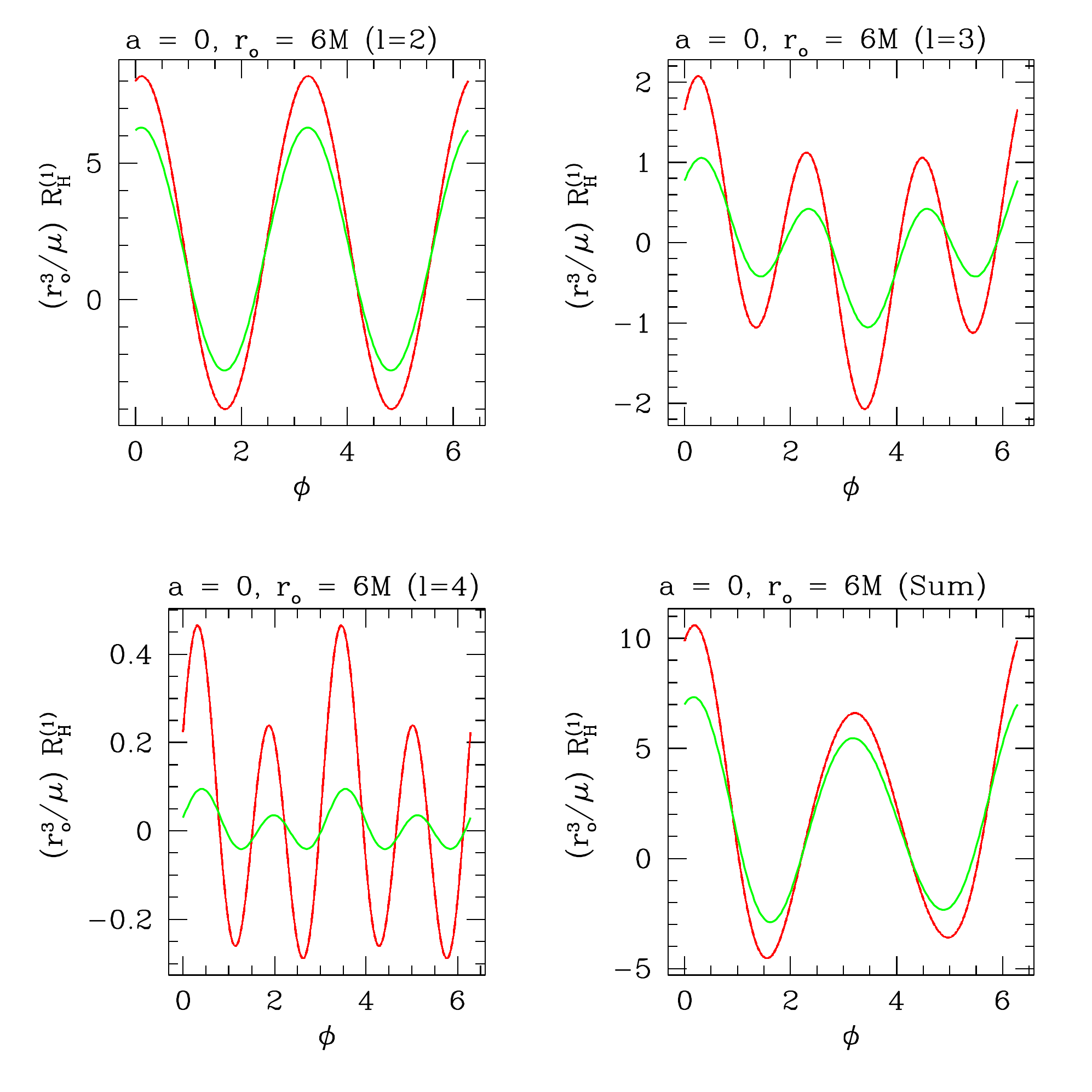}
\caption{Comparison of numerically computed scalar curvature
  perturbation $R^{(1)}_{\rm H}$ for Schwarzschild with the analytic
  expansion given in Eqs.\ (\ref{eq:R20schw})--(\ref{eq:R44schw}).
  The four panels on the left compare numerical (dark gray curves; red
  in color) and analytic (light gray; green in color) results for an
  orbit at $r_{\rm o} = 50M$.  Panels on the right are for $r_{\rm o}
  = 6M$.  In both cases, we plot $(r_{\rm o}^3/\mu)R^{(1)}_{\rm H}$,
  scaling out the leading dependence on orbital radius and the
  orbiting body's mass.  We show contributions for $l = 2$, $l = 3$,
  and $l = 4$, plus the sum of these modes.  For $r_{\rm o} = 50M$, we
  have $u = 0.14$, and we see very good agreement between the
  numerical and analytic formulas.  In several cases, the numerical
  data lie on top of the analytic curves.  For $r_{\rm o} = 6M$, $u =
  0.41$, and the agreement is not as good.  Although the amplitudes
  disagree in the strong field (especially for large $l$), the two
  computations maintain good phase agreement well into the strong
  field.}
\label{fig:schw_Rvsphi}
\end{figure*}

\begin{figure}[ht]
\includegraphics[width = 0.48\textwidth]{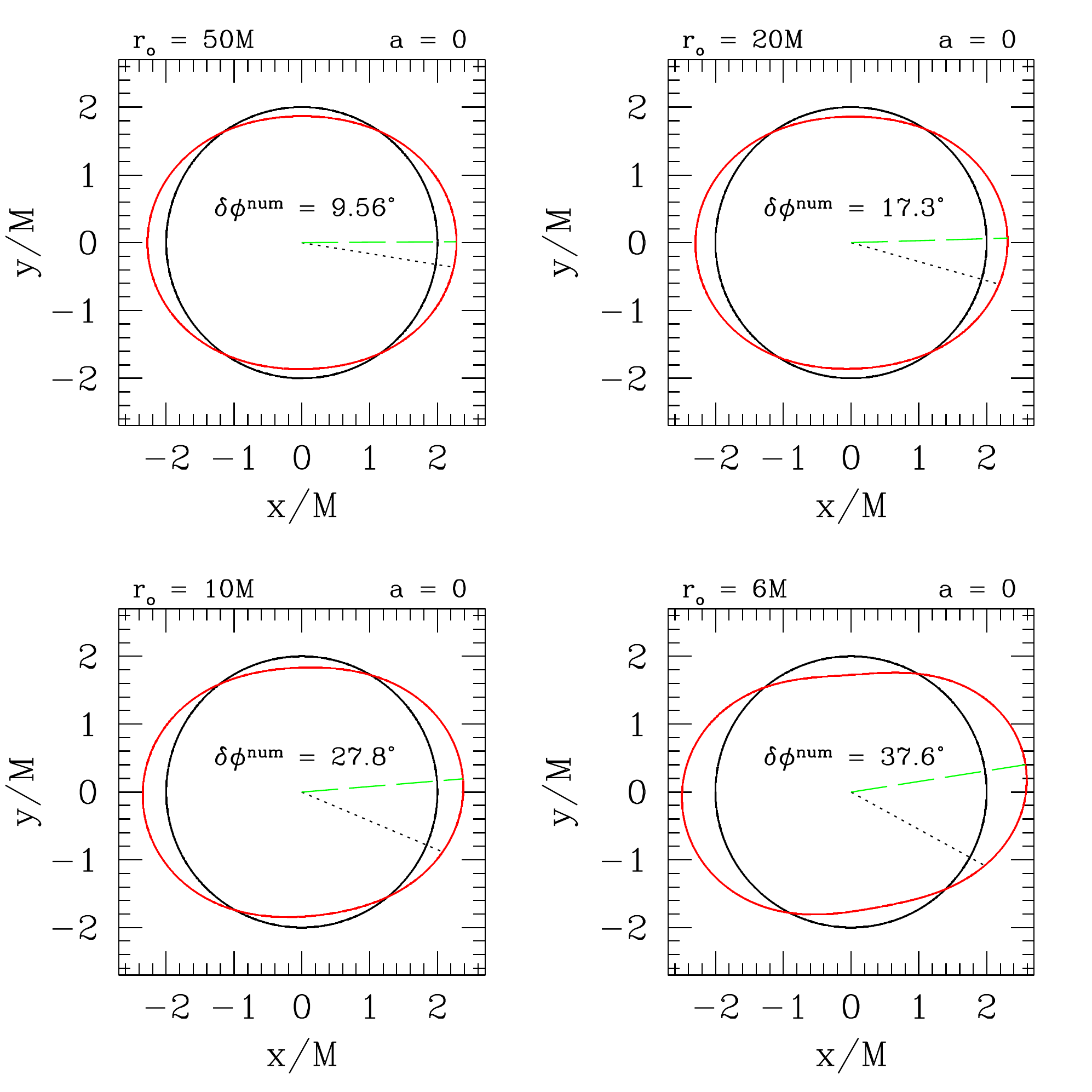}
\caption{Equatorial section of the embedding of a distorted
  Schwarzschild horizon.  Each panel shows the distortion for a
  different orbital radius, varying from $r_{\rm o} = 50M$ to $r_{\rm
    o} = 6M$.  The black circles are the undistorted black hole, and
  the red curves are the distorted horizons, embedded with
  Eq.\ (\ref{eq:schw_spheroid}).  These plots are in a frame that
  corotates with the orbit, and are for a slice of constant ingoing
  time $v$.  The green dashed line in each panel shows the angle at
  which the tidal distortion is largest; the black dotted line shows
  the orbit's position.  Notice that the bulge leads the orbit in all
  cases, with the lead angle growing as the orbit moves to smaller
  orbital radius.  We have rescaled the horizon's tidal distortion by
  a factor $\propto r_{\rm o}^3/\mu$ so that, at leading order, the
  magnitude of the distortion is the same in all plots.}
\label{fig:a0.0_shapes}
\end{figure}

Our numerical results for Schwarzschild black holes are summarized by
Figs.\ {\ref{fig:schw_converge}}, {\ref{fig:schw_Rvsphi}}, and
{\ref{fig:a0.0_shapes}}.  We compute $R^{(1)}_{\rm H}$ by solving for
$Z^{\rm H}_{lm}$ numerically as described in Sec.\ {\ref{sec:ZH}}, and
then applying Eq.\ (\ref{eq:gauss_schw}).  All of our results
illustrate quantities computed in the black hole's equatorial plane,
$\theta = \pi/2$.  We include all contributions up to $l = 15$ in the
sum.  Figure {\ref{fig:schw_converge}} shows that contributions to the
horizon's scalar curvature converge quite rapidly.  The contributions
from $l = 15$ are about $10^{-9}$ of the total for the most extreme
case we consider here, $r_{\rm o} = 6M$.

Figure {\ref{fig:schw_Rvsphi}} compares the analytic predictions for
$R^{\rm H}$ [Eqs.\ (\ref{eq:R20schw})--(\ref{eq:R44schw})] with
numerical results for $l = 2$, $l = 3$, and $l = 4$, and for two
different orbital radii ($r_{\rm o} = 50M$ and $6M$).  The agreement
is outstanding for the large radius orbit.  Our numerical and analytic
predictions can barely be distinguished at $l = 2$ and $l = 3$, and
differ by about $10\%$ at maximum for $l = 4$ (where our analytic
formula includes only the leading contribution to the curvature).  The
agreement is much poorer at small radius.  At $r_{\rm o} = 6M$,
disagreement is several tens of percent for $l = 2$, rising to a
factor $\sim 5$ for $l = 4$.  For both the large and small radius
cases we show, the sum over modes is dominated by the contribution
from $l = 2$.  The phase agreement between analytic and numerical
formulas is quite good all the way into the strong field, even when
the amplitudes differ significantly.

Figure {\ref{fig:a0.0_shapes}} shows distorted black holes by
embedding the horizon in a 3-dimensional Euclidean space, as discussed
in Sec.\ {\ref{sec:l=2embed}}.  Now, we do not truncate at $l = 2$,
but include all moments that we calculate.  We show the equatorial
slices of our embeddings for several different circular orbits
($r_{\rm o} = 50M$, $20M$, $10M$, and $6M$).  In all of our plots, we
scale the horizon distortion $\varepsilon_{lm}$ by a factor
proportional to $r_{\rm o}^3/\mu$ so that the tide's impact is of
roughly the same magnitude for all orbital separations.

The embeddings are shown in a frame that corotates with the orbit at
an instant $v = {\rm constant}$.  The $x$-axis is at $\phi = 0$, so
the orbiting body sits at $\phi = \Delta\phi(r_{\rm o}) = -\Omega_{\rm
  orb}r^*_{\rm o}$.  In each panel, we have indicated where the radius
of the embedding is largest (green dashed line, showing the angle of
greatest tidal distortion) and the angular position of the orbiting
body (black dotted line).  In all cases, the bulge leads the orbiting
body's position, just as predicted in Sec.\ {\ref{sec:schw_bulge}}.
The numerical value of the bulge's position relative to the orbit,
$\delta\phi^{\rm num}$, agrees quite well with $\delta\phi^{\rm
  OB}_{22}$, Eq.\ (\ref{eq:phi_ob_2}) From
Fig.\ {\ref{fig:a0.0_shapes}}, we have
\begin{eqnarray}
\delta\phi^{\rm num} &=& 9.56^\circ\quad r_{\rm o} = 50M\;,
\nonumber\\
&=& 17.3^\circ\quad r_{\rm o} = 20M\;,
\nonumber\\
&=& 27.8^\circ\quad r_{\rm o} = 10M\;,
\nonumber\\
&=& 37.6^\circ\quad r_{\rm o} = 6M\;.
\end{eqnarray}
Equation (\ref{eq:phi_ob_2}) tells us
\begin{eqnarray}
\delta\phi^{\rm OB}_{22}
&=& 9.54^\circ\quad r_{\rm o} = 50M\;,
\nonumber\\
&=& 17.1^\circ\quad r_{\rm o} = 20M\;,
\nonumber\\
&=& 26.8^\circ\quad r_{\rm o} = 10M\;,
\nonumber\\
&=& 35.0^\circ\quad r_{\rm o} = 6M\;.
\end{eqnarray}
In all cases, the true position of the bulge is slightly larger than
$\delta\phi^{\rm OB}_{22}$.  This appears to be due in large part to
the contribution of modes other than $l = |m| = 2$; the agreement
improves if we calculate $\delta\phi^{\rm num}$ using only the $l = 2$
contribution to the embedding.

\section{Results II: Kerr}
\label{sec:kerr_results}

Now consider non-zero black hole spin.  We begin with slow motion and
small black hole spin, expanding Eq.\ (\ref{eq:gauss_kerr}) using $u
\equiv (M/r_{\rm o})^{1/2} \ll 1$ and $q \equiv a/M \ll 1$, and derive
analytic results which are useful points of comparison to the general
case.  We then show numerical results which illustrate tidal
deformations for strong-field orbits.

\subsection{Slow motion: Analytic results}
\label{sec:kerr_analytic}

Here we present analytic results, expanding in powers of $u =
(M/r_{\rm o})^{1/2}$ and $q = a/M$.  We take all relevant quantities
to order $u^5$ and $q$ beyond the leading term; this is far enough to
see how quantities behave for $l \le 4$.  We compare with strong-field
numerical results in the following subsection.

\begin{widetext}

Begin again with ${\cal C}_{lm}$.  Neglect the $k$ and $n$ indices
which are irrelevant for circular, equatorial orbits, and expand
$\lambda = \lambda_0 + (a\omega_m)\lambda_1$, with $\lambda_0$ and
$\lambda_1$ given by Eqs.\ (\ref{eq:lambda0}) and (\ref{eq:lambda1})
for $s = -2$ [recall that $\lambda$ comes from the spheroidal harmonic
  $S^-_{lm}(\theta)$].  Finally, expand to $O(u^5)$ and $O(q)$.  Doing
so, Eq.\ (\ref{eq:Clmkn}) yields
\begin{eqnarray}
{\cal C}_{2m} &=& -\frac{16i}{3}M^2\left(1 - \frac{13}{3}qm^2u^3\right)
\exp\left[-im\left(\frac{13}{2}u^3 - \frac{3}{2}q\right)\right]\;,
\label{eq:C2m_kerr}\\
{\cal C}_{3m} &=& -\frac{16i}{15}M^2\left(1 - \frac{14}{3}qm^2u^3\right)
\exp\left[-im\left(\frac{61}{10}u^3 - \frac{3}{2}q\right)\right]\;,
\label{eq:C3m_kerr}\\
{\cal C}_{4m} &=& -\frac{16i}{45}M^2\left(1 - \frac{24}{5}qm^2u^3\right)
\exp\left[-im\left(\frac{181}{30}u^3 - \frac{3}{2}q\right)\right]\;.
\label{eq:C4m_kerr}
\end{eqnarray}
These reduce to the Schwarzschild results, Eqs.\ (\ref{eq:C2m_schw})
-- (\ref{eq:C4m_schw}), when $q \to 0$.

Next, the amplitudes $Z^{\rm H}_{lm}$, again following the algorithm
described in Sec.\ {\ref{sec:ZH}}.  These results should be understood
to neglect contributions of $O(u^6)$, $O(q^2)$ and higher.  As
elsewhere, $\mu$ is the mass of the smaller body.  For $l = 2$, we
have
\begin{eqnarray}
Z^{\rm H}_{20} &=& \sqrt{\frac{3 \pi }{10}}\frac{\mu}{r_{\rm
    o}^3}\left(1 + \frac{7}{2}u^2 - 4 q u^3 + \frac{561}{56} u^4 - 18
q u^5\right)\;,
\label{eq:ZHkerr20}\\
Z^{\rm H}_{21} &=& -3 i \sqrt{\frac{\pi }{5}}\frac{\mu}{r_{\rm o}^3}
\biggl\{\left(1 - \frac{i}{2}q\right)u - \frac{2}{3}qu^2 +
\left(\frac{8}{3}-\frac{4i}{3}q\right)u^3 + \left[\frac{10i}{3} +
  \left(\frac{1}{6} - \frac{\pi^2}{3}\right)q\right]u^4 +
\left(\frac{152}{21} - \frac{368}{63}q\right)u^5\biggr\}
\nonumber\\
&=& -3i\sqrt{\frac{\pi}{5}}\frac{\mu}{r_{\rm o}^3} \biggl[u -
  \frac{2}{3}qu^2 + \frac{8}{3} u^3 - \left(\frac{3}{2} +
  \frac{\pi^2}{3}\right)qu^4 +
  \frac{152}{21}u^5\biggr]\exp\left[i\left(\frac{10}{3}u^3 -
  \frac{q}{2}\right)\right]\;,
\label{eq:ZHkerr21}\\
Z^{\rm H}_{22} &=& -\frac{3}{2} \sqrt{\frac{\pi}{5}}\frac{\mu}{r_{\rm
    o}^3} \biggl\{1 - i q + \left(\frac{3}{2} -
  \frac{3i}{2}q\right)u^2 + \left[\frac{23i}{3} + \left(15 -
    \frac{4\pi^2}{3}\right)q\right]u^3 + \left(\frac{1403}{168} -
  \frac{1403i}{168}q\right)u^4
\nonumber\\
& & + \left[\frac{473i}{30} + \left(\frac{2449}{90} -
2\pi^{2}\right)q\right]u^5\biggr\}
\nonumber\\
&=& -\frac{3}{2} \sqrt{\frac{\pi}{5}}\frac{\mu}{r_{\rm o}^3} \biggl[1
  + \frac{3}{2}u^2 + \left(\frac{22}{3} -\frac{4\pi ^2}{3}\right)q u^3
  + \frac{1403}{168}u^4 + \left(\frac{103}{9} -
  2\pi^2\right)qu^5\biggr]\exp\left[i\left(\frac{23}{3}u^3 +
  \frac{64}{15}u^5 - q\right)\right]\;.
\nonumber\\
\label{eq:ZHkerr22}
\end{eqnarray}
For $l = 3$,
\begin{eqnarray}
Z^{\rm H}_{30} &=& -i\sqrt{\frac{30\pi}{7}}\frac{\mu}{r_{\rm o}^3}
\left(u^3 - \frac{3}{4}qu^4 + 4u^5\right)\;,
\label{eq:ZHkerr30}
\\
Z^{\rm H}_{31} &=&
-\frac{3}{2}\sqrt{\frac{5\pi}{14}}\frac{\mu}{r_{\rm o}^3}
\left\{\left(1 - \frac{i}{6}q\right)u^2 + \left(\frac{13}{3} -
  \frac{13i}{18}q\right)u^4 + \left[\frac{43i}{30} -
  \left(\frac{247}{180} + \frac{\pi^2}{3}\right)q\right]u^5\right\}
\nonumber\\
&=& -\frac{3}{2}\sqrt{\frac{5\pi}{14}}\frac{\mu}{r_{\rm o}^3}
\left[u^2 + \frac{13}{3}u^4 - \left(\frac{29}{18} +
  \frac{\pi^2}{3}\right)qu^5\right]\exp\left[i\left(\frac{43}{30}u^3 -
  \frac{q}{6}\right)\right]\;,
\label{eq:ZHkerr31}
\\
Z^{\rm H}_{32} &=& 5i\sqrt{\frac{\pi}{7}}\frac{\mu}{r_{\rm o}^3}
\left[\left(1 - \frac{i}{3}q\right)u^3 - \frac{3}{4}qu^4 + 4\left(1 -
  \frac{i}{3}q\right)u^5\right]\nonumber\\
&=& 5i\sqrt{\frac{\pi}{7}}\frac{\mu}{r_{\rm o}^3}\left(u^3 -
\frac{3}{4}qu^4 + 4 u^5\right)\exp\left(-iq/3\right)\;,
\label{eq:ZHkerr32}
\\
Z^{\rm H}_{33} &=& \frac{5}{2}\sqrt{\frac{3\pi}{14}}\frac{\mu}{r_{\rm
    o}^3} \left\{\left(1 - \frac{i}{2}q\right)u^2 + \left(3 -
  \frac{3i}{2}q\right)u^4 + \left[\frac{43i}{10} +
  \left(\frac{393}{20} - 3\pi^2\right)q\right]u^5\right\}
\nonumber\\
&=& \frac{5}{2}\sqrt{\frac{3\pi}{14}}\frac{\mu}{r_{\rm o}^3} \left[u^2
  + 3u^4 + \left(\frac{35}{2} - 3\pi^2\right)qu^5\right]
\exp\left[i\left(\frac{43}{10}u^3 - \frac{q}{2}\right)\right]\;.
\label{eq:ZHkerr33}
\end{eqnarray}
And for $l = 4$,
\begin{eqnarray}
& &Z^{\rm H}_{40} =
-\frac{9}{14}\sqrt{\frac{5\pi}{2}}\frac{\mu}{r_{\rm o}^3} u^4\;,
\label{eq:ZHkerr40}
\\
& &Z^{\rm H}_{41} = \frac{45i}{14}\sqrt{\frac{\pi}{2}}
\frac{\mu}{r_{\rm o}^3} \left[\left(1 + \frac{i}{12}q\right)u^5\right]
= \frac{45i}{14}\sqrt{\frac{\pi}{2}}\frac{\mu}{r_{\rm
    o}^3}u^5\exp\left(iq/12\right)\;,
\label{eq:ZHkerr41}
\\
& &Z^{\rm H}_{42} = \frac{15}{14}\sqrt{\pi}\frac{\mu}{r_{\rm o}^3}
\left[\left(1 + \frac{i}{6}q\right)u^4\right]
= \frac{15}{14}\sqrt{\pi}\frac{\mu}{r_{\rm o}^3}u^4
\exp\left(iq/6\right)\;,
\label{eq:ZHkerr42}
\\ 
& &Z^{\rm H}_{43} = -\frac{15i}{2} \sqrt{\frac{\pi}{14}}
\frac{\mu}{r_{\rm o}^3}\left[\left(1 + \frac{i}{4}q\right)u^5\right]
= -\frac{15i}{2}\sqrt{\frac{\pi}{14}}\frac{\mu}{r_{\rm o}^3} u^5
\exp\left(iq/4\right)\;,
\label{eq:ZHkerr43}
\\
& &Z^{\rm H}_{44} = -\frac{15}{4}\sqrt{\frac{\pi}{7}}\frac{\mu}{r_{\rm
    o}^3} \left[\left(1 + \frac{i}{3}q\right)u^4\right] =
-\frac{15}{4}\sqrt{\frac{\pi}{7}}\frac{\mu}{r_{\rm
    o}^3}u^4\exp\left(iq/3\right)\;.
\label{eq:ZHkerr44}
\end{eqnarray}
Equations (\ref{eq:ZHkerr20}) -- (\ref{eq:ZHkerr44}) reduce to
Eqs.\ (\ref{eq:ZHschw20}) -- (\ref{eq:ZHschw44}) when $q \to 0$.
Modes for $m < 0$ can be obtained using the rule $Z^{\rm H}_{l-m} =
(-1)^l\bar Z^{\rm H}_{lm}$, with overbar denoting complex conjugate.

Lastly, we need the angular function $\bar\eth\bar\eth S^+_{lm}$ to
leading order in $q$.  Using Eqs.\ (\ref{eq:Lminus_s_costheta}),
(\ref{eq:Lminus_s_sintheta}), (\ref{eq:barethbareth_small_a}), and the
condition $q \ll 1$, we have
\begin{equation}
\bar\eth\bar\eth S^+_{lm} = \frac{1}{8M^2}L_-^sL_-^s\left(1 +
iq\cos\theta\right)S^+_{lm} = \frac{1}{8M^2} \left[(1 +
  iq\cos\theta)L_-^sL_-^sS^+_{lm} - 2iq\sin\theta L_-^s
  S^+_{lm}\right]\;.
\label{eq:barethbarethexpand}
\end{equation}
Following the analysis in Appendix {\ref{app:spheroidal_lin}}, the
spheroidal harmonic to this order is
\begin{equation}
S^+_{lm} = {_2}Y_{lm} + qM\omega_m\left[c^{l+1}_{lm} {_2}Y_{(l+1)m}
  + c^{l-1}_{lm} {_2}Y_{(l-1)m}\right]\;,
\label{eq:spheroidalexpand}
\end{equation}
where
\begin{eqnarray}
c^{l+1}_{lm} &=& -\frac{2}{(l+1)^2}\sqrt{\frac{(l + 3)(l - 1)(l + m + 1)(l
    - m + 1)}{(2l + 3)(2l + 1)}}\;,
\\
c^{l-1}_{lm} &=& \frac{2}{l^2}\sqrt{\frac{(l + 2)(l - 2)(l + m)(l -
    m)}{(2l + 1)(2l - 1)}}\;.
\end{eqnarray}
Using Eq.\ (\ref{eq:sphericalharmoniclower}) with
Eqs.\ (\ref{eq:barethbarethexpand}) and (\ref{eq:spheroidalexpand})
and expanding to leading order in $q$, we find
\begin{eqnarray}
\bar\eth\bar\eth S^+_{lm} &=& \frac{1}{8M^2}\biggl[(1 +
  iq\cos\theta)\sqrt{(l+2)(l+1)l(l-1)}\,{_0}Y_{lm} -
  2iq\sin\theta\sqrt{(l + 2)(l - 1)}\,{_1}Y_{lm}
\nonumber\\
& &\qquad + qM\omega_m\left(c^{l+1}_{lm}\sqrt{(l + 3)(l + 2)(l +
    1)l}\,{_0}Y_{(l+1)m} + c^{l-1}_{lm}\sqrt{(l + 1)l(l - 1)(l - 2)}\,
  {_0}Y_{(l-1)m}\right)\biggr]\;.
\label{eq:barethbarethSexpand}
\end{eqnarray}

As in Sec.\ {\ref{sec:schw_analytic}}, it is convenient to combine
modes in pairs.  For $l = 2$, we find
\begin{eqnarray}
R^{(1)}_{{\rm H},20} &=& -\frac{\mu}{r_{\rm o}^3} \left(3\cos^2\theta
- 1\right) \left(1 + \frac{7}{2}u^2 - 4qu^3 + \frac{561}{56}u^4 -
18qu^5\right)\;,
\label{eq:R20kerr}\\
R^{(1)}_{{\rm H},2-1} + R^{(1)}_{{\rm H},21} &=& \frac{4\mu}{r_{\rm
    o}^3} \left(5\cos^2\theta - 1\right)\sin\theta\, q\left(u +
\frac{8}{3}u^3 + \frac{152}{21}u^5\right) \cos\left(\psi - \Omega_{\rm
  orb}v - \frac{3}{2}u^3 + 6M\Omega_{\rm H}\right)\;,
\label{eq:R21kerr}\\
R^{(1)}_{{\rm H},2-2} + R^{(1)}_{{\rm H},22} &=&
\frac{3\mu}{r_{\rm o}^3}\sin^2\theta \left[1 + \frac{3}{2}u^2 -
  \left(10 + \frac{4\pi^2}{3}\right)qu^3 + \frac{1403}{168}u^4 -
  \left(\frac{131}{9} + 2\pi^2\right)qu^5\right]
\times
\nonumber\\
& &\quad
 \cos\left[2\left(\psi - \Omega_{\rm orb} v - \frac{8}{3}u^3 +
   \frac{32}{5}u^5 + \frac{14}{3}M\Omega_{\rm H}\right)\right]\;.
\label{eq:R22kerr}
\end{eqnarray}
For $l = 3$,
\begin{eqnarray}
R^{(1)}_{{\rm H},30} &=& -\frac{\mu}{r_{\rm o}^3}\left(1 -
12\cos^2\theta + 15\cos^4\theta\right)qu^3(1 + 4u^2)\;,
\label{eq:R30kerr}\\
R^{(1)}_{{\rm H},3-1} + R^{(1)}_{{\rm H},31} &=&
\frac{3}{2}\frac{\mu}{r_{\rm o}^3} \sin\theta\left(1 -
5\cos^2\theta\right) u^2\left[1 + \frac{13}{3}u^2
  -\left(\frac{113}{12} + \frac{\pi^2}{2}\right)qu^3\right]
\cos\left[\psi - \Omega_{\rm orb} v - \frac{14}{3}u^3 +
  \frac{20}{3}M\Omega_{\rm H}\right]\;,
\nonumber\\
\label{eq:R31kerr}\\
R^{(1)}_{{\rm H},3-2} + R^{(1)}_{{\rm H},32} &=&
\frac{5}{3}\frac{\mu}{r_{\rm o}^3}q(u^3 + 4u^5)\left\{9\cos^2\theta
\cos\left[2\left(\psi - \Omega_{\rm orb} v - \frac{56}{10}u^3 +
  \frac{22}{3}M\Omega_{\rm H}\right)\right]
\right.\nonumber\\
& &\qquad\qquad\qquad\qquad\left.
 - \cos\left[2\left(\psi - \Omega_{\rm orb} v - \frac{158}{30}u^3 +
   \frac{22}{3}M\Omega_{\rm H}\right)\right]\right\}\;,
\label{eq:R32kerr}\\
R^{(1)}_{{\rm H},3-3} + R^{(1)}_{{\rm H},33} &=&
\frac{5}{2}\frac{\mu}{r_{\rm o}^3}\sin^3\theta u^2\left[1 + 3u^2 -
  \left(\frac{49}{2} + 3\pi^2\right)qu^3\right]
\cos\left[3\left(\psi - \Omega_{\rm orb} v - \frac{14}{3}u^3 +
  \frac{20}{3}M\Omega_{\rm H}\right)\right]\;.
\nonumber\\
\label{eq:R33kerr}
\end{eqnarray}
And for $l = 4$,
\begin{eqnarray}
R^{(1)}_{{\rm H},40} &=& \frac{9}{56}\frac{\mu}{r_{\rm o}^3}
\left(3 - 30\cos^2\theta + 35\cos^4\theta\right)u^4\;,
\label{eq:R40kerr}\\
R^{(1)}_{{\rm H},4-1} + R^{(1)}_{{\rm H},41} &=&
-\frac{9}{28}\frac{\mu}{r_{\rm
    o}^3}\sin\theta\,qu^5\left[98\cos^4\theta\cos\left(\psi -
  \Omega_{\rm orb}v - \frac{169}{30}u^3 + \frac{25}{3}M\Omega_{\rm
    H}\right)
\right.
\nonumber\\
& &\quad\left.
- 57\cos^2\theta\cos\left(\psi - \Omega_{\rm orb}v -
\frac{1568}{285}u^3 + \frac{25}{3}M\Omega_{\rm H}\right)
+ 3\cos\left(\psi - \Omega_{\rm orb}v - \frac{77}{15}u^3 +
\frac{25}{3}M\Omega_{\rm H}\right)\right]\;,
\nonumber\\
\label{eq:R41kerr}\\
R^{(1)}_{{\rm H},4-2} + R^{(1)}_{{\rm H},42} &=&
\frac{15}{14}\frac{\mu}{r_{\rm o}^3}\sin^2\theta \left(1 -
7\cos^2\theta\right)u^4 \cos\left[2\left(\psi - \Omega_{\rm orb} v -
  \frac{181}{30}u^3 + \frac{119}{15}M\Omega_{\rm H}\right)\right]\;,
\label{eq:R42kerr}\\
R^{(1)}_{{\rm H},4-3} + R^{(1)}_{{\rm H},43} &=&
\frac{3}{4}\frac{\mu}{r_{\rm o}^3}\sin^3\theta\,qu^5
\left\{14\cos^2\theta\cos\left[3\left(\psi - \Omega_{\rm orb}v -
  \frac{169}{30}u^3 + \frac{25}{3}M\Omega_{\rm H}\right)\right]
\right.
\nonumber\\
& &\quad\qquad\qquad\qquad\left.
- \cos\left[3\left(\psi - \Omega_{\rm orb}v - \frac{77}{15}u^3 +
  \frac{25}{3}M\Omega_{\rm H}\right)\right]\right\}\;,
\label{eq:R43kerr}\\
R^{(1)}_{{\rm H},4-4} + R^{(1)}_{{\rm H},44} &=&
\frac{15}{8}\frac{\mu}{r_{\rm o}^3}\sin^4\theta u^4
\cos\left[4\left(\psi - \Omega_{\rm orb} v - \frac{181}{30}u^3 +
  \frac{119}{15}M\Omega_{\rm H}\right) \right]\;.
\label{eq:R44kerr}
\end{eqnarray}
\end{widetext}

In writing these formulas, we have used the fact that $\Omega_{\rm H}
= q/4M$ in the $q \ll 1$ limit to rewrite certain terms in the phases
using $\Omega_{\rm H}$ rather than $q$.  For example, in
Eq.\ (\ref{eq:R22kerr}) our calculation yields a term $7q/6$ in the
argument of the cosine, which we rewrite $14M\Omega_{\rm H}/3$.  We
have found that this improves the match of Eqs.\ (\ref{eq:R20kerr}) --
(\ref{eq:R44kerr}) with the numerical results we discuss in
Sec.\ {\ref{sec:kerr_numerical}}.

\subsubsection{Phase of the tidal bulge: Null map}

We begin by examining the bulge-orbit offset using the null map,
Eq.\ (\ref{eq:bulge_vs_orbit_null}).  The horizon's geometry is
dominated by contributions for which $l + m$ is even; modes with $l +
m$ odd are suppressed by $qu$ relative to these dominant modes (thus
vanishing in the Schwarzschild limit).  The dominant modes peak at
$\psi^{\rm bulge}_{lm} = \Omega_{\rm orb}v + \delta\psi_{lm}(u) +
\delta\psi_{lm}(q)$, where $\delta\psi_{lm}(u)$ and
$\delta\psi_{lm}(q)$ can be read out of Eqs.\ (\ref{eq:R20kerr}) --
(\ref{eq:R44kerr}).  The orbit mapped onto the horizon in the null map
is given by Eq.\ (\ref{eq:psi_o_nullmap}).  Following discussion in
Sec.\ {\ref{sec:nullmap}}, the offset phases in the null map for the
dominant modes, to $O(u^5)$ and $O(q)$, are
\begin{eqnarray}
\delta\psi^{\rm OB-NM}_{22} &=& \frac{8}{3}\left(u^3 - M\Omega_{\rm H}\right)
- \frac{32}{5}u^5 - \frac{4M^2\Omega_{\rm H}}{r_{\rm o}}
\nonumber\\
& & - \Delta\psi(r_{\rm o})\;,
\label{eq:psi_obnm_2}\\
\delta\psi^{\rm OB-NM}_{31} &=& \delta\psi^{\rm OB-NM}_{33}
\nonumber\\
&=& \frac{14}{3}\left(u^3 - M\Omega_{\rm H}\right) -
\frac{4M^2\Omega_{\rm H}}{r_{\rm o}} - \Delta\psi(r_{\rm o})\;,
\nonumber\\
\label{eq:psi_obnm_3}\\
\delta\psi^{\rm OB-NM}_{42} &=& \delta\psi^{\rm OB-NM}_{44}
\nonumber\\
&=& \frac{181}{30}u^3 - \frac{89}{15}M\Omega_{\rm H}
- \frac{4M^2\Omega_{\rm H}}{r_{\rm o}} - \Delta\psi(r_{\rm o})\;.
\nonumber\\
\label{eq:psi_obnm_4}
\end{eqnarray}
We again see agreement with Fang and Lovelace for $l = m = 2$, who
correct a sign error in Hartle's {\cite{hartle74}} treatment of the
bulge phase; compare Eq.\ (61) and footnote 6 of Ref.\ {\cite{fl05}}
and associated discussion.  In contrast to the Schwarzschild case, the
Kerr offset phases can be positive or negative, depending on the
values of $r_{\rm o}$ and $q$.  To highlight this further, let us
examine Eq.\ (\ref{eq:psi_obnm_2}) for very large $r_{\rm o}$: we drop
the term in $u^5$, and expand $\Delta\psi(r_{\rm o})$.  The result is
\begin{equation}
\delta\psi^{\rm OB-NM}_{22} \simeq \frac{8}{3}\left(u^3 - M\Omega_{\rm
  H}\right) + \sqrt{\frac{M}{r_{\rm o}}}\;.
\label{eq:psi_obnm_2_v2}
\end{equation}
As $r_{\rm o} \to \infty$, we see that this bulge lags the orbit by
$\delta^{\rm OB-NM}_{22} = -8M\Omega_{\rm H}/3$, which reproduces
Hartle's finding for a stationary moon orbiting a slowly rotating Kerr
black hole [Eq.\ (4.34) of Ref.\ {\cite{hartle74}}, correcting the
  sign error discussed in footnote 6 of Ref.\ {\cite{fl05}}].  We
discuss this point further in Sec.\ {\ref{sec:leadlag}}.

\subsubsection{Phase of the tidal bulge: Instantaneous map}

Consider next the instantaneous-in-$v$ map discussed in
Sec.\ {\ref{sec:instmap}}.  The position of the orbit on the horizon
in this mapping is given by Eq.\ (\ref{eq:bulge_vs_orbit_inst}).  To
$O(u^5)$ and $O(q)$, the offset phase for the dominant modes in this
map is
\begin{eqnarray}
\delta\psi^{\rm OB-IM}_{22} &=& \frac{8}{3}u^3 -
\frac{14}{3}M\Omega_{\rm H} - \frac{32}{5}u^5 - \Delta\psi(r_{\rm
  o})\;,
\nonumber\\
\label{eq:psi_obim_2}\\
\delta\psi^{\rm OB-IM}_{31} &=& \delta\psi^{\rm OB-IM}_{33}
\nonumber\\
&=& \frac{14}{3}u^3 - \frac{20}{3}M\Omega_{\rm H} - \Delta\psi(r_{\rm
  o})\;,
\label{eq:psi_obim_3}\\
\delta\psi^{\rm OB-IM}_{42} &=& \delta\psi^{\rm OB-IM}_{44}
\nonumber\\
&=& \frac{181}{30}u^3 - \frac{119}{15}M\Omega_{\rm H} -
\Delta\psi(r_{\rm o})\;.
\label{eq:psi_obim_4}
\end{eqnarray}
As in the null map, these phases can be positive or negative,
depending on the values of $r_{\rm o}$ and $q$.  As we'll see when we
examine numerical results for the horizon geometry,
Eq.\ (\ref{eq:psi_obim_2}) does a good job describing the angle of the
peak horizon bulge for small values of $q$.

\subsubsection{Phase of the tidal bulge: Tidal field versus tidal
response}

Finally, let us examine the relative phase of tidal field modes
$\psi^{\rm HH}_{0,lm}$ and the horizon's response $R^{(1)}_{{\rm
    H},lm}$.  For $q \ll 1$, we have $\kappa^{-1} = 4M + O(q^2)$.
Expanding in the weak-field limit, Eq.\ (\ref{eq:deltapsiTB}) becomes
\begin{equation}
\delta\psi^{\rm TB}_{lm} = 4m(u^3 - M\Omega_{\rm H}) + {\cal
  S}_{lm}(\pi/2)\;.
\label{eq:deltaTB_weak}
\end{equation}
For the modes with $l + m$ even which dominate the horizon's response,
it is not difficult to compute ${\cal S}_{lm}(\pi/2)$ to leading order
in $q$.  Equation (\ref{eq:barethbarethSexpand}) and the definition
(\ref{eq:spheroidratio}) yield
\begin{equation}
{\cal S}_{lm}(\pi/2) = \frac{2q}{\sqrt{l(l+1)}}
  \frac{{_1}Y_{lm}(\pi/2)}{{_0}Y_{lm}(\pi/2)} + O(q^2)\;.
\label{eq:Sphase_smallq}
\end{equation}
We also know [cf.\ Eq.\ (A8) of Ref.\ {\cite{h2000}}] that
\begin{equation}
{_1}Y_{lm}(\theta) = -\frac{1}{\sqrt{l(l+1)}}\left(\partial_\theta -
m\csc\theta\right){_0}Y_{lm}(\theta)\;.
\end{equation}
For $l + m$ even, $\partial_\theta\,{_0}Y_{lm} = 0$ at $\theta =
\pi/2$.  Plugging the resulting expression for ${_1}Y_{lm}(\pi/2)$
into Eq.\ (\ref{eq:Sphase_smallq}), we find
\begin{equation}
{\cal S}_{lm}(\pi/2) = \frac{2mq}{l(l+1)} = \frac{8mM\Omega_{\rm
    H}}{l(l+1)}\;,
\end{equation}
where in the last step we again used $q = 4M^2\Omega_{\rm H}$,
accurate for $q \ll 1$.  With this, Eq.\ (\ref{eq:deltaTB_weak})
becomes
\begin{eqnarray}
\delta\psi^{\rm TB}_{lm} &=& 4m\left[u^3 - M\Omega_{\rm H}\left(1 -
  \frac{2}{l(l+1)}\right)\right]
\nonumber\\
&=& 4m\left[u^3 - M\Omega_{\rm H}\frac{(l+2)(l-1)}{l(l+1)}\right]\;.
\label{eq:deltaTB}
\end{eqnarray}
Just as with the offset phases of the bulge and the orbit for Kerr,
this tidal bulge phase can be either positive or negative depending on
$r_{\rm o}$ and $q$, and so the horizon's response can lead or lag the
applied tidal field.

\subsection{Fast motion: Numerical results}
\label{sec:kerr_numerical}

\begin{figure*}[ht]
\includegraphics[width = 0.48\textwidth]{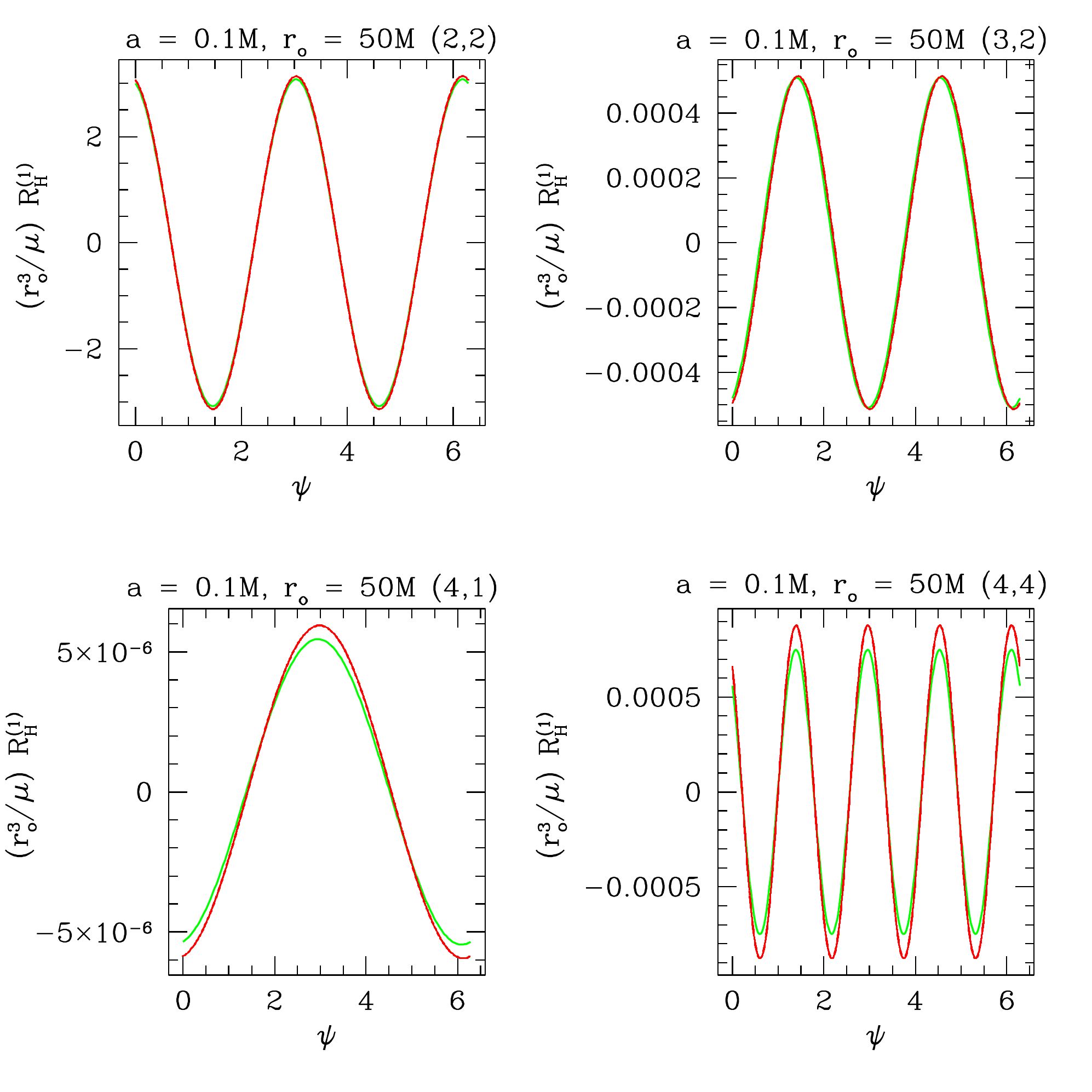}
\includegraphics[width = 0.48\textwidth]{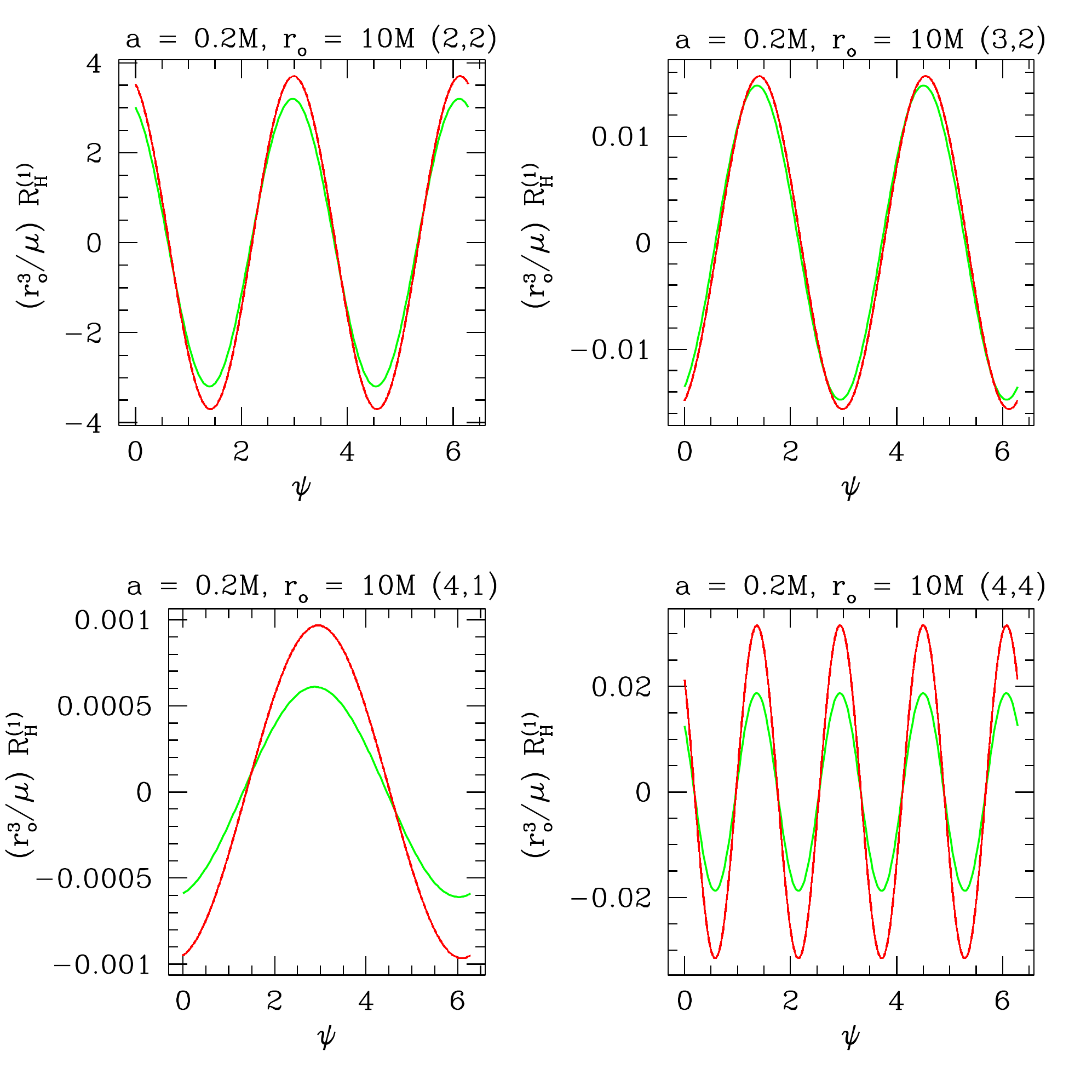}
\caption{Comparison of selected modes for the numerically computed
  scalar curvature perturbation $R^{(1)}_{{\rm H},lm}$ with the
  analytic expansion given in Eqs.\ (\ref{eq:R20kerr}) --
  (\ref{eq:R44kerr}).  The four panels on the left are for orbits of a
  black hole with $a = 0.1M$ at $r_{\rm o} = 50M$; those on the right
  are for orbits of a black hole with $a = 0.2M$ at $r_{\rm o} = 10M$.
  The mode shown is indicated by $(l,m)$ in the upper right corner of
  each panel [we actually show the contributions from $(l,m)$ and
    $(l,-m)$].  In all cases, we plot $(r_{\rm o}^3/\mu)R^{(1)}_{\rm
    H}$, scaling out the leading dependence on orbital radius and the
  orbiting body's mass.  Curves in light gray (green in color) are the
  analytic results, those in dark gray (red in color) are our
  numerical data.  Agreement for the large radius, low spin cases is
  extremely good, especially for small $l$ where the numerical data
  lies practically on top of the analytic predictions.  As we increase
  $q$ and decrease $r_{\rm o}$, the amplitude agreement becomes less
  good, though the analytic formulas still are within several to
  several tens of percent of the numerical data.  The phase agreement
  is outstanding in all of these cases.  }
\label{fig:kerr_Rvspsi}
\end{figure*}

\begin{figure*}[ht]
\includegraphics[width = 0.48\textwidth]{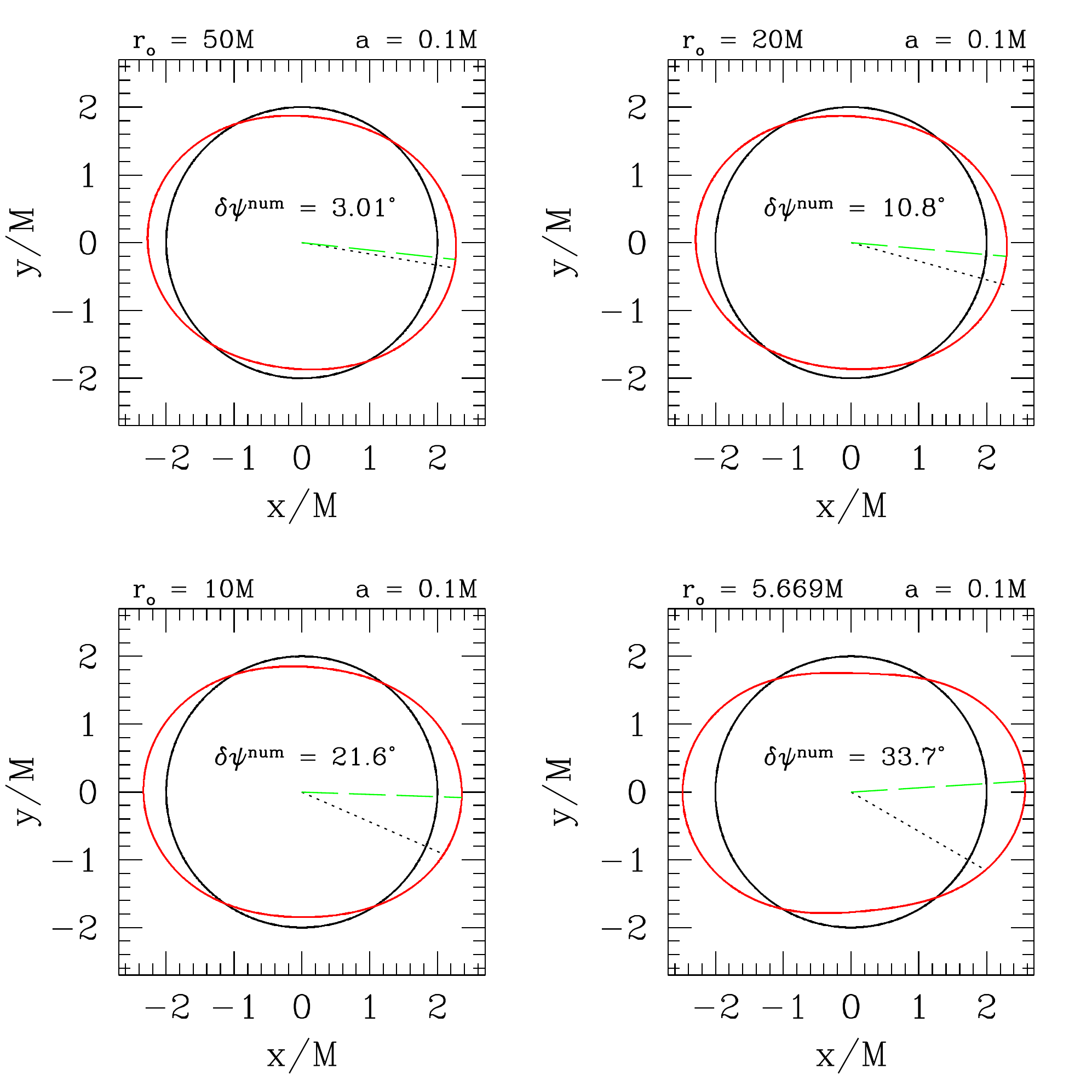}
\includegraphics[width = 0.48\textwidth]{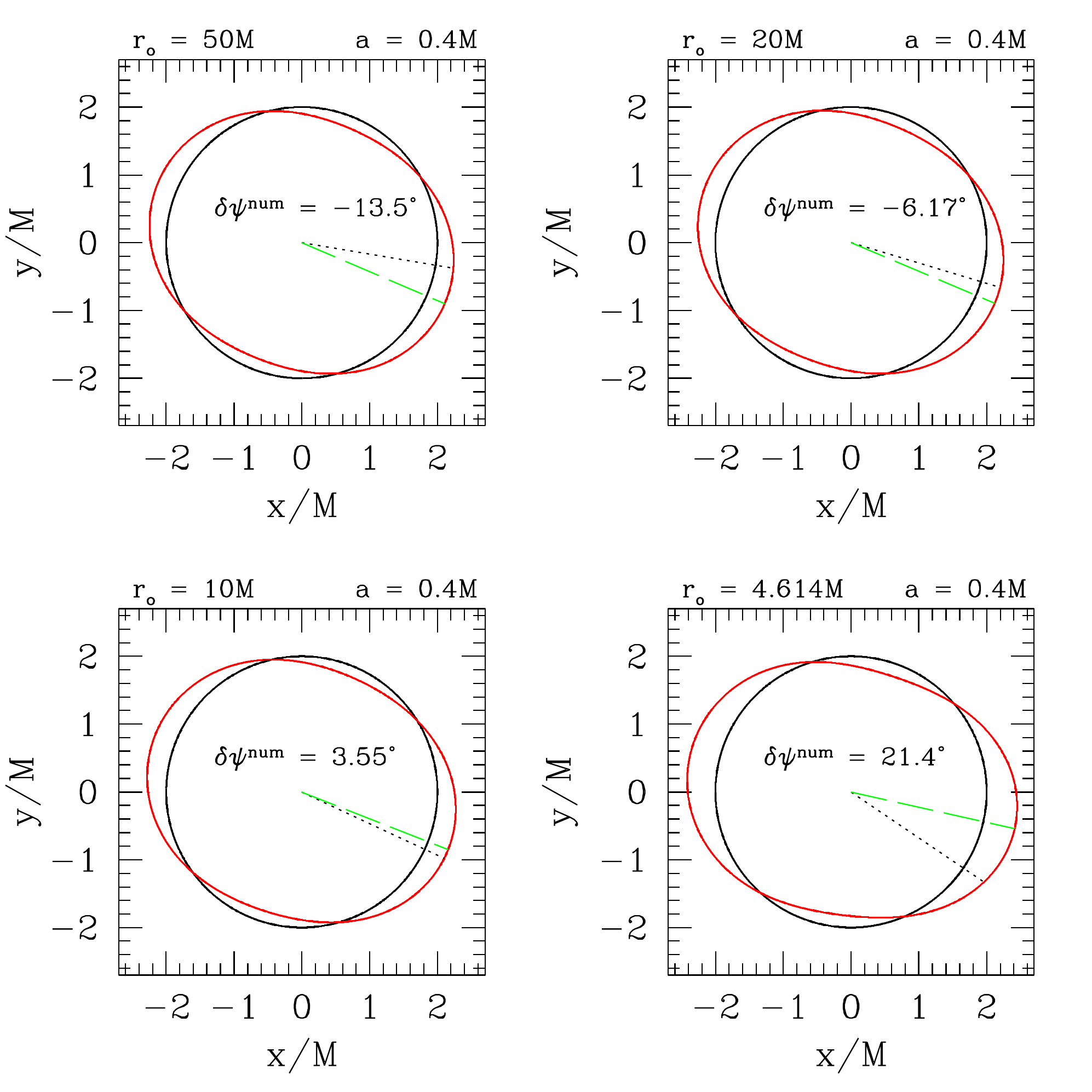}
\caption{Equatorial section of the embedding of distorted Kerr black
  hole event horizons, $a = 0.1M$ and $a = 0.4M$.  Each panel
  represents the distortion for a different radius of the orbiting
  body, varying from $r_{\rm o} = 50M$ to the innermost stable
  circular orbit ($r_{\rm o} = 5.669M$ for $a = 0.1M$, $r_{\rm o} =
  4.614M$ for $a = 0.4M$).  As in Fig.\ {\ref{fig:a0.0_shapes}}, the
  green dashed line shows the angle at which the tidal distortion is
  largest, and the black dotted line shows the position of the orbit.
  As in Fig.\ {\ref{fig:a0.0_shapes}}, we have rescaled by a factor
  $\propto r_{\rm o}^3/\mu$ to account for the leading dependence of
  the tide on mass and orbital separation.  In contrast to the
  Schwarzschild results, the bulge does not lead the orbit in all
  cases here.  The amount by which the bulge leads the orbit grows as
  the orbit moves to small orbital radius (in some cases, changing
  from a lag to a lead as part of this trend).}
\label{fig:a0.1_a0.4_shapes}
\end{figure*}

\begin{figure*}[ht]
\includegraphics[width = 0.48\textwidth]{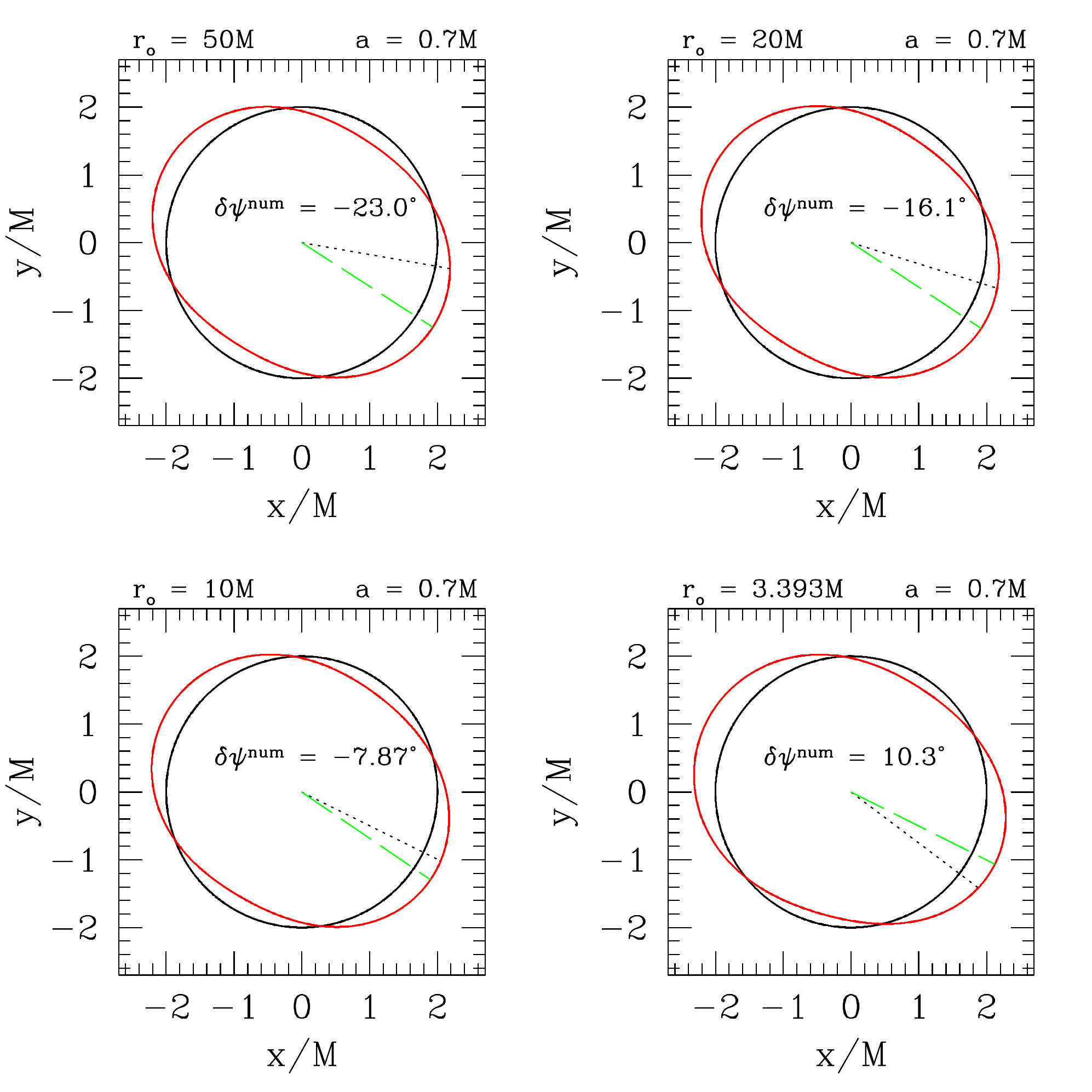}
\includegraphics[width = 0.48\textwidth]{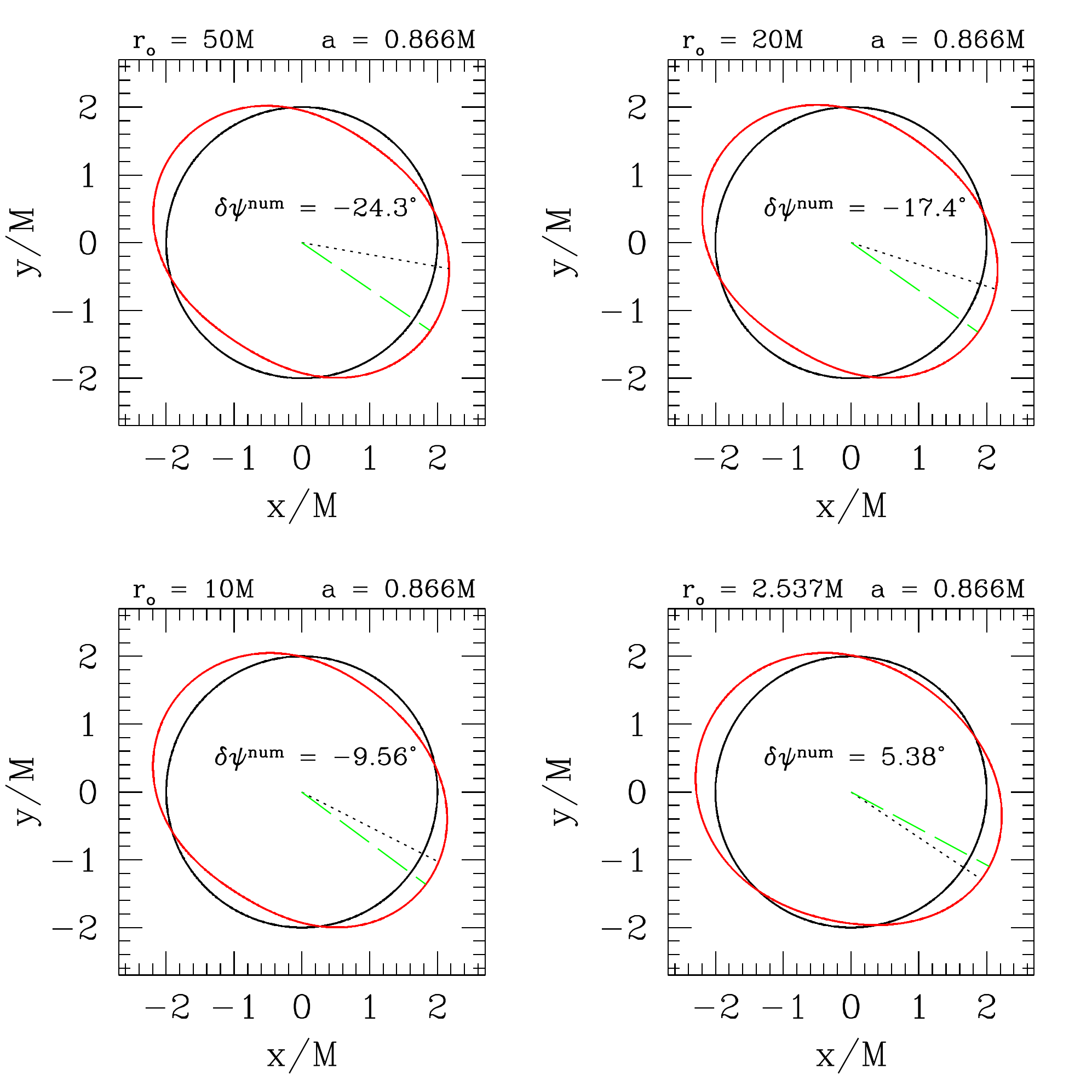}
\caption{Equatorial section of the embedding of distorted Kerr black
  hole event horizons, $a = 0.7M$ and $a = 0.866M$.  Each panel
  represents the distortion for a different radius of the orbiting
  body, varying from $r_{\rm o} = 50M$ to the innermost stable
  circular orbit ($r_{\rm o} = 3.393M$ for $a = 0.7M$, $r_{\rm o} =
  2.537M$ for $a = 0.866M$), with the green dashed and black dotted
  lines labeling the locations of maximal distortion and position of
  the orbit, respectively, and with the distortion rescaled by a
  factor $\propto r_{\rm o}^3/\mu$.  The bulge lags the orbit in most
  cases we show here, with the lag angle getting smaller and
  converting to a small lead as the orbit moves to smaller and smaller
  orbital radius.}
\label{fig:a0.7_a0.866_shapes}
\end{figure*}

Figures {\ref{fig:kerr_Rvspsi}}, {\ref{fig:a0.1_a0.4_shapes}}, and
{\ref{fig:a0.7_a0.866_shapes}} present summary data for our numerical
calculations of tidally distorted Kerr black holes.  Just as in
Sec.\ {\ref{sec:schw_numerical}}, we compute $R^{(1)}_{\rm H}$ by
solving for $Z^{\rm H}_{lm}$ as described in Sec.\ {\ref{sec:ZH}}, and
then apply Eq.\ (\ref{eq:gauss_kerr}).  As in the Schwarzschild case,
we find rapid convergence with mode index $l$.  All the data we show
are for the equatorial plane, $\theta = \pi/2$, and are rescaled by
$(r_{\rm o}^3/\mu)$.  We typically include all modes up to $l = 15$
(increasing this to $20$ and $25$ in a few very strong field cases).
Contributions beyond this are typically at the level of $10^{-9}$ or
smaller, which is accurate enough for this exploratory analysis.

Figure {\ref{fig:kerr_Rvspsi}} is the Kerr analog of
Fig.\ {\ref{fig:schw_Rvsphi}}, comparing numerical results for $R^{\rm
  H}_{lm}$ with analytic predictions for selected black hole spins,
mode numbers, and orbital radii.  For all modes we show here, we see
outstanding agreement in both phase and amplitude for $q = 0.1$ and
$r_{\rm o} = 50M$; in some cases, the numerical data lies almost
directly on top of the analytic prediction.  The amplitude agreement
is not quite as good as we increase the spin to $q = 0.2$ and move to
smaller radius ($r_{\rm o} = 10M$), though the phase agreement remains
quite good for all modes.

Figures {\ref{fig:a0.1_a0.4_shapes}} and
{\ref{fig:a0.7_a0.866_shapes}} show equatorial slices of the embedding
of distorted Kerr black holes for a range of orbits and black hole
spins.  These embeddings are similar to those we used for distorted
Schwarzschild black holes (as described in Sec.\ {\ref{sec:l=2embed}}),
with a few important adjustments.  The embedding surface we use has the
form
\begin{equation}
r_{\rm E} = r_{\rm E}^{0}(\theta) + r_+\sum_{\ell m}\varepsilon_{\ell
  m}(\theta,\psi)\;.
\end{equation}
Both the undistorted radius $r_{\rm E}^{(0)}(\theta)$ and the tidal
distortion $\varepsilon_{\ell m}(\theta,\psi)$ are described in
Appendix {\ref{app:embed}}; see also Ref.\ {\cite{smarr}}.  The
background embedding reduces to a sphere of radius $2M$ when $a = 0$,
but is more complicated in general.  The embedding's tidal distortion
is linearly related to the curvature $R^{(1)}_{{\rm H},lm}$, but in a
way that is more complicated than the Schwarzschild relation
(\ref{eq:schw_spheroid}).  In particular, mode mixing becomes
important: Different angular basis functions are needed to describe
the curvature $R^{(1)}_{{\rm H},lm}$ and the embedding distortion
$\varepsilon_{\ell m}$ when $a \ne 0$.  Hence, the $\ell = 2$
contribution to the horizon's shape has contributions from all $l$
curvature modes, not just $l = 2$.  See Appendix {\ref{app:embed}} for
detailed discussion.

In this paper, we only generate embeddings for $a/M \le \sqrt{3}/2$.
For spins greater than this, the horizon cannot be embedded in a
global 3-dimensional Euclidean space.  A ``belt'' from $\pi -
\theta_{\rm E} \le \theta \le \theta_{\rm E}$ can always be embedded
in 3-dimensional Euclidean space, but the ``polar cones'' $0 \le
\theta < \theta_{\rm E}$ and $\pi - \theta_{\rm E} < \theta \le \pi$
must be embedded in a Lorentzian geometry (where $\theta_{\rm E}$ is
related to the root of a function used in the embedding; see Appendix
{\ref{app:embed}} for details).  Alternatively, one can embed the
entire horizon in a different space, as discussed in
Refs.\ {\cite{frolov,gibbons}}.  We defer detailed discussion of
embeddings that can handle the case $a/M > \sqrt{3}/2$ to a later
paper.

As with the Schwarzschild embeddings shown in
Fig.\ {\ref{fig:a0.0_shapes}}, the Kerr embeddings we show are all
plotted in a frame that corotates with the orbit at a moment $v = {\rm
  constant}$.  The $x$-axis is at $\psi = 0$, and the orbiting body
sits at $\psi = \Delta\psi(r_{\rm o}) = \bar r_{\rm o} - \Omega_{\rm
  orb}r^*_{\rm o}$.  As in Fig.\ {\ref{fig:a0.0_shapes}}, the green
dashed line labels the horizon's peak bulge, and the black dotted line
shows the position of the orbiting body.

For small $q$, we find that the numerically computed bulge offset
agrees quite well with the $l = 2$ analytic expansion in the
instantaneous map, Eq.\ (\ref{eq:psi_obim_2}).  For $q = 0.1$, our
numerical results are
\begin{eqnarray}
\delta\psi^{\rm num} &=& 3.01^\circ\quad r_{\rm o} = 50M\;,
\nonumber\\
&=& 10.8^\circ\quad r_{\rm o} = 20M\;,
\nonumber\\
&=& 21.6^\circ\quad r_{\rm o} = 10M\;,
\nonumber\\
&=& 33.7^\circ\quad r_{\rm o} = 5.669M\;.
\label{eq:q=0.1_num}
\end{eqnarray}
These are within a few percent of predictions based on the weak-field,
slow spin expansion:
\begin{eqnarray}
\delta\psi^{\rm OB-IM}_{22}
&=& 2.95^\circ\quad r_{\rm o} = 50M\;,
\nonumber\\
&=& 10.7^\circ\quad r_{\rm o} = 20M\;,
\nonumber\\
&=& 20.6^\circ\quad r_{\rm o} = 10M\;,
\nonumber\\
&=& 30.0^\circ\quad r_{\rm o} = 5.669M\;.
\end{eqnarray}
As we move to larger spin, the agreement rapidly becomes worse.  Terms
which we neglect in our expansion become important, and the mode
mixing described above becomes very important.  For $q = 0.4$, the
agreement degrades to a few tens of percent in most cases:
\begin{eqnarray}
\delta\psi^{\rm num} &=& -13.5^\circ\quad r_{\rm o} = 50M\;,
\nonumber\\
&=& -6.17^\circ\quad r_{\rm o} = 20M\;,
\nonumber\\
&=& 3.55^\circ\quad r_{\rm o} = 10M\;,
\nonumber\\
&=& 21.4^\circ\quad r_{\rm o} = 4.614M\;;
\label{eq:q=0.4_num}
\end{eqnarray}
and
\begin{eqnarray}
\delta\psi^{\rm OB-IM}_{22}
&=& -17.9^\circ\quad r_{\rm o} = 50M\;,
\nonumber\\
&=& -9.76^\circ\quad r_{\rm o} = 20M\;,
\nonumber\\
&=& 0.82^\circ\quad r_{\rm o} = 10M\;,
\nonumber\\
&=& 13.6^\circ\quad r_{\rm o} = 4.614M\;.
\end{eqnarray}
The agreement gets significantly worse as $q$ is increased further.
Presumably, $q \sim 0.3$ is about as far as the leading order
expansion in $q$ can reasonably be taken.

\begin{figure*}[ht]
\includegraphics[width = 0.497\textwidth]{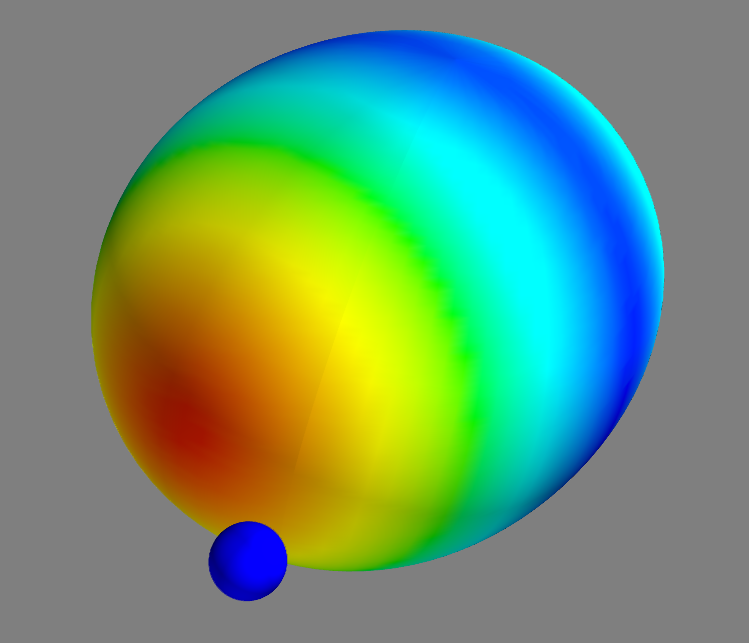}
\includegraphics[width = 0.463\textwidth]{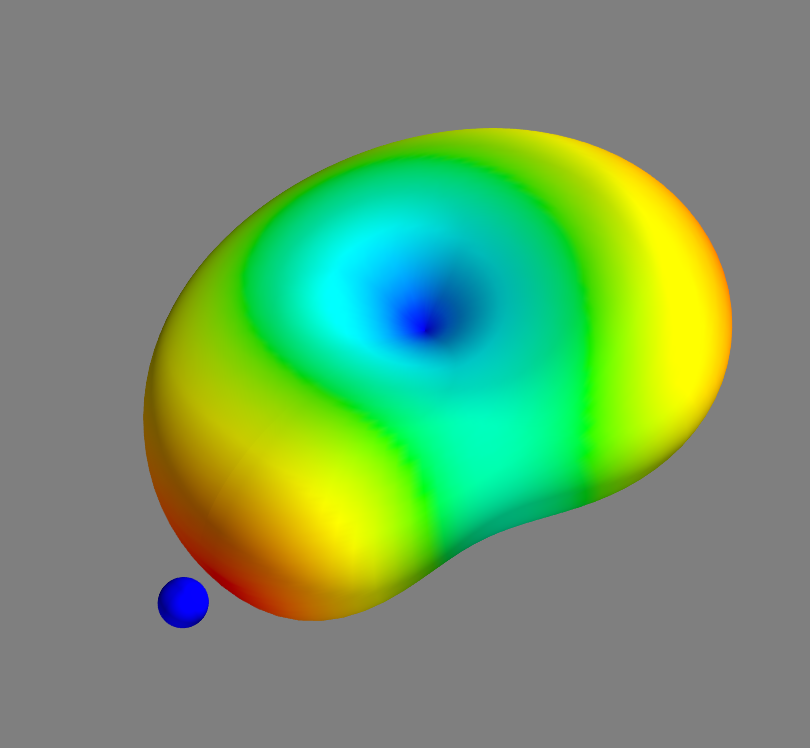}
\caption{Two example embeddings of the tidally distorted horizon's
  surface.  Both panels show the 3-dimensional Euclidean embedding
  surface, $r_{\rm E}(\theta,\psi)$; the shading (or color scale)
  indicates the horizon's distortion relative to an isolated Kerr
  black hole.  The hole is stretched (i.e., $r_{\rm E}$ increased by
  the tides relative to an isolated hole; red in color) at the end
  near to and opposite from the orbiting body.  It is squeezed
  ($r_{\rm E}$ decreased by tides; blue in color) in a band between
  these two ends.  As in other figures illustrating the embedded
  distorted horizon, we have rescaled the distortion by a factor
  $\propto r_{\rm o}^3/\mu$.  On the left, we show a relatively gentle
  deformation around a moderately spinning black hole: $a = 0.3M$,
  $r_{\rm o} = 20M$.  The distortion here is dominated by a
  quadrupolar deformation of the horizon (lagging the orbiting body,
  whose angular position is indicated by the small blue ball).  On the
  right, we show a rather extreme case: $a = 0.866M$, $r_{\rm o} =
  1.75M$.  The deformation here is much more complicated, as many
  multipoles beyond $l = 2$ contribute to the shape of the horizon.}
\label{fig:surfaces}
\end{figure*}

To conclude this section, we show two examples of embeddings for the
entire horizon surface, rather than just the equatorial slice.  The
left-hand panel of Fig.\ {\ref{fig:surfaces}} is an example of a
relatively mild tidal distortion.  The black hole has spin $a = 0.3M$,
and the orbiting body is at $r_{\rm o} = 20M$.  The distortion is
strongly dominated by the $\ell = 2$ contribution, and we see a fairly
simple prolate ellipsoid whose bulge lags the orbit.  The right-hand
panel shows a much more extreme example.  The black hole here has $a =
0.866M$, and the orbiting body is at $r_{\rm o} = 1.75M$.  The
horizon's shape has strong contributions from many multipoles, and so
is bent in a rather more complicated way than in the mild case.  The
connection between the orbit and the horizon geometry is quite unusual
here.  Note that this extreme case corresponds to an {\it unstable}
circular orbit, and so one might question whether this figure is
physically relevant.  We include it because we expect similar horizon
distortions for very strong field orbits of black holes with $a/M >
\sqrt{3}/2$, and that such a horizon geometry will be produced
transiently from the closest approach of eccentric orbits around black
holes with $a/M \lesssim \sqrt{3}/2$.  Both of these cases will be
investigated more thoroughly in later papers.

\section{Lead or lag?}
\label{sec:leadlag}

We showed in Sec.\ {\ref{sec:downhoriz}} that the orbital energy
evolves due to horizon coupling according to $dE^{\rm H}/dt \propto
(\Omega_{\rm orb} - \Omega_{\rm H})$.  As discussed in the
Introduction, it is simple to build an intuitive picture of this in
Newtonian physics.  For a Newtonian tide acting on a fluid body, when
$\Omega_{\rm H} > \Omega_{\rm orb}$ tidal forces raise a bulge on the
body which leads the orbit's position.  This bulge exerts a torque
which transfers energy from the body's spin to the orbit.  When
$\Omega_{\rm H} < \Omega_{\rm orb}$, the bulge lags the orbit, and the
torque transfers energy from the orbit to the body's spin.  When
$\Omega_{\rm H} = \Omega_{\rm orb}$, $dE^{\rm H}/dt = 0$.  The
Newtonian fluid expectation is thus that there should be no offset
between the bulge and the orbit.  The tidal bulge should point
directly at the orbiting body, locking the body's tide to the orbit.

Consider now a fully relativistic calculation of tides acting on a
black hole.  When $\Omega_{\rm orb} \gg \Omega_{\rm H}$ (e.g., the
Schwarzschild limit) and $\Omega_{\rm H} \gg \Omega_{\rm orb}$ (large
radius orbits of Kerr black holes), the Newtonian fluid intuition is
consistent with our results, modulo the switch of ``lead'' and ``lag''
thanks to the teleological nature of the event horizon.  However, it
is not so clear if this intuition holds up when $\Omega_{\rm orb}$ and
$\Omega_{\rm H}$ are comparable in magnitude.

Let us investigate this systematically.  Begin with the weak-field $l
= m = 2$ offset angles in the null and instantaneous maps,
Eqs.\ (\ref{eq:psi_obnm_2}) and (\ref{eq:psi_obim_2}).  Dropping terms
of $O(u^5)$ and noting that $u^3 = M\Omega_{\rm orb} + O(qu^6)$, we
solve for the conditions under which $\delta\psi^{\rm OB-NM}_{22}$ and
$\delta\psi^{\rm OB-IM}_{22}$ are zero.  In the null map, we find
\begin{equation}
\Omega_{\rm orb} = \Omega_{\rm H} + \frac{3M\Omega_{\rm H}}{2r_{\rm o}} +
\frac{3\Delta\psi_{\rm o}}{8M}\;.
\label{eq:NMoffsetzero}
\end{equation}
The bulge leads the orbit when the equals in the above equation is
replaced by greater than, and lags when replaced by less than.  In the
instantaneous map,
\begin{equation}
\Omega_{\rm orb} = \frac{7}{4}\Omega_{\rm H} + \frac{3\Delta\psi_{\rm
    o}}{8M}\;,
\label{eq:IMoffsetzero}
\end{equation}
with the same replacements indicating lead or lag.

{\it Neither of these conditions are consistent with $\Omega_{\rm orb}
  = \Omega_{\rm H}$ indicating zero bulge-orbit offset.}  In both the
null and instantaneous maps, we find $\Omega_{\rm orb} \ll \Omega_{\rm
  H}$ when the bulge angle is zero.  For example, for $a = 0.3M$
(roughly the largest $a$ for which the small spin expansion is
trustworthy), Eq.\ (\ref{eq:NMoffsetzero}) has a root at $r_{\rm o} =
35.9M$, for which $M\Omega_{\rm orb} = 0.00464$, $M\Omega_{\rm H} =
0.0768$.  (A second root exists at $r_{\rm o} = 2.15M$, but this is
inside the photon orbit.)  Using the instantaneous map changes the
numbers, but not the punchline: for $a = 0.3M$, the root moves to
$r_{\rm o} = 16.7M$, with $M\Omega_{\rm orb} = 0.0146$.  Changing the
spin changes the numbers, but leaves the message the same: zero offset
in these maps does not correspond to $\Omega_{\rm orb} = \Omega_{\rm
  H}$.

\begin{figure*}[ht]
\includegraphics[width = 0.48\textwidth]{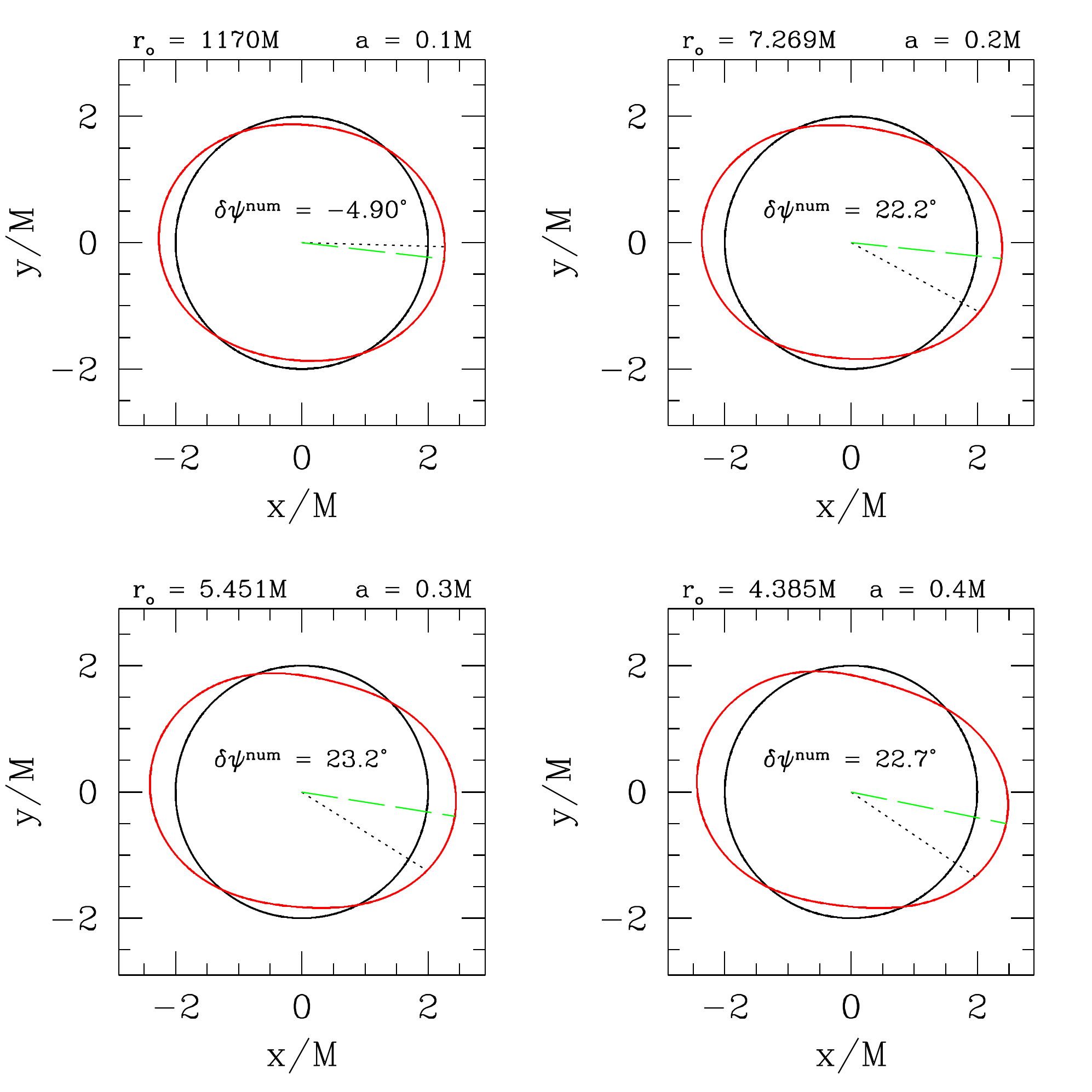}
\includegraphics[width = 0.48\textwidth]{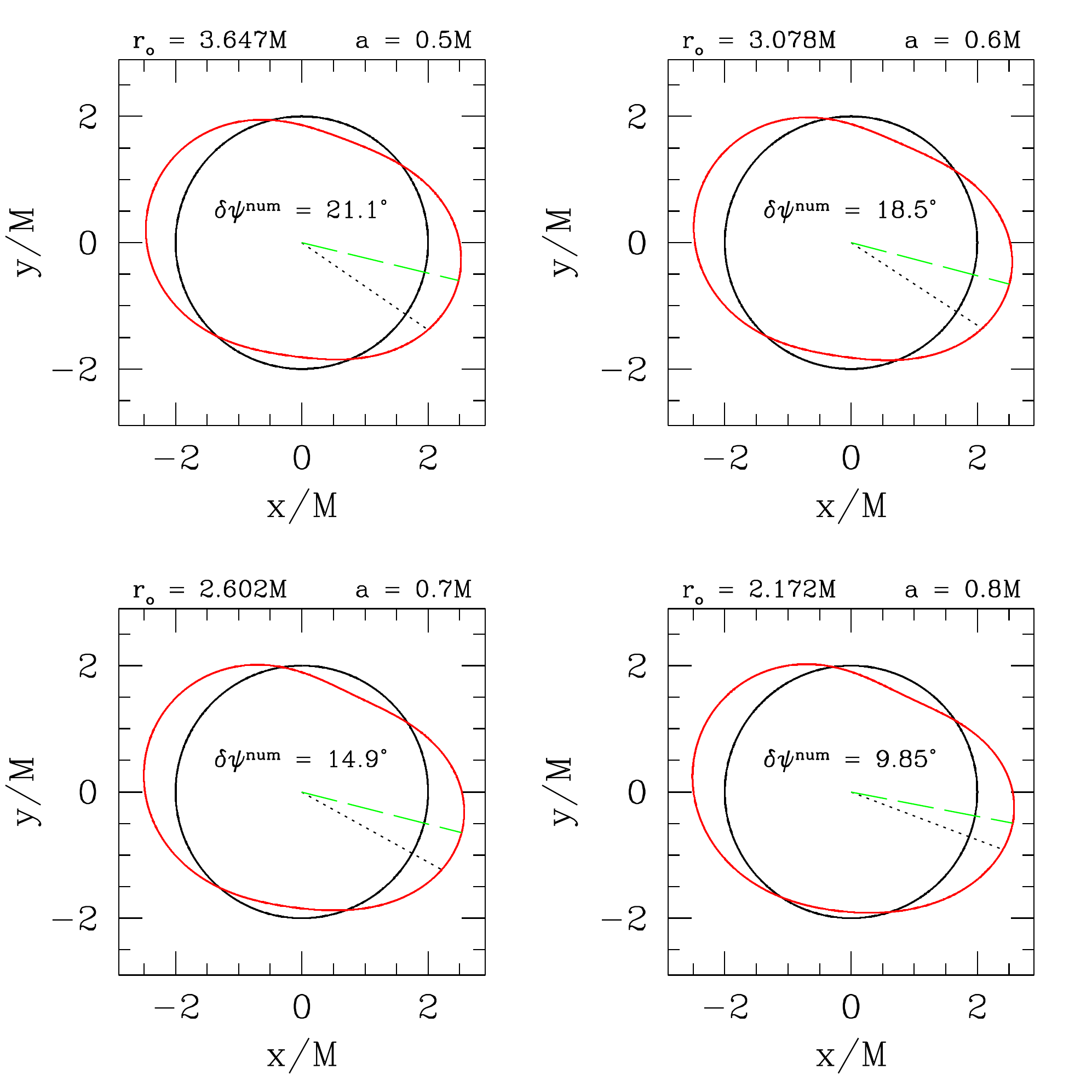}
\caption{Embedding of distorted Kerr black hole event horizons for a
  corotating orbit --- i.e., an orbit for which $\Omega_{\rm orb} =
  \Omega_{\rm H}$.  As in Figs.\ {\ref{fig:a0.0_shapes}},
        {\ref{fig:a0.1_a0.4_shapes}}, and
        {\ref{fig:a0.7_a0.866_shapes}}, the green dashed line points
        along the direction of greatest horizon distortion, and the
        black dotted line points to the orbiting body; the distortions
        are all scaled by a factor $\propto r_{\rm o}^3/\mu$.  At very
        small spins (for which the corotating orbital radius is very
        large), the bulge lags the orbit slightly, but the bulge leads
        for all other spins.}
\label{fig:corotshapes}
\end{figure*}

Equations (\ref{eq:NMoffsetzero}) and (\ref{eq:IMoffsetzero}) were
derived using a small spin expansion.  Before drawing too firm a
conclusion from this, let us examine the situation using numerical
data good for large spin.  In Fig.\ {\ref{fig:corotshapes}}, we
examine a sequence of ``corotating'' orbits --- orbits for which
$\Omega_{\rm H} = \Omega_{\rm orb}$, so that $dE^{\rm H}/dt = 0$.  For
very small spins, the orbit leads the bulge.  As the black hole's spin
increases, the lead becomes a lag.  This lead gets smaller as the spin
gets larger.  Since the lag becomes a lead as the spin is changed from
$a = 0.1M$ to $a = 0.2M$, there must be a spin value between $a =
0.1M$ and $a = 0.2M$ for which the lead angle is zero for the
corotating orbit.  Our data also suggest that the lead angle may
approach zero as the spin gets very large.  But this suggests that the
horizon locks to the orbit for at most only two spin values, in this
map --- a set of measure zero.  We do not find any systematic
connection between the geometry and the horizon for these orbits.

Before concluding, let us examine the relative phase of the tidal
field and the horizon's curvature, Eq.\ (\ref{eq:deltaTB}).  Setting
$\delta\psi^{\rm TB}_{lm} = 0$ yields
\begin{equation}
\Omega_{\rm orb} = \Omega_{\rm H}\,\frac{(l+2)(l-1)}{l(l+1)}\;.
\label{eq:TBoffsetzero}
\end{equation}
We again see $\Omega_{\rm orb} \ne \Omega_{\rm H}$ when the field and
the response are aligned (although $\Omega_{\rm orb} \to \Omega_{\rm
  H}$ as $l$ gets very large).

The analytical expansions and numerical data indicate that the
Newtonian fluid intuition for the geometry of tidal coupling simply
does not work well for strong-field black hole binaries, even
accounting for the teleological swap of ``lag'' and ``lead.''  Only in
the extremes can we make statements with confidence: when $\Omega_{\rm
  H} \gg \Omega_{\rm orb}$, the tidal bulge will lag the orbit; when
$\Omega_{\rm orb} \gg \Omega_{\rm H}$, the bulge will lead the orbit.
But when $\Omega_{\rm orb}$ and $\Omega_{\rm H}$ are of similar
magnitude, we cannot make a clean prediction.

The tidal bulge is {\it not} locked to the orbit when $dE^{\rm H}/dt =
0$, at least using any scheme to define the lead/lag angle that we
have examined.

\section{Conclusions}
\label{sec:conclude}

In this paper, we have presented a formalism for computing tidal
distortions of Kerr black holes.  Using black hole perturbation
theory, our approach is good for fast motion, strong field orbits, and
can be applied to a black hole of any spin parameter.  We have also
developed tools for visualizing the distorted horizon by embedding its
2-dimensional surface in a 3-dimensional Euclidean space.  For now,
our embeddings are only good for Kerr spin parameter $a/M \le
\sqrt{3}/2$, the highest value for which the entire horizon can be
embedded in a globally Euclidean space.  Higher spins require either a
piecewise embedding of an equatorial ``belt'' in a Euclidean space,
and a region near the ``poles'' in a Lorentzian space, or else
embedding in a different space altogether.

Although our formalism is good for arbitrary bound orbits, we have
focused on circular and equatorial orbits for this first analysis.
This allowed us to validate this formalism against existing results in
the literature, and to explore whether there is a simple connection
between the tidal coupling of the hole to the orbit, and the relative
geometry of the orbit and the horizon's tidal bulge.  We find that
there is no such simple connection in general.  Perhaps not
surprisingly, strong-field black hole systems are more complicated
than Newtonian fluid bodies.

We plan two followup analyses to extend the work we have done here.
First, we plan to extend the work on embedding horizons to $a/M >
\sqrt{3}/2$, the domain for which we cannot use a globally Euclidean
embedding.  Work in progress indicates that the simplest and perhaps
most useful approach is to use the globally hyperbolic 3-space $H^3$
{\cite{gibbons}}.  This allows us to treat the entire range of
physical black hole spins, $0 \le a/M \le 1$, using a single global
embedding space.  Second, we plan to examine tidal distortions from
generic --- inclined and eccentric --- Kerr orbits.  The circular
equatorial orbits we have studied in this first paper are stationary,
as are the tidal fields and tidal responses that arise from them.  If
one examines the system and the horizon's response in a frame that
corotates with the orbit, the tide and the horizon will appear static.
This will not be the case for generic orbits.  Even when viewed in a
frame that rotates at the orbit's mean $\phi$ frequency, the orbit
will be dynamical, and so the horizon's response will likewise be
dynamical.  Similar analyses for Schwarzschild have already been
presented by Vega, Poisson, and Massey {\cite{vpm11}}; it will be
interesting to compare with the more complicated and less symmetric
Kerr case.

An extension of our analysis may be useful for improving initial data
for numerical relativity simulations of merging binary black holes.
One source of error in such simulations is that the black holes
typically have the wrong initial geometry --- unless the binary is
extremely widely separated, we expect each hole to be distorted by
their companion's tides.  Accounting for this in the initial data
requires matching the near-horizon geometry to the binary's spacetime
metric; see {\cite{chu14}} for an up-to-date discussion of work to
include tidal effects in a binary's initial data.  Much work has been
done on binaries containing tidally deformed Schwarzschild black holes
{\cite{alvi,yunesetal,mcdanieletal}}, and efforts now focus on the
more realistic case of binaries containing spinning black holes
{\cite{gallouinetal,chu14}}.  With some effort (in order to get the
geometry in a region near the horizon, not just on the horizon), we
believe it should be possible to use this work as an additional tool
for extending the matching procedure to the realistic orbital
geometries of rotating black holes.

\acknowledgments

We thank Eric Poisson for useful discussions and comments, as well as
feedback on an early draft-in-progress on this paper; Robert Penna for
helpful comments and discussion, particularly regarding non-Euclidean
horizon embeddings; Nicol\'as Yunes for suggesting that this technique
might usefully connect to initial data for binary black holes; Daniel
Kennefick for helpful discussions as this project was originally being
formulated; and this paper's referee for very helpful comments and
feedback.  This work was supported by NSF grant PHY-1068720.  SAH
gratefully acknowledges fellowship support by the John Simon
Guggenheim Memorial Foundation, and sabbatical support from the
Canadian Institute for Theoretical Astrophysics and Perimeter
Institute for Theoretical Physics.

\appendix

\section{Details of computing $\bar\eth$}
\label{app:ethdetails}

In this appendix, we present details regarding the operator $\bar\eth$
in the form that we need it for our analysis.

\begin{widetext}

\subsection{The Newman-Penrose tetrad legs}

A useful starting point is to write out the Newman-Penrose tetrad legs
${\bf l}$, ${\bf n}$, and ${\bf m}$.  In much of the literature on
black hole perturbation theory, we use the Kinnersley form of these
tetrad legs in Boyer-Lindquist coordinates:
\begin{eqnarray}
\left(l^\mu\right)_{\rm BL} &\doteq& \frac{1}{\Delta}\left[(r^2 +
  a^2),\Delta,0,a\right]\;,
\\
\left(n^\mu\right)_{\rm BL} &\doteq& \frac{1}{2\Sigma}\left[(r^2 +
  a^2),-\Delta,0,a\right]\;,
\\
\left(m^\mu\right)_{\rm BL} &\doteq& \frac{1}{\sqrt{2}(r +
  ia\cos\theta)} \left[ia\sin\theta, 0, 1, i\csc\theta\right]\;;
\end{eqnarray}
\begin{eqnarray}
\left(l_\mu\right)_{\rm BL} &\doteq& \left[-1, \Sigma/\Delta, 0,
  a\sin^2\theta\right]\;,
\\
\left(n_\mu\right)_{\rm BL} &\doteq& \frac{1}{2\Sigma}\left[-\Delta,
  -\Sigma, 0, a\Delta\sin^2\theta\right]\;,
\\
\left(m_\mu\right)_{\rm BL} &\doteq& \frac{1}{\sqrt{2}(r +
  ia\cos\theta)} \left[-ia\sin\theta, 0, \Sigma, i(r^2 +
  a^2)\sin\theta\right]\;.
\end{eqnarray}
The components of the fourth leg, ${\bf\bar m}$, are related to the
components of ${\bf m}$ by complex conjugation.  The notation
$(b^\mu)_{\rm BL} \doteq (b^t, b^r, b^\theta, b^\phi)$ means ``the
components of the 4-vector ${\bf b}$ in Boyer-Lindquist coordinates
are represented by the array on the right-hand side,'' and similarly
for the 1-form components $(b_\mu)_{\rm BL}$.

Because our analysis focuses on the Kerr black hole event horizon, we
will find it useful to transform to Kerr ingoing coordinates
$(v,r',\theta,\psi)$.  Using Eqs.\ (\ref{eq:vdef}) --
(\ref{eq:psidef}), we transform tetrad components between the two
coordinate systems with the matrix elements
\begin{equation}
\frac{\partial v}{\partial t} = 1\;,
\qquad
\frac{\partial v}{\partial r} = \frac{r^2 + a^2}{\Delta}\;,
\qquad
\frac{\partial\psi}{\partial r} = \frac{a}{\Delta}\;,
\qquad
\frac{\partial\psi}{\partial\phi} = 1\;,
\qquad
\frac{\partial r'}{\partial r} = 1\;.
\end{equation}
All elements which could connect $(t,r,\phi)$ and $(v,r',\psi)$ which
are not explicitly listed here are zero; the angle $\theta$ is the
same in the two coordinate systems.  The matrix elements for the
inverse transformation are
\begin{equation}
\frac{\partial t}{\partial v} = 1\;,
\qquad
\frac{\partial t}{\partial r'} = -\frac{(r^2 + a^2)}{\Delta}\;,
\qquad
\frac{\partial\psi}{\partial r'} = -\frac{a}{\Delta}\;,
\qquad
\frac{\partial\phi}{\partial\psi} = 1\;,
\qquad
\frac{\partial r}{\partial r'} = 1\;.
\end{equation}
As noted in the Introduction, $r$ and $r'$ are identical; we just
maintain a notational distinction for clarity while transforming
between these two different coordinate systems.

With these, it is a simple matter to transform the tetrad components
to their form in Kerr ingoing coordinates:
\begin{eqnarray}
\left(l^\mu\right)_{\rm IN} &\doteq& \frac{1}{\Delta}\left[2[(r')^2 +
  a^2],\Delta,0,2a\right]\;,
\\
\left(n^\mu\right)_{\rm IN} &\doteq& \frac{1}{2\Sigma}
\left[0,-\Delta,0,0\right]\;,
\\
\left(m^\mu\right)_{\rm IN} &\doteq& \frac{1}{\sqrt{2}(r' +
  ia\cos\theta)} \left[ia\sin\theta, 0, 1, i\csc\theta\right]\;;
\end{eqnarray}
\begin{eqnarray}
\left(l_\mu\right)_{\rm IN} &\doteq& \left[-1, 2\Sigma/\Delta, 0,
  a\sin^2\theta\right]\;,
\\
\left(n_\mu\right)_{\rm IN} &\doteq& \frac{1}{2\Sigma}\left[-\Delta,
  0, 0, a\Delta\sin^2\theta\right]\;,
\\
\left(m_\mu\right)_{\rm IN} &\doteq& \frac{1}{\sqrt{2}(r' +
  ia\cos\theta)} \left[-ia\sin\theta, 0, \Sigma, i[(r')^2 +
  a^2)\sin\theta\right]\;.
\end{eqnarray}
The notation $(b^\mu)_{\rm IN} \doteq (b^v, b^{r'}, b^\theta, b^\psi)$
means ``the components of the 4-vector ${\bf b}$ in Kerr ingoing
coordinates are represented by the array on the right-hand side,'' and
similarly for the 1-form components $(b_\mu)_{\rm IN}$.  In the above
equations, $\Delta$ and $\Sigma$ take their usual forms, but with $r
\to r'$.  At this point, the notational distinction between $r'$ and
$r$ is no longer needed, so we drop the prime on $r$ in what follows.

Changing coordinates is not enough to fix various pathologies
associated with the behavior of quantities on the event horizon.  To
ensure that quantities we examine are well behaved there, we next
change to the Hawking-Hartle tetrad.  This is done in two steps.
First we perform a boost (cf.\ Ref.\ {\cite{stewart}}, Sec.\ 2.6),
putting
\begin{eqnarray}
{\bf l'} &=& \frac{\Delta}{2\varpi^2}\,{\bf l}\;,
\\
{\bf n'} &=& \frac{2\varpi^2}{\Delta}\,{\bf n}\;,
\\
{\bf m'} &=& {\bf m}\;,
\end{eqnarray}
where we've introduced $\varpi^2 = r^2 + a^2$.  This is followed by a
null rotation around ${\bf l}$:
\begin{eqnarray}
{\bf l}_{\rm HH} &=& {\bf l'}\;,
\\
{\bf m}_{\rm HH} &=& {\bf m'} + \bar c \,{\bf l'}\;,
\\
{\bf n}_{\rm HH} &=& {\bf n'} + c\,{\bf m'} + \bar c\,{\bf\bar m'} +
c\bar c\,{\bf l'}\;,
\end{eqnarray}
with
\begin{equation}
c = \frac{ia\sin\theta}{\sqrt{2}(r - ia\cos\theta)}\;.
\end{equation}
With this, we finally obtain the tetrad elements that we need for this
analysis:
\begin{eqnarray}
\left(l^\mu\right)_{\rm HH,\,IN} &\doteq& \frac{1}{\varpi^2}\left[\varpi^2,
\Delta/2,0,a\right]\;,
\\
\left(m^\mu\right)_{\rm HH,\,IN} &\doteq& \frac{1}{\sqrt{2}(r +
  ia\cos\theta)} \left[0,-\frac{ia\Delta\sin\theta}{2\varpi^2}, 1,
  i\csc\theta -\frac{ia^2\sin\theta}{\varpi^2}\right]\;,
\\
\left(n^\mu\right)_{\rm HH,\,IN} &\doteq& \frac{1}{4\varpi^2\Sigma}
\left[-2a^2\varpi^2\sin^2\theta,-4\varpi^4 +
  a^2\Delta\sin^2\theta,0,-4a\varpi^2 + 2a^3\sin^2\theta\right]\;;
\end{eqnarray}
\begin{eqnarray}
\left(l_\mu\right)_{\rm HH,\,IN} &\doteq& \frac{1}{2\varpi^2}
\left[-\Delta, 2\Sigma, 0, a\Delta\sin^2\theta\right]\;,
\\
\left(m_\mu\right)_{\rm HH,\,IN} &\doteq&
  \frac{1}{\sqrt{2}(r + ia\cos\theta)}\times
  \frac{1}{2(r^2 + a^2)}\times
\nonumber\\
& &\left[
        -ia(2\varpi^2 - \Delta)\sin\theta,
        -2ia\Sigma\sin\theta,
        2\varpi^2\Sigma,
        i\left(2\varpi^4 - a^2\Delta\sin^2\theta\right)\sin\theta
        \right]\;,
\\
\left(n_\mu\right)_{\rm HH,\,IN} &\doteq& \frac{1}{4\varpi^2\Sigma}
\left[-4\varpi^4 + a^2\left(4\varpi^2 - \Delta\right)\sin^2\theta,
2a^2\Sigma\sin^2\theta, 0, a^3\Delta\sin^4\theta\right]\;.
\end{eqnarray}
In the remainder of this appendix, we will use the Hawking-Hartle
components in ingoing coordinates, and will drop the ``HH, IN''
subscript.

\subsection{Constructing $\bar\eth$}

Here we derive the form of the operator $\bar\eth$, acting at the
radius of the Kerr event horizon, $r = r_+$.  Following Hartle
{\cite{hartle74}}, $\bar\eth$ acting upon a quantity $\eta$ of
spin-weight $s$ is given by
\begin{equation}
\bar\eth\eta = \left[\bar\delta - s(\alpha - \bar\beta)\right]\eta\;.
\end{equation}
The operator $\bar\delta = \bar m^\mu\partial_\mu$.  Evaluating this
at $r = r_+$ [using the fact that $\Delta = 0$ there, and
that $a/(r_+^2 + a^2) = a/(2Mr_+) = \Omega_{\rm H}$] we find
\begin{equation}
\bar\delta = \frac{1}{\sqrt{2}(r_+ - ia\cos\theta)}
\left[\partial_\theta - i(\csc\theta -
  a\Omega_{\rm H}\sin\theta)\partial_\psi\right]\;.
\label{eq:deltabar}
\end{equation}
Next consider the Newman-Penrose spin coefficients $\alpha$ and
$\beta$.  With the metric signature we use ($-+++$), they are given by
\begin{eqnarray}
\alpha &=& \frac{1}{2}\bar m^\nu\left(\bar m^\mu\nabla_\nu m_\mu
- n^\mu\nabla_\nu l_\mu\right)\;,
\\
\beta &=& \frac{1}{2} m^\nu\left(\bar m^\mu\nabla_\nu m_\mu -
n^\mu\nabla_\nu l_\mu\right)\;.
\end{eqnarray}
This means that
\begin{equation}
\alpha - \bar\beta = \frac{1}{2}\bar m^\nu \left(\bar m^\mu\nabla_\nu
m_\mu - m^\mu\nabla_\nu \bar m_\mu\right)\;.
\end{equation}
Using ingoing coordinates, we find
\begin{eqnarray}
\left(\alpha - \bar\beta\right)|_{r \to r_+} &=& \frac{(a^2 -
  2Mr_+)\cot\theta + iar_+\csc\theta}{\sqrt{2}r_+(r_+ -
  ia\cos\theta)^2}
\nonumber\\
&=& \frac{1}{\sqrt{2}(r_+ -
  ia\cos\theta)^2}\left[\frac{(a^2 - 2Mr_+)}{r_+}\cot\theta +
  ia\csc\theta\right]\;.
\label{eq:alpha_minus_betabar}
\end{eqnarray}

Finally, we combine Eqs.\ (\ref{eq:deltabar}) and
(\ref{eq:alpha_minus_betabar}) to build $\bar\eth$.  Assume that
$\eta$ is a function of spin-weight $s$ with an axial dependence
$e^{im\psi}$:
\begin{eqnarray}
\bar\eth\eta &=& \frac{1}{\sqrt{2}(r_+ - ia\cos\theta)}
\left[\partial_\theta - i(\csc\theta -
  a\Omega_{\rm H}\sin\theta)\partial_\psi - \frac{s}{(r_+ -
    ia\cos\theta)}\left[\frac{(a^2 - 2Mr_+)}{r_+}\cot\theta +
    ia\csc\theta\right]\right]\eta
\nonumber\\
&=& \frac{1}{\sqrt{2}(r_+ - ia\cos\theta)} \biggl[\partial_\theta +
  s\cot\theta + m\csc\theta - am\Omega_{\rm H}\sin\theta
\nonumber\\
& &
\qquad\qquad\qquad\qquad \left. - s\cot\theta - \frac{s}{(r_+ -
  ia\cos\theta)}\left[\frac{(a^2 - 2Mr_+)}{r_+}\cot\theta +
  ia\csc\theta\right]\right]\eta
\nonumber\\
&=& \frac{1}{\sqrt{2}(r_+ - ia\cos\theta)} \left[L^s_- -
  am\Omega_{\rm H}\sin\theta - \frac{s}{(r_+ -
    ia\cos\theta)}\left[\frac{(a^2 + r_+^2 - 2Mr_+ -
      iar_+\cos\theta)\cot\theta}{r_+} +
    ia\csc\theta\right]\right]\eta
\nonumber\\
&=& \frac{1}{\sqrt{2}(r_+ - ia\cos\theta)} \left[L^s_- -
  am\Omega_{\rm H}\sin\theta - \frac{s}{(r_+ -
    ia\cos\theta)}\left(ia\csc\theta -
    ia\cos\theta\cot\theta\right)\right]\eta
\nonumber\\
&=& \frac{1}{\sqrt{2}(r_+ - ia\cos\theta)} \left[L^s_- -
  am\Omega_{\rm H}\sin\theta - \frac{ias\sin\theta}{(r_+ -
    ia\cos\theta)}\right]\eta
\nonumber\\
&=& \frac{1}{\sqrt{2}r_+}
\left(1 - \frac{ia\cos\theta}{r_+}\right)^{s-1}
\left[L^s_- - am\Omega_{\rm H}\sin\theta\right]
\left(1 - \frac{ia\cos\theta}{r_+}\right)^{-s}\eta
\label{eq:bareth1}
\end{eqnarray}

\end{widetext}

In going from the first to the second equality in
Eq.\ (\ref{eq:bareth1}), we used the fact that $\eta \propto
e^{im\psi}$; we also added and subtracted $s\cot\theta$ inside the
square brackets.  In going from the second to the third equality, we
recognized that the first three terms inside the brackets are just the
operator $L^s_-$; cf.\ Eq.\ (\ref{eq:sphericalharmoniclower}).  We
also moved the negative $s\cot\theta$ term inside the second set of
square brackets.  In going from the third to the fourth equality, we
used the fact that $r_+^2 + a^2 = 2Mr_+$.  We then used $\csc\theta -
\cot\theta\cos\theta = \sin\theta$ to go from the fourth to the fifth,
and finally used Eq.\ (\ref{eq:usefulidentity}) to obtain our final
form for this operator.  This last line is identical to
Eq.\ (\ref{eq:kerr_eth}).

\section{Visualizing a distorted horizon}
\label{app:embed}

Following Hartle {\cite{hartle73,hartle74}}, we visualize distorted
horizons by embedding the two-surface of the horizon on a constant
time surface in a flat three-dimensional space.  The embedding is a
surface $r_{\rm E}(\theta,\psi)$ that has the same Ricci scalar
curvature as the distorted horizon.  For unperturbed Schwarzschild
black holes, $r_{\rm E} = 2M$; for an unperturbed Kerr hole, $r_{\rm
  E}$ is a more complicated function that varies with $\theta$.  In
the general case, we write
\begin{equation}
r_{\rm E}(\theta,\psi) = r_{\rm E}^{(0)}(\theta) + r_{\rm E}^{(1)}(\theta,\psi)\;.
\label{eq:generalembedding}
\end{equation}
In this paper, we focus on cases where the entire horizon can be
embedded in a Euclidean space, which means that we require $a/M \le
\sqrt{3}/2$.  (We briefly discuss considerations for $a/M >
\sqrt{3}/2$ at the end of this appendix.)  To generate the embedding,
we define Cartesian coordinates on the horizon as usual:
\begin{eqnarray}
X(\theta,\psi) &=& r_{\rm E}(\theta,\psi)\sin\theta\cos\psi\;,
\\
Y(\theta,\psi) &=& r_{\rm E}(\theta,\psi)\sin\theta\sin\psi\;,
\\
Z(\theta,\psi) &=& r_{\rm E}(\theta,\psi)\cos\theta\;.
\end{eqnarray}
We compute the line element
\begin{eqnarray}
ds^2 &=& dX^2 + dY^2 + dZ^2
\nonumber\\
&\equiv& g^{\rm E}_{\theta\theta}d\theta^2 + 2g^{\rm
  E}_{\theta\psi}d\theta\,d\psi + g^{\rm E}_{\psi\psi}d\psi^2\;,
\end{eqnarray}
and then the Ricci scalar corresponding to the embedding metric
$g^{\rm E}_{\alpha\beta}$ to linear order in $r_{\rm E}^{(1)}$.  We
require this to equal the scalar curvature computed using
Eq.\ (\ref{eq:gauss_kerr}), and then read off the distortion $r_{\rm
  E}^{(1)}(\theta,\psi)$.

\subsection{Schwarzschild}
\label{app:embed_schw}

Thanks to the spherical symmetry of the undistorted Schwarzschild
black hole, results for this limit are quite simple.  The metric on
an embedded surface of radius
\begin{equation}
r_{\rm E} = 2M + r_{\rm E}^{(1)}(\theta,\phi)
\end{equation}
is given by
\begin{equation}
ds^2 = (2M)^2\left[1 + \frac{r_{\rm
      E}^{(1)}(\theta,\phi)}{M}\right](d\theta^2 +
\sin^2\theta\,d\phi^2)\;.
\label{eq:schw_horizon_metric}
\end{equation}
(Recall that $\psi = \phi$ for $a = 0$.)  It is a straightforward
exercise to compute the scalar curvature associated with the metric
(\ref{eq:schw_horizon_metric}); we find
\begin{equation}
R_{\rm E} = \frac{1}{2M^2} - \left[2 +
  \frac{1}{\sin\theta}\frac{\partial}{\partial\theta}
  \left(\sin\theta\frac{\partial}{\partial\theta}\right) -
  \frac{m^2}{\sin^2\theta}\right]\frac{r_{\rm E}^{(1)}}{4M^3}\;.
\label{eq:schw_horizon_curv1}
\end{equation}
Let us expand $r_{\rm E}^{(1)}$ in spherical harmonics:
\begin{equation}
r_{\rm E}^{(1)}(\theta,\phi) = 2M\sum_{lm}\varepsilon_{lm} \,
{_0}Y_{lm}(\theta)e^{im\phi}\;.
\end{equation}
Using this, Eq.\ (\ref{eq:schw_horizon_curv1}) simplifies further:
\begin{equation}
R_{\rm E} = \frac{1}{2M^2}\left[1 + \sum_{lm}\varepsilon_{lm}
(l+2)(l-1){_0}Y_{lm}(\theta)e^{im\phi}\right]\;.
\label{eq:schw_horizon_curv}
\end{equation}

The scalar curvature we compute using black hole perturbation theory
takes the form
\begin{equation}
R_{\rm H} = R^{(0)}_{\rm H} + \sum_{lmkn}R^{(1)}_{{\rm H},lmkn}\;,
\end{equation}
where $R^{(0)}_{\rm H} = 1/2M^2$.  Equating this to $R_{\rm E}$, we
find
\begin{equation}
\varepsilon_{lm}\,{_0}Y_{lm}(\theta)e^{im\phi} =
\sum_{kn}\frac{2M^2R^{(1)}_{{\rm H},lmkn}}{(l+2)(l-1)}\;,
\end{equation}
or
\begin{equation}
r_{\rm E}^{(1)}(\theta,\phi) = \sum_{lmkn}\frac{4M^3R^{(1)}_{{\rm
      H},lmkn}}{(l+2)(l-1)}\;.
\label{eq:schw_embed_spheroid}
\end{equation}
Equation (\ref{eq:schw_embed_spheroid}) is identical (modulo a slight
change in notation) to the embedding found in Ref.\ {\cite{vpm11}};
compare their Eqs.\ (4.33) and (4.34).  We use $r_{\rm
  E}^{(1)}(\theta,\phi)$ to visualize distorted Schwarzschild black
holes in Sec.\ {\ref{sec:schw_numerical}} (dropping the indices $k$
and $n$ since we only present results for circular, equatorial orbits
in this paper).

\subsection{Kerr}
\label{app:embed_kerr}

Embedding a distorted Kerr black hole is rather more complicated.
Indeed, embedding an {\it undistorted} Kerr black hole is not trivial:
as discussed in Sec.\ {\ref{sec:geometry}}, the scalar curvature
$R_{\rm H}$ of an undistorted Kerr black hole changes sign near the
poles for spins $a/M > \sqrt{3}/2$.  A hole with this spin cannot be
embedded in a global Euclidean space, and one must instead use a
Lorentzian embedding near the poles {\cite{smarr}}.  We briefly
describe how to embed a tidally distorted black hole with $a/M >
\sqrt{3}/2$ at the end of this appendix, but defer all details to a
later paper.  For now, we focus on the comparatively simple case $a/M
\le \sqrt{3}/2$.

\subsubsection{Undistorted Kerr}

We begin by reviewing embeddings of the undistorted case.  Working in
ingoing coordinates, the metric on the horizon is given by
\begin{eqnarray}
& &ds^2 = g_{xx}\,dx^2 + g_{\psi\psi}\,d\psi^2\;,\qquad{\rm with}
\nonumber\\
& &g_{xx} = \frac{r_+^2 + a^2x^2}{1 - x^2}\;,\quad
g_{\psi\psi} = \frac{4M^2r_+^2(1 - x^2)}{r_+^2 + a^2x}\;.
\nonumber\\
\label{eq:Kerronhoriz}
\end{eqnarray}
We have introduced $x\equiv\cos\theta$.  Equation
(\ref{eq:Kerronhoriz}) is the metric on a spheroid of radius
\begin{equation}
r^{(0)}(x) = \sqrt{r_\perp(x)^2 + Z(x)^2}\;,
\label{eq:Kerr_embed}
\end{equation}
where
\begin{eqnarray}
r_\perp(x) &=& \eta \sqrt{f(x)}\;,
\label{eq:rperp_embed}
\\
Z(x) &=& \eta\int_0^x\sqrt{\frac{4 -
    (df/dx)^2}{4f(x')}}dx'\;,
\label{eq:Z_embed}
\end{eqnarray}
with
\begin{eqnarray}
f(x) &=& \frac{1 - x^2}{1 - \beta^2(1 - x^2)}\;,
\label{eq:f_embed}
\\
\eta &=& \sqrt{r_+^2 + a^2}\;,
\label{eq:eta_embed}
\\
\beta &=& a/\eta\;.
\label{eq:beta_embed}
\end{eqnarray}
Using Eqs.\ (\ref{eq:f_embed}) -- (\ref{eq:beta_embed}), we can
rewrite $Z(x)$ as
\begin{equation}
Z(x) = \int_0^x \frac{H(x')}{\left[r_+^2 + a^2
    (x')^2\right]^{3/2}}dx'\;,
\end{equation}
where
\begin{eqnarray}
H(x) &=& \bigl[r_+^8 - 6a^4 r_+^4 x^2 - 4a^6r_+^2x^2(1 + x^2)
\nonumber\\
& &\qquad - a^8x^2(1 + x^2 + x^4)\bigr]^{1/2}\;.
\label{eq:H}
\end{eqnarray}
For $a/M > \sqrt{3}/2$, $H(x) = 0$ at some value $|x| = x_{\rm E}$.
This means that $H(x)$ is imaginary for $|x| > x_{\rm E}$ for this
spin; $Z$ is imaginary over this range as well.  The horizon can be
embedded in a Euclidean space over the range $-x_{\rm E} \le x \le
x_{\rm E}$.  For all $a$, the equator ($x = 0$) is a circle of radius
$2M$.  The scalar curvature associated with this metric is
\begin{equation}
R^{(0)}_{\rm E} = R^{(0)}_{\rm H} = \frac{2}{r_+^2}\frac{(1 + a^2/r_+^2)(1 -
  3a^2x^2/r_+^2)}{(1 + a^2x^2/r_+^2)^3}\;.
\label{eq:Kerr_curvature2}
\end{equation}

\subsubsection{Distorted Kerr: $a/M \le \sqrt{3}/2$}

For this calculation, it will be convenient to use Dirac notation to
describe the dependence on $x$.  We write the spin-weighted spherical
harmonics as a ket,
\begin{equation}
{_s}Y_{lm}(x) \longrightarrow |slm\rangle \;,
\end{equation}
and define the inner product
\begin{equation}
\langle skm | f(x) | slm \rangle = 2\pi\int_{-1}^1
{_s}Y_{km}(x) f(x) {_s}Y_{lm}(x) dx\;.
\label{eq:innerprod}
\end{equation}
These harmonics are normalized so that
\begin{eqnarray}
\delta_{kl}\delta_{nm} &=& 
\int_0^{2\pi}d\psi\int_{-1}^1 dx\,{_s}Y_{kn}(x){_s}Y_{lm}(x)e^{i(m - n)\psi}
\nonumber\\
&\equiv& \delta_{nm}\langle skn|slm\rangle\;.
\end{eqnarray}
The $2\pi$ prefactor in Eq.\ (\ref{eq:innerprod}) means that
\begin{equation}
\langle skm | slm \rangle = \delta_{kl}\;.
\label{eq:spherical_orthog}
\end{equation}

Using this notation, let us now consider the curvature of a tidally
distorted Kerr black hole.  Begin with the curvature from black hole
perturbation theory, Eq.\ (\ref{eq:gauss_kerr}).  Translating into
Dirac notation, we have
\begin{eqnarray}
|R^{(1)}_{\rm H}\rangle &=& {\rm Im}\sum_{lmkn}\left[{\cal C}_{lmkn}
  Z^{\rm H}_{lmkn} e^{i\Phi_{mkn}}|\bar\eth\bar\eth
  S^+_{lmkn}\rangle\right]
\nonumber\\
&\equiv&
{\rm Im}\,R^{(1)}_{\rm H, c}\;,
\label{eq:RH_BHPT}
\end{eqnarray}
where $|\bar\eth\bar\eth S^+_{lmkn}\rangle$ is given by
Eq.\ (\ref{eq:barethbarethS}).

We now must assume a functional form for the embedding surface.  A key
issue is what basis functions we should use to describe the angular
dependence of this surface.  The basis functions for the angular
sector, $\bar\eth\bar\eth S^+_{lmkn}$, depend on mode frequency, and
so are not useful for describing the embedding surface.  Since
spherical harmonics are complete functions on the sphere, we
use
\begin{equation}
r_{\rm E}(x,\psi) = r_{\rm E}^{0}(x) + r_{\rm E}^{(1)}(x,\psi)\;,
\label{eq:rembed_kerr}
\end{equation}
where
\begin{equation}
r_{\rm E}^{(1)}(x,\psi) = r_+\sum_{\ell m}\varepsilon_{\ell m}\,{_0}Y_{\ell
  m}(x)e^{im\psi}\;,
\label{eq:r1}
\end{equation}
and where $r_{\rm E}^{0}(\theta)$ is given by
Eq.\ (\ref{eq:Kerr_embed}).  This quantity must be real, so the
expansion coefficients must satisfy the symmetry
\begin{equation}
\varepsilon_{\ell -m} = (-1)^\ell \bar\varepsilon_{\ell m}\;,
\label{eq:rE_symmetry}
\end{equation}
where as before overbar denotes complex conjugate.  Note that the
index $\ell$ used in Eq.\ (\ref{eq:r1}) is not the same as the index
$l$ used in Eq.\ (\ref{eq:RH_BHPT}).  It is important to maintain a
distinction between the indices that are used on the spheroidal and
the spherical harmonics.

\begin{widetext}

Using Eqs.\ (\ref{eq:rembed_kerr}) and (\ref{eq:r1}), we find that the
embedding surface yields a metric on the horizon given by
\begin{equation}
ds^2 = (g_{xx} + h_{xx})dx^2 + 2h_{x\psi}\,dx\,d\psi + (g_{\psi\psi} +
h_{\psi\psi})d\psi^2\;,
\label{eq:distort_Kerronhoriz}
\end{equation}
with $g_{xx}$ and $g_{\psi\psi}$ given by Eq.\ (\ref{eq:Kerronhoriz}),
and
\begin{eqnarray}
h_{xx} &=& \frac{2}{(r_+^2 + a^2x^2)^{3/2}}
\left[\left(H + \frac{4M^2r_+^2x^2}{1 - x^2}\right)r^{(1)} + \left(H -
  4M^2r_+^2\right)x\frac{\partial r^{(1)}}{\partial x}\right]\;,
\label{eq:hxx}\\
h_{x\psi} &=& \frac{(H - 4M^2r_+^2)x}{(r_+^2 + a^2
  x^2)^{3/2}}\frac{\partial r^{(1)}}{\partial\psi}\;,
\label{eq:hxpsi}\\
h_{\psi\psi} &=& 4\frac{Mr_+(1 - x^2)}
{\sqrt{r_+^2 + a^2x^2}}\,r^{(1)}\;.
\label{eq:hpsipsi}
\end{eqnarray}
The function $H = H(x)$ was introduced in the embedding of the
undistorted Kerr hole, Eq.\ (\ref{eq:H}).  By restricting ourselves to
$a/M \le \sqrt{3}/2$, we are guaranteed that $H$ is real for this
analysis.

Computing the embedding curvature from this metric, we find
\begin{equation}
R_{\rm E} = R^{(0)}_{\rm E} + R^{(1)}_{\rm E}\;,
\end{equation}
where $R^{(0)}_{\rm E}$ is given by Eq.\ (\ref{eq:Kerr_curvature2}),
and
\begin{eqnarray}
R^{(1)}_{\rm E} \longrightarrow |R^{(1)}_{\rm E}\rangle &=& \sum_{\ell
  m}\varepsilon_{\ell m}\left[C(x)|0\ell m\rangle + D(x)\,
  \frac{d}{dx} |0\ell m\rangle\right]e^{im\psi}
\nonumber\\
&\equiv& \sum_{\ell m}\varepsilon_{\ell m} e^{im\psi}{\cal E}|0\ell
m\rangle\;.
\label{eq:RH_embed}
\end{eqnarray}
We've introduced the operator ${\cal E} = C(x) + D(x)\,d/dx$ for later
notational convenience.  The functions $C(x)$ and $D(x)$ which appear
in it are given by
\begin{eqnarray}
C(x) &=& \frac{1}{2HM^2r_+(r_+^2 + a^2x^2)^{11/2}}
\left[\sum_{j = 0}^8 c_{0,j}\,a^{2j}
+ \sum_{j = 0}^5 \frac{c_{1,j}\,a^{2j}}{1 - x^2}\right]\;,
\label{eq:C}\\
D(x) &=& \frac{1}{HM(r_+^2 + a^2x^2)^{11/2}}\sum_{j = 0}^7 d_j\,a^{2j}\;,
\label{eq:D}
\end{eqnarray}
where
\begin{eqnarray}
c_{0,0} &=& 2 r_+^5 \left\{\ell(\ell + 1) H M\left[r_+^6 + 4M^2(H -
    4M^2r_+^2)x^2\right] - r_+^{11} + 2r_+^5 (r_+^6 - 4 H
  M^2)x^2\right\}\;,
\\
c_{0,1} &=& 8 \ell(\ell + 1)H^2M^3r_+^3 x^4 - r_+^{14} (6 - 23x^2 +
9x^4)\nonumber\\
& &\quad - 2HMr_+^5\left\{16 \ell(\ell + 1) M^4x^4 + 6Mr_+^3x^2(4 -
3x^2) + r_+^4\left[1 - (4 + 5\ell(\ell + 1)) x^2\right]\right\} \;,
\\
c_{0,2} &=& r_+^6 \left\{4 \left[6 + 5\ell(\ell + 1)\right] HMr_+x^4 -
  12HM^2x^2 (4 - 9x^2) - r_+^6 (6 - 63x^2 + 57x^4)\right\}\;,
\\
c_{0,3} &=& r_+^4 \left\{4HMr_+x^4\left[3 + \left[6 + 5\ell(\ell +
    1)\right]x^2\right] - 4HM^2x^2(4 - 27x^2) - r_+^6 (2 - 103x^2 +
181x^4 - 24x^6)\right\}\;,
\\
c_{0,4} &=& r_+^2 x^2 \left\{36 H M^2 x^2+2 H M r_+ x^4 \left[8 +
  \left[4 + 5 \ell(\ell + 1)\right] x^2\right] + r_+^6 (104 - 332x^2 +
67x^4 + 21x^6)\right\}\;,
\\
c_{0,5} &=& r_+ x^2 \left\{2 H M x^6 \left[3 + \ell(\ell + 1)
    x^2\right] + r_+^5 (63 - 355x^2 + 56x^4 + 62x^6 + 6x^8)\right\} \;,
\\
c_{0,6} &=& r_+^4 x^2(21 - 217x^2 + 6x^4 + 60x^6 + 18x^8) \;,
\\
c_{0,7} &=& r_+^2x^2(3 - 71x^2 - 8x^4 + 18x^6 + 18x^8)\;,
\\
c_{0,8} &=& -x^4 (10 + x^2 + x^4 - 6x^6)\;;
\end{eqnarray}
\begin{eqnarray}
c_{1,0} &=& 2 m^2 M r_+^3(2HMr_+^7 - Hr_+^8 - 4H^2M^2r_+^2x^2 +
16HM^4r_+^4x^2 - 2HMr_+^7x^2 + 8M^3r_+^9x^2)\;,
\\
c_{1,1} &=& -2 H m^2 M r_+^3 x^2\left[5r_+^6 + 4M^2(H -
  4M^2r_+^2)x^2\right]\;,
\\
c_{1,2} &=& -4m^2Mr_+^6x^4\left[H(6M + 5r_+) - 6M(H -
  4M^2r_+^2)x^2\right]\;,
\\
c_{1,3} &=& -4m^2Mr_+^4x^6\left[H(8M + 5r_+) - 8M(H -
  4M^2r_+^2)x^2\right]\;,
\\
c_{1,4} &=& -2m^2Mr_+^2x^8\left[H(6M + 5r_+) - 6M(H -
  4M^2r_+^2)x^2\right]\;,
\\
c_{1,5} &=& -2Hm^2 M r_+ x^{10}\;;
\end{eqnarray}
\begin{eqnarray}
d_0 &=& 2 r_+^8 x\left\{-2r_+^6 (1 - x^2) + HM\left[r_+ + M(6 -
  8x^2)\right]\right\}\;,
\\
d_1 &=& r_+^6x\left\{-r_+^6(8 - 17x^2 + 9x^4) + 4HM\left[6M - 2(9M -
  r_+)x^2 + 9Mx^4\right]\right\}\;,
\\
d_2 &=& -4r_+^4x\left\{r_+^6(1 - 13x^2 + 12x^4) - 3HM\left[M - 8Mx^2 +
  (6M + r_+)x^4\right]\right\}\;,
\\
d_3 &=& r_+^2 x^3\left\{r_+^6(89 - 113x^2 + 24x^4) + 4HM\left[2r_+x^4
  - M(10 - 9x^2)\right]\right\}\;,
\\
d_4 &=& 2HMr_+x^9 + r_+^6x^3(75 - 149x^2 + 53x^4 + 21x^6) \;,
\\
d_5 &=& r_+^4 x^3(30 - 114x^2 + 35x^4 + 43x^6 + 6x^8)\;,
\\
d_6 &=& r_+^2 x^3(5 - 47x^2 + 7x^4 + 23x^6 + 12x^8)\;,
\\
d_7 &=& -x^5(8 - x^2 - x^4 - 6 x^6)\;.
\end{eqnarray}
The term in $C(x)$ that is proportional to $1/(1 - x^2)$ is written so
that $C(x)$ is well behaved in the limit $x \to \pm 1$:
\begin{eqnarray}
\lim_{x \to \pm 1} \sum_{j = 0}^5 \frac{c_{1,j}\,a^{2j}}{1 - x^2} =
128a^2m^2M^6\biggl[64M^7r_+ - 16a^2M^5(3r_+ + 2M) - 8a^4M^3(r_+ - 2M)
+ a^6M(5r_+ + 6M) - a^8\biggr]\;.
\nonumber\\
\end{eqnarray}
This ensures that this function is well-behaved in all of our
numerical applications.

For small spin, the functions $C$ and $D$ become
\begin{eqnarray}
C(x) &=& \frac{(\ell + 2)(\ell - 1)}{2M^2} + \frac{\ell(\ell + 1)(5x^2
  - 2) + 2(18x^2 + m^2 - 7)}{16M^2}\left(\frac{a}{M}\right)^2
\nonumber\\
& & - \frac{498x^4 - 764x^2 + 168 - 4m^2(5x^2 + 6) - \ell(\ell + 1)
  (51x^4 - 100x^2 + 16)}{256M^2}\left(\frac{a}{M}\right)^4 + \ldots
\\
D(x) &=& \frac{5x(1 - x^2)}{8M^2}\left(\frac{a}{M}\right)^2 +
\frac{x(1 - x^2)(58 - 67x^2)}{64M^2}\left(\frac{a}{M}\right)^4 +
\ldots \;.
\end{eqnarray}

\end{widetext}

The two expressions for the deformed hole's curvature,
Eqs.\ (\ref{eq:RH_BHPT}) and (\ref{eq:RH_embed}), must equal one
another.  To facilitate comparing the two expressions, let us
introduce a complex embedding curvature, $R^{(1)}_{{\rm E, c}}$.  In
Dirac notation, we write this quantity
\begin{equation}
|R^{(1)}_{\rm E, c}\rangle = \sum_{\ell m}\varepsilon^{\rm c}_{\ell
  m}e^{im\psi}{\cal E}|0\ell m\rangle\;,
\label{eq:RH_embed_complex}
\end{equation}
introducing new coefficients $\varepsilon^{\rm c}_{\ell m}$.  We
require that
\begin{equation}
{\rm Im}\,R^{(1)}_{\rm E, c} = R^{(1)}_{\rm E}\;.
\label{eq:RH_embed_real_from_complex}
\end{equation}
Referring to Eq.\ (\ref{eq:RH_BHPT}), we see that we can enforce
$R^{(1)}_{\rm E} = R^{(1)}_{\rm H}$ by requiring
\begin{equation}
R^{(1)}_{\rm E, c} = R^{(1)}_{\rm H, c}\;.
\label{eq:complex_curvatures}
\end{equation}
As we will now show, Eq.\ (\ref{eq:complex_curvatures}) gives us a
simple expression for the complex embedding coefficients
$\varepsilon_{\ell m}^{\rm c}$.  Once those coefficients are known, it
is straightforward to extract the embedding coefficients
$\varepsilon_{\ell m}$.

Both $R^{(1)}_{\rm E, c}$ and $R^{(1)}_{\rm H, c}$ vary with $\psi$ as
$e^{im\psi}$, so we can examine them $m$-mode by $m$-mode.  To
facilitate this comparison, we break up the phase function
$\Phi_{mkn}(v,\psi)$ [Eq.\ (\ref{eq:Phi_mkn})] as
\begin{equation}
\Phi_{mkn}(v,\psi) = m\psi + \delta\Phi_m + \delta\Phi_{kn}\;,
\end{equation}
with
\begin{eqnarray}
\delta\Phi_m &=& -m\left[\Omega_\phi v + K(a)\right]\;,
\nonumber\\
\delta\Phi_{kn} &=& -(k\Omega_\theta + n\Omega_r)v\;.
\end{eqnarray}
Recall $K(a)$ is defined in Eq.\ (\ref{eq:K_of_a}).

With all of this in hand, let us now compare the two curvature
expressions for each $m$:
\begin{eqnarray}
|{_mR}^{(1)}_{\rm H,c}\rangle &=&
e^{i\delta\Phi_m}\sum_{lkn} {\cal C}_{lmkn} Z^{\rm H}_{lmkn}
e^{i\delta\Phi_{kn}}|\bar\eth\bar\eth S^+_{lmkn}\rangle\;,
\nonumber\\
|{_mR}^{(1)}_{\rm E,c}\rangle &=& \sum_{\ell}
\varepsilon^{\rm c}_{\ell m} {\cal E}|0\ell m\rangle\;.
\label{eq:embeddingcondition}
\end{eqnarray}
The sums over $l$ and $\ell$ are taken from ${\rm min}(2,|m|)$ to
$\infty$; the sums over $k$ and $n$ are both taken from $-\infty$ to
$\infty$.  Multiply these expressions by $e^{im\psi}$ and sum over $m$
from $-\infty$ to $\infty$ to recover our expressions for
$R^{(1)}_{\rm E,c}$ and $R^{(1)}_{\rm H,c}$.

Left multiply both expressions in Eq.\ (\ref{eq:embeddingcondition})
by $\langle 0qm|$.  Define the vector ${_m}\vec R^{\rm c}$ as the
object with components
\begin{equation}
{_m}R^{\rm c}_q = e^{i\delta\Phi_m}\sum_{lkn} {\cal C}_{lmkn} Z^{\rm
  H}_{lmkn} e^{i\delta\Phi_{kn}} \langle 0qm |\bar\eth\bar\eth
S^+_{lmkn}\rangle\;.
\label{eq:distortKerr_BHPT}
\end{equation}
Likewise define the embedding matrix ${_m}\mathsf{E}$ as the object
whose components are
\begin{equation}
{_m}\mathsf{E}_{q\ell} = \langle 0qm|{\cal E}|0\ell m\rangle\;.
\label{eq:distortKerr_matrix}
\end{equation}
The function $C(x)$ which appears in the operator ${\cal E}$ is an
even function of $x$, and $D(x)$ [which appears in ${\cal E}$ in the
  combination $D(x)\,d/dx$] is odd.  It follows that the only non-zero
elements of ${_m}\mathsf{E}_{q\ell}$ are those for which $q$ and
$\ell$ are either both even or both odd.

Finally, define ${_m}\vec{\boldsymbol{\varepsilon}}^{\rm c}$ as the
vector whose components are $\varepsilon^{\rm c}_{\ell m}$.  Requiring
$\langle0qm|{_m}R^{(1)}_{\rm H,c}\rangle = \langle0qm|{_m}R^{(1)}_{\rm
  E,c}\rangle$ yields the matrix equation
\begin{equation}
{_m}{\vec R}^{\rm c} =
{_m}\mathsf{E}\cdot{_m}\vec{\boldsymbol{\varepsilon}}^{\rm c}\;,
\label{eq:distortKerr_complexembed}
\end{equation}
which is easily solved:
\begin{equation}
{_m}\vec{\boldsymbol{\varepsilon}}^{\rm c} = {_m}\mathsf{E}^{-1}\cdot
{_m}{\vec R}^{\rm c}\;.
\label{eq:solve_distortKerr_complexembed}
\end{equation}

Equation (\ref{eq:solve_distortKerr_complexembed}) yields the complex
embedding coefficients $\varepsilon^{\rm c}_{\ell m}$.  From this, we
must extract the true embedding coefficients $\varepsilon_{\ell m}$
which appear in Eq.\ (\ref{eq:r1}).  We do this by considering the
symmetries of $\varepsilon^{\rm c}_{\ell m}$ and $\varepsilon_{\ell
  m}$, and by enforcing Eq.\ (\ref{eq:RH_embed_real_from_complex}).
We have already presented the symmetry of $\epsilon_{\ell m}$ in
Eq.\ (\ref{eq:rE_symmetry}).  For $\varepsilon^{\rm c}_{\ell m}$,
first write Eq.\ (\ref{eq:solve_distortKerr_complexembed}) in index
notation,
\begin{equation}
\varepsilon^{\rm c}_{\ell m} = \sum_q {_m}\mathsf{E}^{-1}_{\ell q}\,\,
           {_m}R^{\rm c}_q\;.
\end{equation}
Carefully examining their definitions and the symmetries of the
quantities which go into them, we find that
\begin{eqnarray}
{_{-m}}\mathsf{E}_{q\ell} &=& (-1)^\ell\,\,{_m}\mathsf{E}_{q\ell}\;,
\nonumber\\
{_{-m}}R^{\rm c}_q &=& -({_m}\bar R^{\rm c}_q)\;,
\label{eq:symmetries}
\end{eqnarray}
so
\begin{equation}
\varepsilon^{\rm c}_{\ell-m} = (-1)^{\ell + 1}\bar\varepsilon^{\rm
  c}_{\ell m}\;.
\end{equation}
Now enforce Eq.\ (\ref{eq:RH_embed_real_from_complex}): We
require
\begin{equation}
{\rm Im}\sum_{\ell m} \varepsilon^{\rm c}_{\ell m}\,e^{im\psi} {\cal
  E}{_0}Y_{\ell m}(x) = \sum_{\ell m} \varepsilon_{\ell
  m}\,e^{im\psi}{\cal E}{_0}Y_{\ell m}(x)\;.
\label{eq:RH_embed_real_from_complex2}
\end{equation}
The operator ${\cal E}$ acting on the spherical harmonic ${_0}Y_{\ell
  m}$ yields a real function.  With this in mind, and recalling that
the sum must be real, we see that
\begin{eqnarray}
\varepsilon_{\ell m} &=& -i\varepsilon^{\rm c}_{\ell m}\quad\qquad m \ne 0\;,
\nonumber\\
&=& -{\rm Im}\,\varepsilon^{\rm c}_{\ell m}\qquad m = 0\;.
\end{eqnarray}
We then assemble $r_{\rm E}^{(1)}(x,\psi)$ using Eq.\ (\ref{eq:r1}).

At least for the circular, equatorial orbits we have studied so far,
we find that both the vector ${_m}R^{\rm c}_q$ and the matrix
${_m}\mathsf{E}_{q\ell}$ converge quickly.  Consider first convergence
of the terms which contribute to ${_m}R^{\rm c}_q$.  Strictly
speaking, the sum over $l$ in Eq.\ (\ref{eq:distortKerr_BHPT}) goes to
infinity.  We find that this sum is dominated by the term with $q =
l$; other terms are reduced from this peak term by a factor
$\sim\epsilon^{|q - l|}$, with $\epsilon$ ranging from $0$ for
Schwarzschild (only terms with $q = l$ are non-zero in that case), to
about $0.1 - 0.2$ for orbits near the innermost stable circular orbit
for spin $a/M = \sqrt{3}/2$.  We have found that taking the sum to
$l_{\rm max} = 15$ is sufficient to ensure fractional accuracy of
about $10^{-9}$ or better in the components ${_m}R^{\rm c}_q$ for
small spins ($a \lesssim 0.4M$) for all the orbits we have considered;
we take the sums to $l_{\rm max} = 20$ or $l_{\rm max} = 25$ to
achieve this accuracy for small radius orbits at spins $a/M = 0.7$ and
$\sqrt{3}/2$.

Next consider the components of ${_m}\vec R^{\rm c}$ and
${_m}\mathsf{E}$ themselves.  Formally, we should treat both ${_m}\vec
R^{\rm c}$ and ${_m}\mathsf{E}$ as infinite dimensional objects.
However, their contributions to the tidal distortion falls off quite
rapidly as $q$ and $\ell$ become large.  We find that ${_m}R^{\rm
  c}_q$ is dominated by the $q \equiv q_{\rm peak} = {\rm max}(2,
|m|)$ component.  Components beyond this peak fall off as
$\epsilon^{|q - q_{\rm peak}|}$, with $\epsilon \sim 0.1$ across a
wide range of spins.  The matrix components ${_m}\mathsf{E}_{q\ell}$
are dominated by those with $q = \ell$, but fall off with a similar
power law form as we move away from the diagonal.  We have found
empirically that our results are accurate to about $10^{-9}$ including
terms out to $q = \ell = 15$ for small spin, but need to go as high as
$q = \ell = 25$ for large spin, strong-field orbits.

\subsubsection{Distorted Kerr: Considerations for $a/M > \sqrt{3}/2$}

The techniques described above do not work when $a/M > \sqrt{3}/2$.
For these spins, $H(x) = 0$ at $|x| = x_{\rm E}$, and is imaginary for
$|x| > x_{\rm E}$.  One way to handle this spin range would be to
introduce separate embeddings to cover the domains $|x| \le x_{\rm E}$
and $|x| > x_{\rm E}$.  Special care must be taken at the boundaries
$|x| = x_{\rm E}$, since factors of $1/H$ in the embedding curvature
$R^{(1)}_{\rm E}$ introduce singularities there.  The basis functions
used to expand the embedding function $r_{\rm E}^{(1)}(\theta,\psi)$
must be chosen so that these singularities are canceled out, leaving
the embedding curvature smooth and well behaved.  One could also
simply work in a different embedding space; work in progress indicates
that a 3-dimensional hyperbolic space $H^3$ is particularly useful,
since it can handle all black hole spins {\cite{gibbons}}.

Although straightforward to do in principle, these other embeddings do
not add substantially to the core physics we wish to present
(although, at least for some embeddings, they do add substantially to
the already rather large number of long equations in this paper).  We
defer a detailed analysis of horizon embeddings for $a/M > \sqrt{3}/2$
in a later paper.

\section{Spin-weighted spheroidal harmonics to linear order in $a/M$}
\label{app:spheroidal_lin}

In Sec.\ {\ref{sec:kerr_analytic}}, we derive analytic results for the
tidal distortion to leading order in $q \equiv a/M$, and to order
$u^5$ (where $u \equiv \sqrt{M/r}$).  As part of that analysis, we
need analytic expressions for the spin-weighted spheroidal harmonics
${_{+2}}S_{lm}$ to leading order in $q$.  We also need the eigenvalue
$\lambda$ for $s = -2$ to the same order.  Here we derive the relevant
results for arbitrary spin-weight $s$.  Similar results for $s = -2$
can be found in Refs.\ {\cite{minoetal,poisson93_2}}; much of this
approach is laid out (and intermediate steps provided) in
Ref.\ {\cite{pt73}}.

The equation governing the spin-weighted spheroidal harmonics for
spin-weight $s$ and black hole spin $a$ is
\begin{eqnarray}
&&\frac{1}{\sin\theta}\frac{d}{d\theta}\left(\sin\theta
  \frac{dS}{d\theta}\right) + \biggl[\lambda - a^2\omega^2\sin^2\theta
\nonumber\\    
&&
+ 2a\omega(m - s\cos\theta) - \frac{(m + s\cos\theta)^2}{\sin^2\theta}
+ s\biggr]{_s}S_{lm}(\theta) = 0\;.
\nonumber\\
\label{eq:spheroidal_gen}
\end{eqnarray}
The parameter $\lambda$ appearing here is one form of the eigenvalue
for this equation; another common form is ${\cal E} = \lambda +
2am\omega - a^2\omega^2 + s(s+1)$; still another (which appears in at
least one of Teukolsky's original papers {\cite{teuk73}}) is $A =
{\cal E} - s(s+1)$.  We write both $\lambda$ and the harmonic as
expansions in $a\omega$:
\begin{eqnarray}
\lambda &=& \lambda_0 + (a\omega)\lambda_1\;,
\\
{_s}S_{lm}(\theta) &=& {_{s}Y}_{lm}(\theta) + (a\omega)
{_s}S^{1}_{lm}(\theta)\;.
\end{eqnarray}
This could be taken to higher order (for example,
Ref.\ {\cite{minoetal}} does so to $O(a^2\omega^2)$ for $s = -2$), but
linear order is enough for our purposes.

Begin by defining the operator
\begin{equation}
{\cal L}_0 \equiv
\frac{1}{\sin\theta}\frac{d}{d\theta}\left(\sin\theta
\frac{d}{d\theta}\right) + \left[s - \frac{(m +
    s\cos\theta)^2}{\sin^2\theta}\right]\;.
\end{equation}
Equation (\ref{eq:spheroidal_gen}) can then be decomposed order by
order, becoming
\begin{eqnarray}
&&\left({\cal L}_0 + \lambda_0\right){_sY}_{lm} = 0\;,
\label{eq:spheroidal_0}\\
&&\left({\cal L}_0 + \lambda_0\right){_s}S^{1}_{lm} =
\left(2s\cos\theta - 2m - \lambda_1\right){_s}Y_{lm}\;.
\label{eq:spheroidal_1}
\end{eqnarray}
Equation (\ref{eq:spheroidal_0}) tells us that
\begin{equation}
\lambda_0 = (l - s)(l + s + 1)\;.
\label{eq:lambda0}
\end{equation}
Multiply Eq.\ (\ref{eq:spheroidal_1}) by $2\pi{_s}Y_{lm}\sin\theta$
and integrate both sides with respect to $\theta$ from $0$ to $\pi$.
Integrating by parts, using Eqs.\ (\ref{eq:spherical_orthog}) and
(\ref{eq:spheroidal_0}) and the fact that
\begin{equation}
2\pi\int_0^\pi {_s}Y_{lm}(\theta)\cos\theta\,{_s}Y_{lm}(\theta)
\sin\theta\,d\theta = -\frac{sm}{l(l+1)}\;,
\end{equation}
we find
\begin{equation}
\lambda_1 = -2m\left[1 + \frac{s^2}{l(l+1)}\right]\;.
\label{eq:lambda1}
\end{equation}

To compute $_sS^1_{lm}$, put
\begin{equation}
_sS^1_{lm} = \sum_{l' = {\rm min}(|s|,|m|)}^\infty c^{l'}_{lm}\,{_s}Y_{l'm}\;.
\end{equation}
Inserting this into Eq.\ (\ref{eq:spheroidal_1}), multiplying by
$2\pi\,{_s}Y_{l'm}\sin\theta$ and integrating, we find
\begin{eqnarray}
c^{l'}_{lm} &=& \frac{4\pi s}{\lambda_0(l) - \lambda_0(l')}
\int_0^\pi {_s}Y_{l'm}(\theta)\cos\theta\,{_s}Y_{lm}(\theta)
\sin\theta\,d\theta
\nonumber\\
& &\qquad\qquad\qquad\qquad\qquad\qquad (l' \ne l)\;,
\\
&=& 0\qquad\qquad\qquad\qquad\qquad\quad\;\; (l' = l)\;.
\end{eqnarray}
Using the fact that this integral can be expressed using
Clebsch-Gordan coefficients, we see that $c^{l'}_{lm}$ is non-zero
only for $l' = l \pm 1$.  We find
\begin{eqnarray}
c^{l+1}_{lm} &=& -\frac{s}{(l + 1)^2}\times
\nonumber\\
& &\sqrt{\frac{(l + s + 1)(l - s + 1)(l + m +1)(l - m + 1)}{(2l +
    3)(2l + 1)}}\;,
\nonumber\\
\label{eq:clp1lm}\\
c^{l-1}_{lm} &=&
\frac{s}{l^2}\sqrt{\frac{(l+s)(l-s)(l+m)(l-m)}{(2l+1)(2l-1)}}\;.
\label{eq:clm1lm}
\end{eqnarray}
For $s = -2$, these reproduce the values given in Appendix A of
Ref.\ {\cite{minoetal}}.

\section{Glossary of notation changes}
\label{app:glossary}

Previous work by one of the present authors and various collaborators
(e.g., Ref.\ {\cite{dh2006}}) has used notation for various quantities
related to the Teukolsky equation and its solutions which differs from
that used by Fujita and Tagoshi and their collaborators
{\cite{st03,ft04,ft05}}.  We have recently switched our core numerical
engine to one that is based on the Fujita-Tagoshi method, and as such
have found it to be much more convenient to follow their conventions
in our work.

Begin by examining how Eqs.\ (\ref{eq:Rinr+})--(\ref{eq:Rupinf})
appear in the previous notation:
\begin{eqnarray}
R^{\rm H}_{lm\omega}(r \to r_+) &=& B^{\rm hole}_{lm\omega}\Delta^2 e^{-ipr^*}\;,
\\
R^{\rm H}_{lm\omega}(r \to \infty) &=& B^{\rm out}_{lm\omega}r^3e^{i\omega r^*}
+ \frac{B^{\rm in}_{lm\omega}}{r}e^{-i\omega r^*}\;,
\nonumber\\
\\
R^{\rm \infty}_{lm\omega}(r \to r_+) &=& D^{\rm out}_{lm\omega}e^{ip r^*}
+ {D^{\rm in}_{lm\omega}}\Delta^2e^{-ip r^*}\;,
\nonumber\\
\\
R^{\rm \infty}_{lm\omega}(r \to \infty) &=& D^{\rm
  \infty}_{lm\omega}r^3 e^{i\omega r^*}\;.
\end{eqnarray}
[These are Eqs.\ (3.15a--d) in Ref.\ {\cite{dh2006}}.]  As discussed
in Sec.\ {\ref{sec:ZH}}, we use these homogeneous solutions to
assemble a Green's function, and then define a general solution
\begin{equation}
R_{lm\omega}(r) = Z^{\rm H}_{lm\omega}(r)R^\infty_{lm\omega}(r) +
Z^\infty_{lm\omega}(r)R^{\rm H}_{lm\omega}(r)\;,
\end{equation}
where
\begin{eqnarray}
Z^{\rm H}_{lm\omega}(r) &=& \frac{1}{\cal W} \int_{r_+}^r
\frac{R^{\rm H}_{lm\omega}(r'){\cal T}_{lm\omega}(r')}{\Delta(r')^2}dr'\;,
\label{eq:OldZH1}
\\
Z^\infty_{lm\omega}(r) &=& \frac{1}{\cal W} \int_{r}^\infty
\frac{R^\infty_{lm\omega}(r'){\cal T}_{lm\omega}(r')}{\Delta(r')^2}dr'\;,
\label{eq:OldZInf2}
\end{eqnarray}
where ${\cal W}$ is the Wronskian associated with $R^{\rm
  H}_{lm\omega}$, $R^\infty_{lm\omega}$.  We then define
\begin{eqnarray}
Z^{\rm H}_{lm\omega} \equiv Z^{\rm H}_{lm\omega}(r \to \infty)\;,
\label{eq:OldZH}
\\
Z^\infty_{lm\omega} \equiv Z^\infty_{lm\omega}(r \to r_+)\;.
\label{eq:OldZInf}
\end{eqnarray}
These amplitudes define the fluxes of energy and angular momentum into
the black hole's event horizon and carried to infinity.
Unfortunately, they have the rather annoying property that their
connection to these fluxes is ``backwards'': $Z^{\rm H}_{lm\omega}$
encodes information about the fluxes at infinity, and
$Z^\infty_{lm\omega}$ encodes fluxes on the horizon.  Although the
labels defined by Eqs.\ (\ref{eq:OldZH}) and (\ref{eq:OldZInf}) follow
logically from their connection to the homogeneous solutions $R^{\rm
  H}_{lm\omega}$ and $R^\infty_{lm\omega}$, they connect rather
illogically to the fluxes that they ultimately encode.

To switch to the notation that is used in
Refs.\ {\cite{st03,ft04,ft05}}, we rename various functions and
coefficients.  For the fields that are regular on the horizon, we put
\begin{eqnarray}
R^{\rm H}_{lm\omega} &\to& R^{\rm in}_{lm\omega}\;,
\\
B^{\rm hole}_{lm\omega} &\to& B^{\rm trans}_{lm\omega}\;,
\\
B^{\rm out}_{lm\omega} &\to& B^{\rm ref}_{lm\omega}\;,
\\
B^{\rm in}_{lm\omega} &\to& B^{\rm inc}_{lm\omega}\;;
\end{eqnarray}
and for fields that are regular at infinity,
\begin{eqnarray}
R^{\rm \infty}_{lm\omega} &\to& R^{\rm up}_{lm\omega}\;,
\\
D^{\rm \infty}_{lm\omega} &\to& C^{\rm trans}_{lm\omega}\;,
\\
D^{\rm out}_{lm\omega} &\to& C^{\rm up}_{lm\omega}\;,
\\
D^{\rm in}_{lm\omega} &\to& C^{\rm ref}_{lm\omega}\;.
\end{eqnarray}
The general solution which follows from this is our
Eq.\ (\ref{eq:inhomogeneous}), with functions $Z^{\rm
  in}_{lm\omega}(r)$ and $Z^{\rm up}_{lm\omega}(r)$ defined in
Eqs.\ (\ref{eq:Zin1}) and (\ref{eq:Zup1}).  As described in
Sec.\ {\ref{sec:ZH}}, we then define
\begin{eqnarray}
Z^{\rm H}_{lm\omega} &\equiv& Z^{\rm up}_{lm\omega}(r \to r_+)\;,
\\
Z^\infty_{lm\omega} &\equiv& Z^{\rm in}_{lm\omega}(r \to \infty)\;.
\end{eqnarray}
This definition reverses the labels that were introduced in
Eqs.\ (\ref{eq:OldZH}) and (\ref{eq:OldZInf}), so that fluxes on the
horizon are encoded by $Z^{\rm H}_{lm\omega}$, and those to infinity
by $Z^\infty_{lm\omega}$.  It is also in accord with the notation used
in Refs.\ {\cite{st03,ft04,ft05}}.

\end{document}